\documentclass[useAMS,usenatbib]{mn2e}
\usepackage{subfigure,amsmath,amssymb}
\usepackage{graphicx,aas_macros}
\DeclareGraphicsExtensions{.eps}
\DeclareMathOperator{\erf}{erf}

\title[Substructure in 2D in simulated galaxy clusters]{The relationship between substructure in 2D X-ray surface brightness images and weak lensing mass maps of galaxy clusters: A simulation study}

\author[Leila C. Powell et al.]
{Leila C. Powell$^{1}$\thanks{E-mail:lcp@astro.ox.ac.uk}, Scott T. Kay$^{2}$ and Arif Babul$^{3}$ \\
$^{1}$Oxford Astrophysics, Denys Wilkinson Building, Keble Road, Oxford, OX1 3RH, UK\\
$^{2}$Jodrell Bank Centre for Astrophysics, Alan Turing Building, The University of Manchester, Manchester, M13 9PL, UK\\
$^{3}$Department of Physics and Astronomy, University of Victoria, Elliot Building, 3800 Finnerty Road, Victoria, BC, V8P 5C2, Canada}

\begin{document}


\pagerange{000--000} \pubyear{0000}

\maketitle

\label{firstpage}

\begin{abstract}

Recent X-ray and weak-lensing observations of galaxy clusters have
revealed that the hot gas does not always directly trace the dark matter
within these systems.  Such configurations are extremely interesting. They offer a new vista onto the complex interplay between gravity and
baryonic  physics, and may even be used as indicators of the clusters' dynamical state. In this paper, we undertake a study to determine what insight can be reliably gleaned from the comparison of the X-ray and the weak lensing mass maps of galaxy clusters.  We do this by investigating the 2D substructure within three high-resolution cosmological simulations of galaxy clusters.
Our main results focus on non-radiative gas dynamics, but we also
consider the effects of radiative cooling at high redshift. For our
analysis, we use a novel approach, based on unsharp-masking, to
identify substructures in 2D surface mass density and X-ray surface
brightness maps. At full resolution ($\sim 15 h^{-1}$ kpc), this technique
is capable of identifying almost all self-bound dark matter
subhaloes with $M>10^{12}h^{-1}M_{\odot}$. We also report a correlation
between the mass of a subhalo and the area of its corresponding 2D
detection; such a correlation, once calibrated, could provide a useful
estimator for substructure mass. Comparing our 2D mass and X-ray
substructures, we find a surprising number of cases where the
matching fails: around one third of galaxy-sized substructures have
no X-ray counterpart. Some interesting cases are also found at larger
masses, in particular the cores of merging clusters where the
situation can be complex. Finally,  we degrade our mass maps to what
is currently achievable with weak-lensing observations ($\sim 100
h^{-1}$kpc at $z=0.2$). While the completeness mass limit
increases by around an order of magnitude, a mass-area correlation
remains. Our paper clearly demonstrates that the next generation of lensing surveys should start to reveal a wealth of information
on cluster substructure.
\end{abstract}
 

\section{Introduction}\label{intro}

The advent of the weak lensing technique has allowed observers to directly probe the distribution of mass in galaxy clusters, rather than simply assuming that light provides an accurate tracer of the underlying dark matter (DM) distribution. This allows us to separate out shortfalls in our understanding of baryonic physics from direct challenges to the Cold Dark Matter (CDM) model, providing an excellent test of the predictions of the CDM paradigm itself and a clearer picture of the influence of the baryonic component. 

In recent years, there has been a flurry of papers, with these goals in mind, which compare weak lensing mass reconstructions to X-ray images of galaxy clusters \citep[e.g.][]{smail97:apj,machacek:apj,smithhst:mnras}. Some such observations have highlighted dramatic exceptions to the basic picture that light follows mass, most famously, the bullet cluster \citep{bulletclowe} where the main peaks in the X-ray image are offset from those in the weak lensing mass reconstruction. There have been several follow-up theoretical studies of this unique system \citep[for example,][]{2006PASJ...58..925T, 2007MNRAS.380..911S,  2007arXiv0711.0967M} which conclude that its main features can be reproduced well by idealized, non-radiative merger simulations suggesting the driving factor is ram-pressure. 

There have also been observations of clusters with features in the weak lensing map which are absent in the X-ray image. For example, in MS1054-0321  \citep{jee:apj}  and in Abell 1942 \citep{2000A&A...355...23E, 2001MNRAS.325..111G},  for which several theories are put forward: chance alignments of background galaxies, galaxy clusters that have not yet virialized and so possess little X-ray emitting gas or substructure within the cluster that has somehow been stripped of its gas. Even more puzzling is the recent observation of Abell 520 \citep{2007ApJ...668..806M}, in which an X-ray peak with no corresponding mass concentration and a mass concentration with no galaxies are detected. This is postulated to be a result of either a multiple body interaction, or the collision of weakly self-interacting DM during the merger event. Most recently is the observation of another extreme merger event \citep{bullet2}, in which two clusters with M$\sim 10^{14}{\rm M}_{\odot}$ are both displaced from the single peak in the X-ray emission suggesting even higher mass substructures could be seen to be `dark'. 

On the galaxy-mass scale, studies of X-ray observations of the hot haloes of elliptical galaxies \citep{2006ApJ...644..155M} exhibiting features characteristic of ram pressure stripping were carried out, suggesting we should expect to find galaxy-sized subhaloes that are dark in X-rays. However, a systematic study by \citet{2007ApJ...657..197S} found $60$ per cent of galaxies brighter than 2$L_{*}$ still retained small X-ray coronae, potentially indicating a more complex picture than just hydrodynamics, involving the suppression of heat conduction and viscosity by magnetic fields.

There have been many theoretical studies with the aim of understanding the global properties of purely DM substructure. For example, the systematics \citep{gao:apj}, evolution \citep{ 2004MNRAS.351..410G,2005MNRAS.359.1537R}, effects of the parent halo merger history \citep{taybabul:mnras} and spatial distribution \citep{2004MNRAS.352..535D}  of subhalo populations have all been studied in great depth. Attention is now also being paid to the fate of the gas in subhaloes. \citet{2006ApJ...647..910H} incorporated a hot halo component into an analytical model of ram pressure stripping of galaxies in groups and clusters and found that most galaxies were readily stripped of the majority of this. Inspired by the first observations of cold fronts in {\it Chandra} data \citep[e.g.][]{2000ApJ...541..542M} some authors invoked separations between the hot gas and DM of either the main cluster \citep{2006ApJ...650..102A} or a merging subcluster \citep{2005ApJ...629..791T} as a possible mechanism for their production. There were also many other complimentary studies into the fate of gas in subhaloes on the group or cluster mass scale. For example,  \citet{2002ApJ...578L...9B} report the ablation of gas away from the core of a merging subcluster's DM potential, in a cosmological simulation, resulting in adiabatic cooling and \citet{2003MNRAS.346...13H} use idealized merger simulations to study this process in more detail. More recently \citet{clusmergers_poole06} performed a suite of idealized cluster mergers and found gas in the both cores was often disrupted, leading to additional transient structures in the X-ray emitting gas.  In order to specifically investigate the fate of hot gas in galaxies orbiting in groups and clusters \citet{2008MNRAS.383..593M} studied a suite of hydrodynamic simulations. They find the majority of the hot gas is stripped within a few gigayears but that around 30 per cent is retained even after 10 gigayears.

Much of this work on the gaseous component, uses simulations of idealized mergers in order to reproduce specific observational features of galaxy cluster substructure. What is required now is a similar treatment to that afforded for DM subhaloes; a systematic study of the statistics of hot gas substructure in fully cosmological simulations. Indeed there have only been two studies of this kind already, \citep{2004MNRAS.350.1397T,newsubfind}; The former focusses on the time evolution of subhaloes in non-radiative simulations, while the latter examines how the overall distribution of subhalo masses and compositions differ, depending on the physics incorporated. There are two main issues still to address. Firstly, many  of the interesting substructures seen in X-ray images of clusters (tidal tails, diffuse gas clouds etc) are omitted from simulation studies which simply identify substructure as hot gas bound to subhaloes. Secondly, observationally we can only view the substructure in projection; how does this relate to the substructure in 3D? Both of these issues can be addressed by undertaking an analysis of galaxy cluster substructure in 2D, allowing projection effects to be quantified without restricting the analysis to the bound components. A comprehensive study in this area will help us to construct a framework within which to interpret the surprising results from comparisons between weak lensing and X-ray observations, of which there will undoubtedly be many more in the near future.

In this paper, we use high resolution resimulations of three galaxy clusters to compare the substructure in the hot gas and DM components and examine what factors affect their similarity, or otherwise. We use a technique based on unsharp-masking to identify enhancements to the cluster background in maps of the X-ray surface brightness and total surface mass density, providing us with catalogues of 2D substructures. Our aims are to understand the relationship between 3D DM subhaloes and our 2D total mass substructure catalogues, including the contribution of 3D subhaloes that lie infront of or behind the cluster, yet within the map region. We wish to understand how these 2D mass sources then relate to substructures in the projected X-ray surface brightness, in order that we may place some constraints on the frequency of mismatches between substructure in the hot gas and DM and the mass scales at which these occur. Finally, we investigate how various selection and model parameters influence these two relationships.

The paper is structured as follows. In Section \ref{simsec}, the simulation properties, selection of the cluster sample and generation of the maps are outlined. The detection technique and properties of our 3D subhalo catalogues are included in Section \ref{detectsec}, while Section \ref{2dsec} provides the same information for our 2D substructure catalogues.  In Section \ref{2d3dsec}, the results of a direct comparison between the 2D mass map substructures and the 3D subhaloes are presented. We investigate the likelihood of finding a 2D X-ray counterpart for each 2D mass substructure in Section \ref{massgassec} and explore the effect that redshift, dynamical state, the inclusion of cooling and observational noise have on this in Section \ref{paramsec}. Section \ref{paramsec} also includes several case studies, to illustrate in more detail the fate of a 2D mass substructure's hot gas component when a 2D X-ray counterpart cannot be found. We provide a short summary of our results at the end of Sections 5, 6 and 7, should the reader wish to skip to the end of these sections. Finally, Section \ref{conclusionsec} outlines the main conclusions and implications of this work.

\section[]{Cluster Simulations}\label{simsec}

We use the resimulation technique to study the clusters with high 
resolution. Three clusters were selected from the larger sample 
studied by \citet{gao:apj} and resimulated with gas using the 
publicly-available {\sc gadget2} $N$-body/SPH code \citep{gadget2}. A
$\Lambda$CDM cosmological model was assumed, adopting the following values
for key cosmological parameters: 
$\Omega_{\rm m}=0.3, \Omega_{\Lambda}=0.7, \Omega_{\rm b}=0.045, h=0.7, 
\sigma_8=0.9$. The DM and gas particle masses in the high-resolution 
regions were set to $m_{\rm dark}=4.3\times 10^{8} \, h^{-1} {\rm M}_{\odot}$
and $m_{\rm gas}=7.7 \times 10^{7} \, h^{-1} {\rm M}_{\odot}$ respectively, 
within a comoving box-size of $479 h^{-1}{\rm Mpc}$. The simulations
were evolved from $z=49$ to $z=0$, outputting 50 snapshots equally 
spaced in time. The Plummer gravitational softening length was fixed at 
$\epsilon=10 \, h^{-1} {\rm kpc}$ in the comoving frame until $z=1$, after 
which its proper length ($\epsilon=5 \, h^{-1} {\rm kpc}$) was fixed. 

For our main results, we have chosen not to incorporate the complicating 
effects of non-gravitational physics (particularly radiative cooling and 
heating from galaxies), for two reasons. Firstly, we wish to investigate 
any differences between the hot gas and DM in the simplest 
scenario, i.e. due to ram-pressure stripping and viscous heating of the gas. 
Secondly, a model that successfully reproduces the observed X-ray 
properties of galaxy clusters in detail does not yet exist, and so only 
phenomenological heating models tend to be implemented in cluster simulations. 
Nevertheless, we include a limited analysis of the effects of non-gravitational
physics on our results, namely allowing the gas to cool radiatively at high 
redshift, in Section~\ref{paramsec}. We defer a study of the additional 
effects of heating from stars and active galactic nuclei to future work.

\subsection[]{Cluster identification and general properties}

\begin{figure*}
\centering
\includegraphics[width=0.33\textwidth]{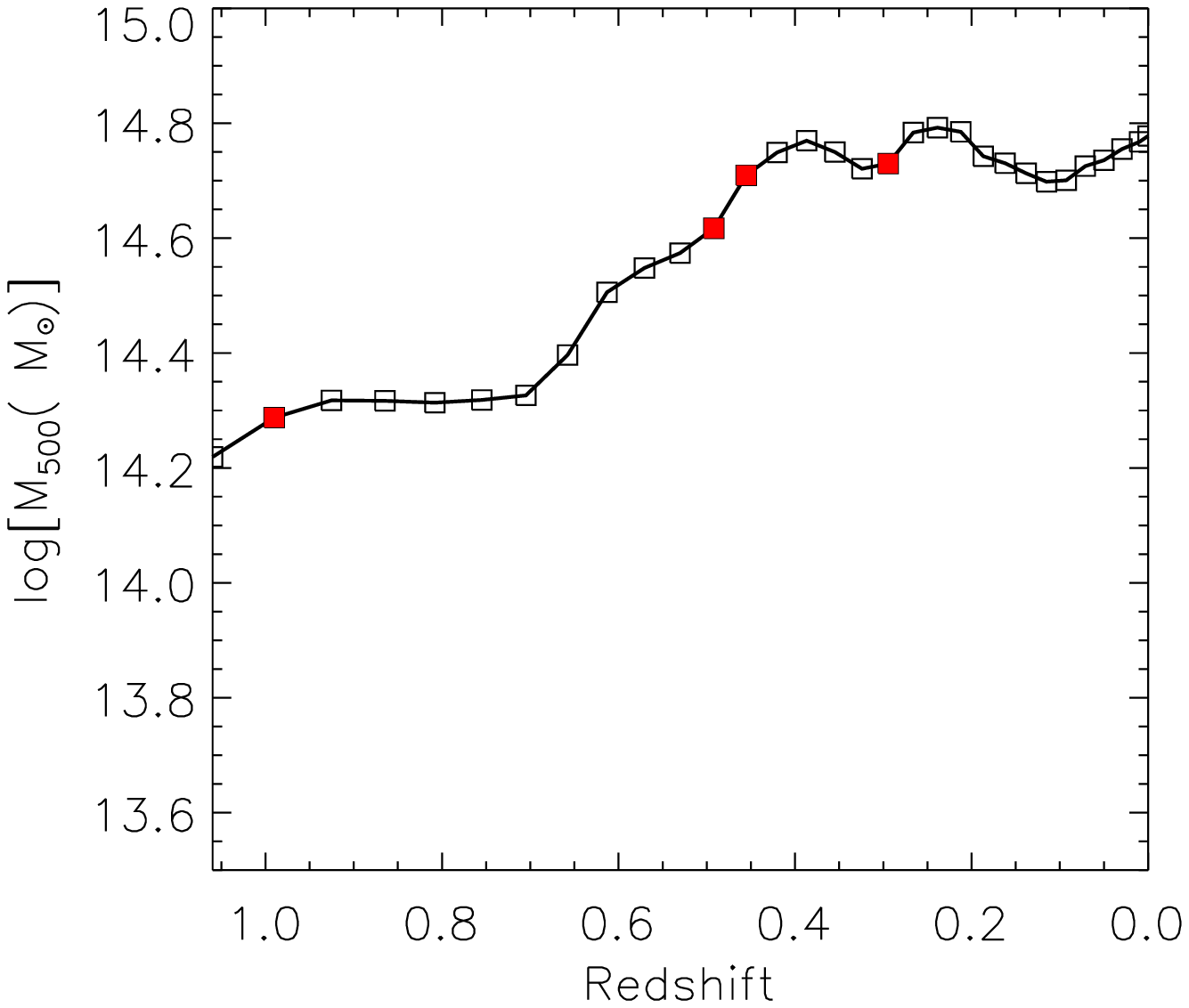}
\includegraphics[width=0.33\textwidth]{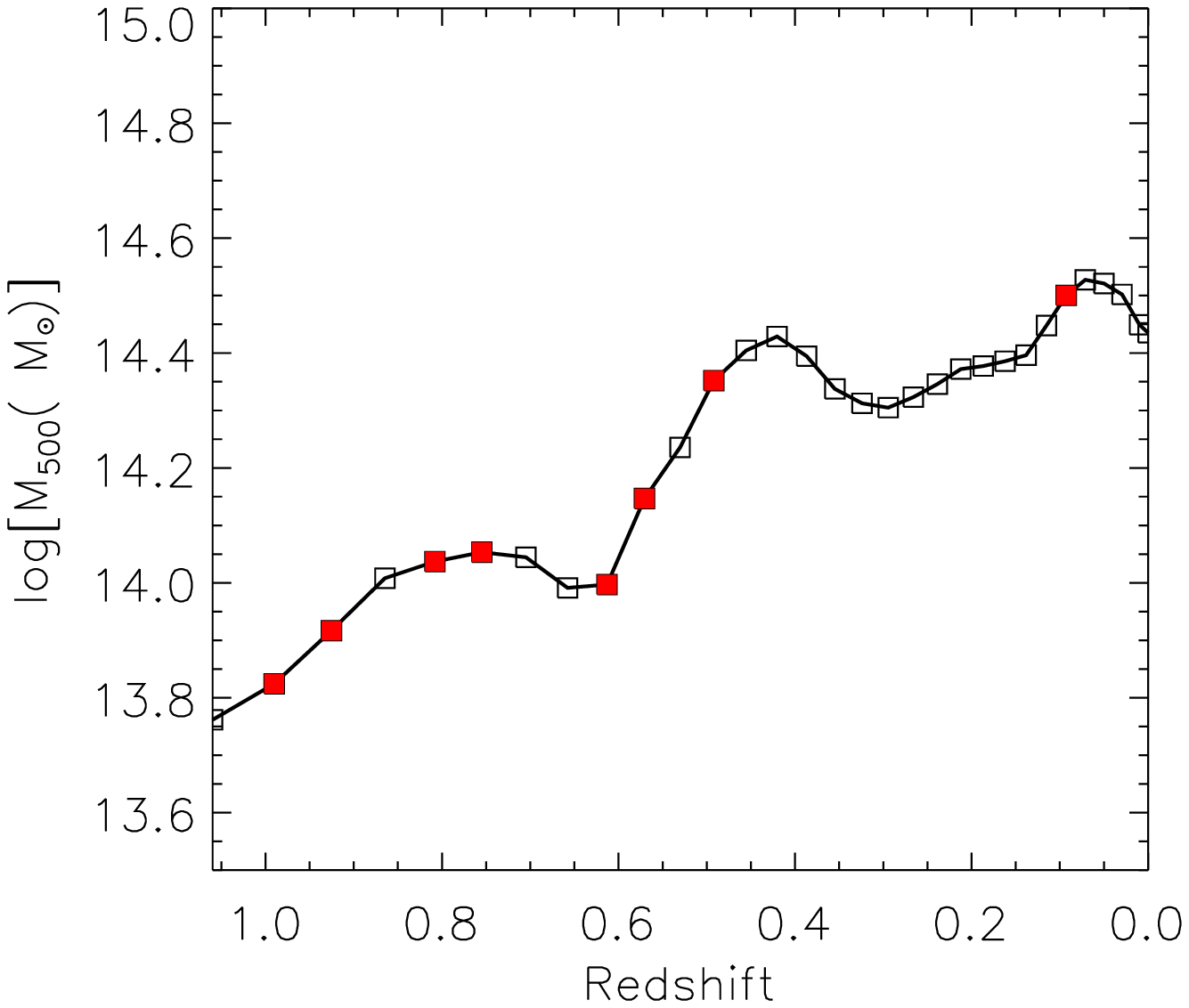}
\includegraphics[width=0.33\textwidth]{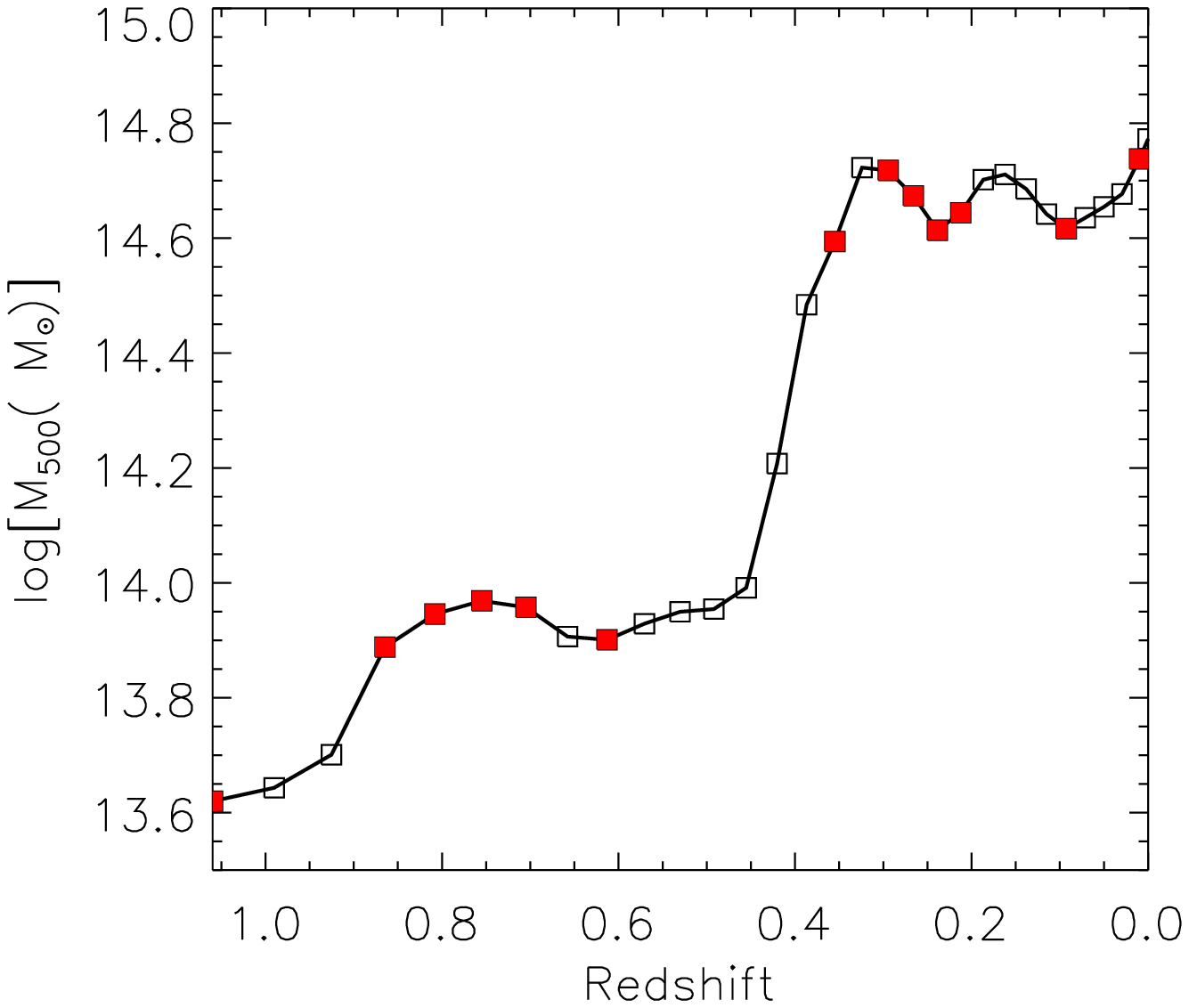}

\includegraphics[width=0.33\textwidth]{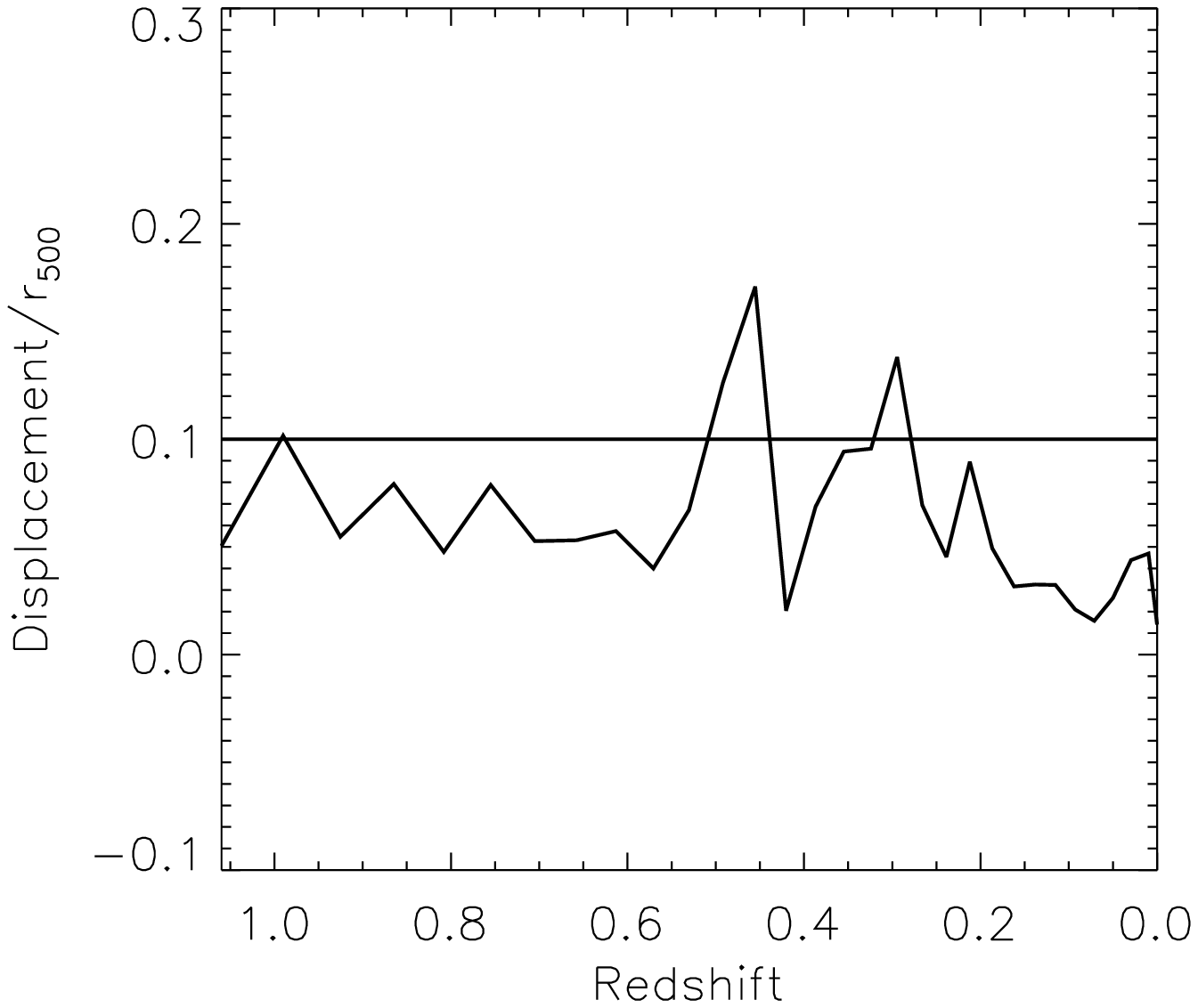}
\includegraphics[width=0.33\textwidth]{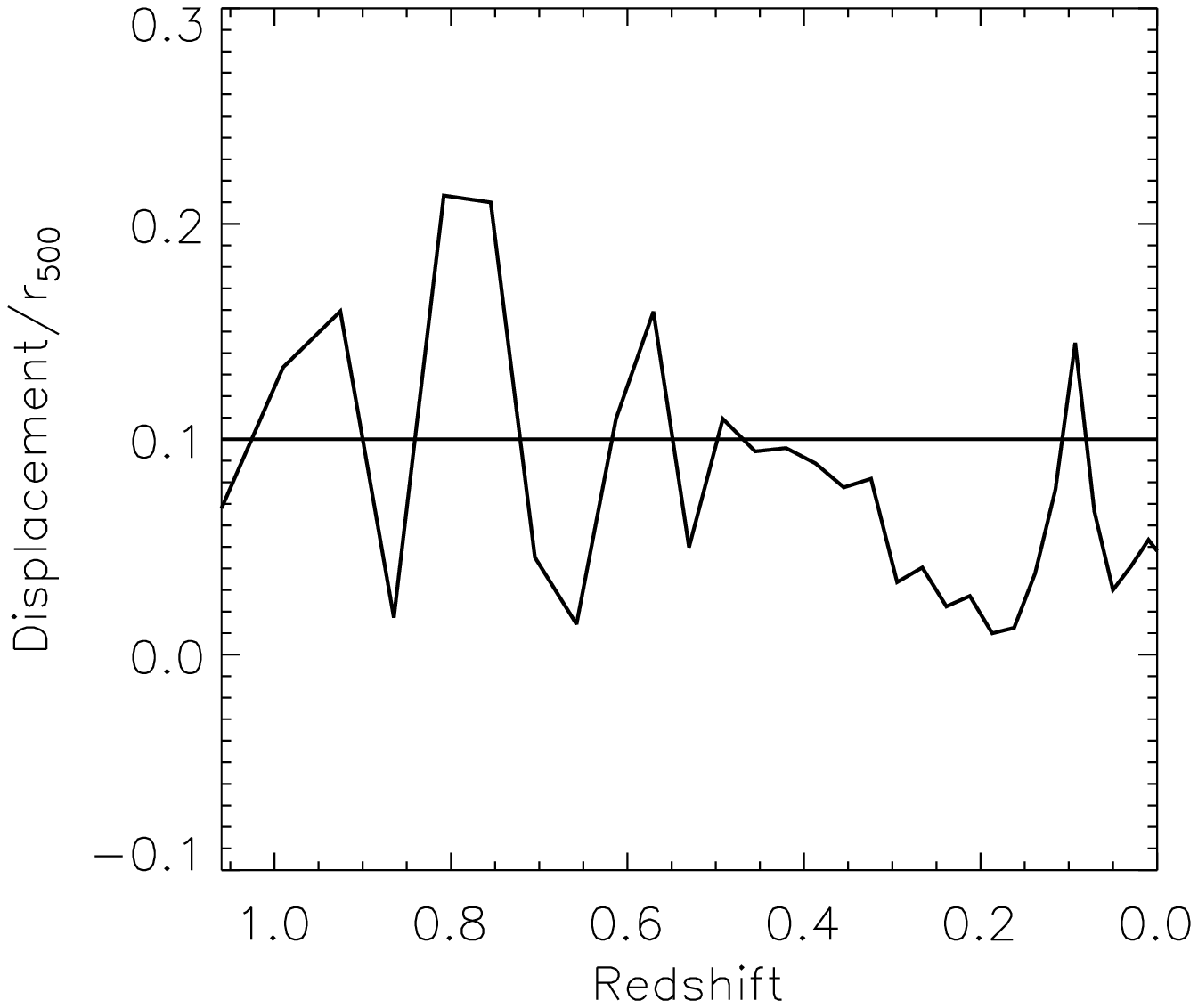}
\includegraphics[width=0.33\textwidth]{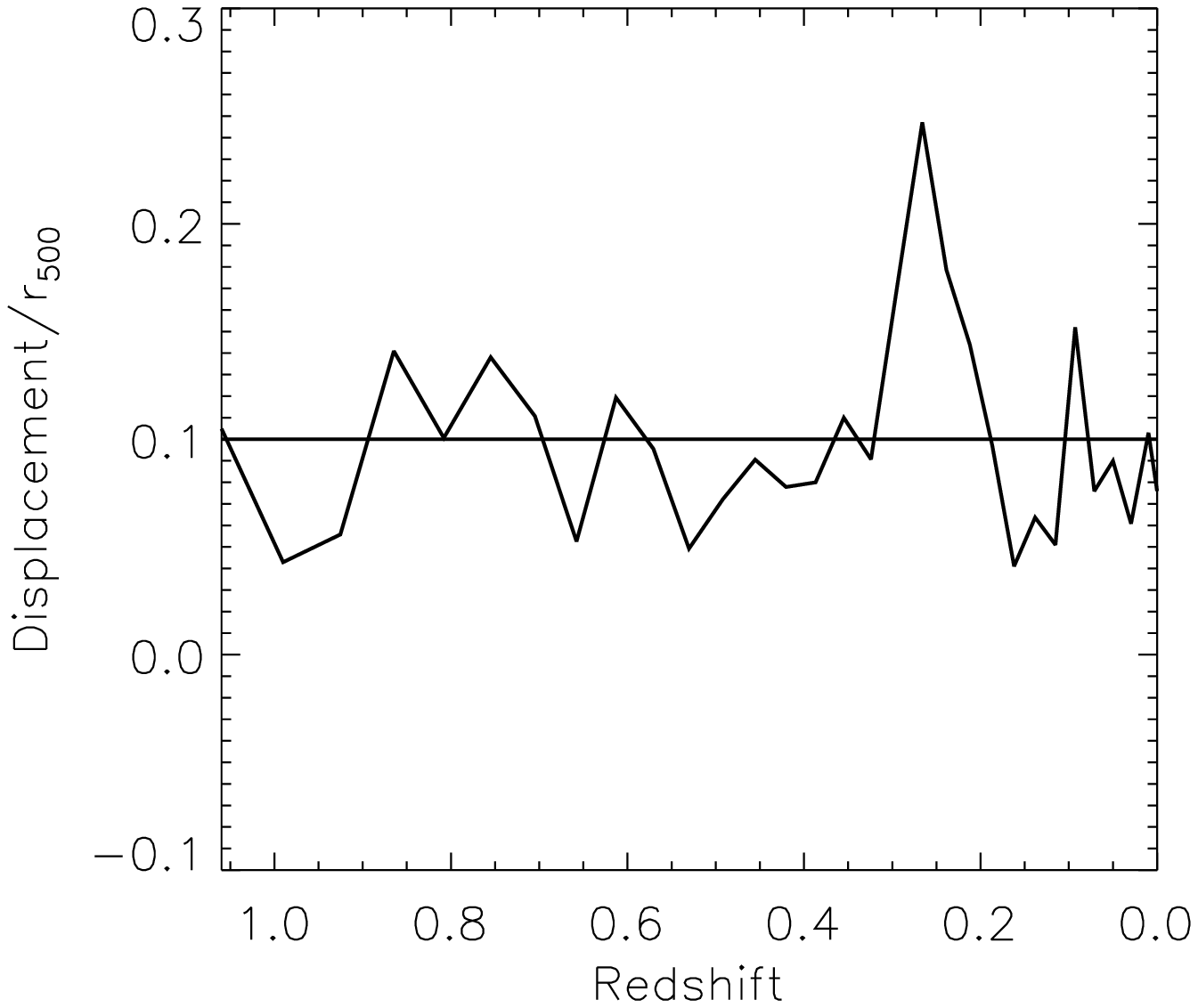}
\\

\includegraphics[width=0.3\textwidth]{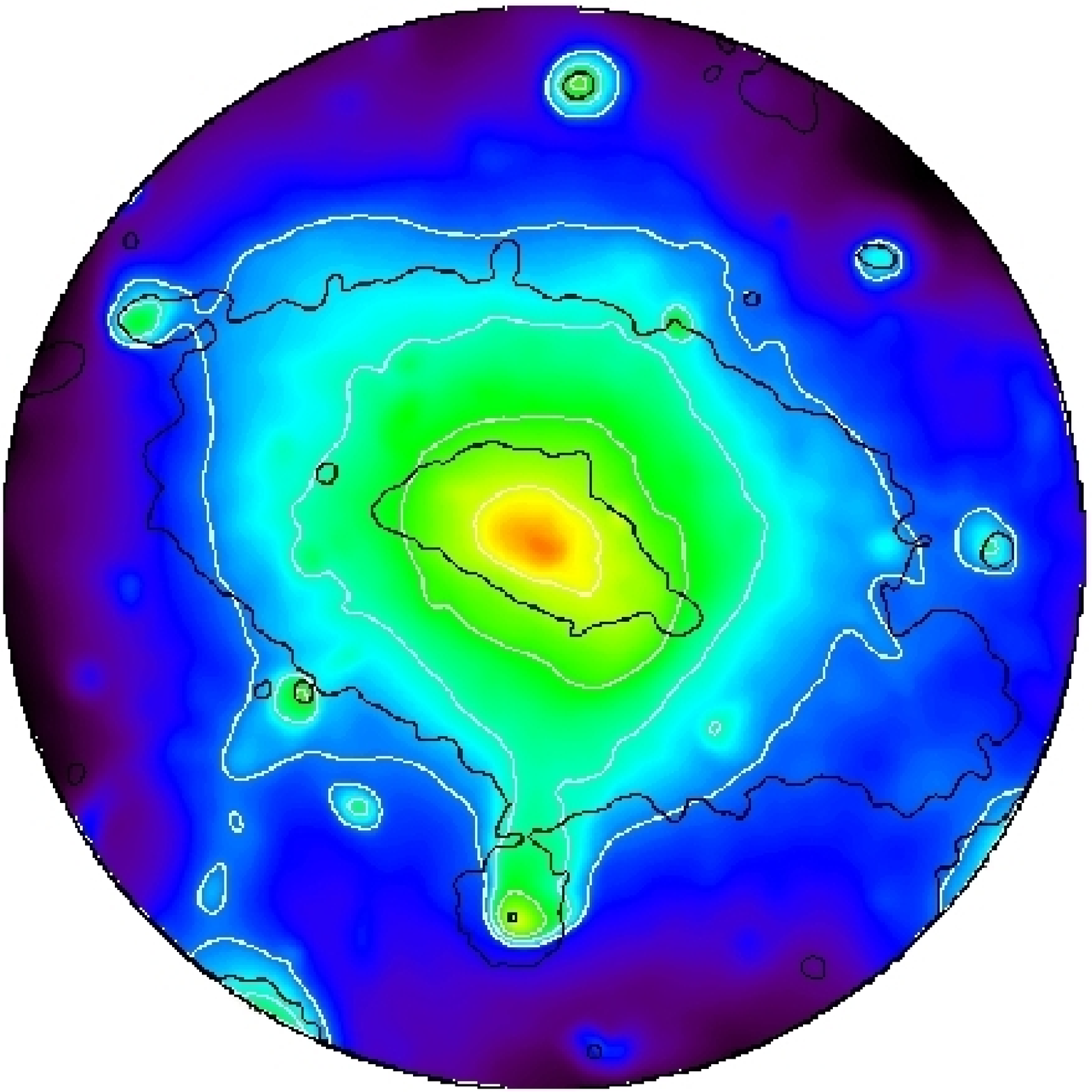}
\hspace{0.04\textwidth}
\includegraphics[width=0.3\textwidth]{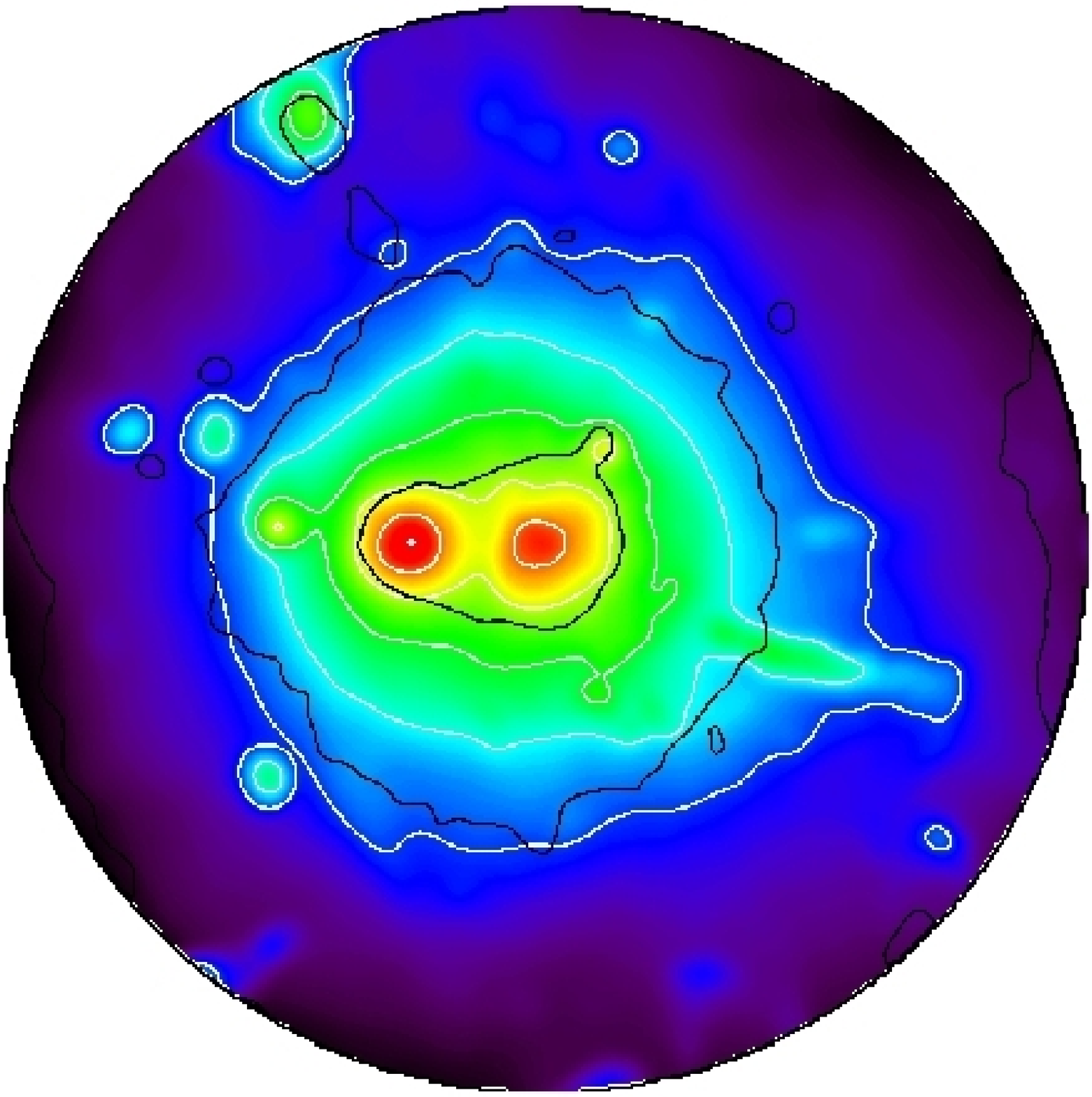}
\hspace{0.04\textwidth}
\includegraphics[width=0.3\textwidth]{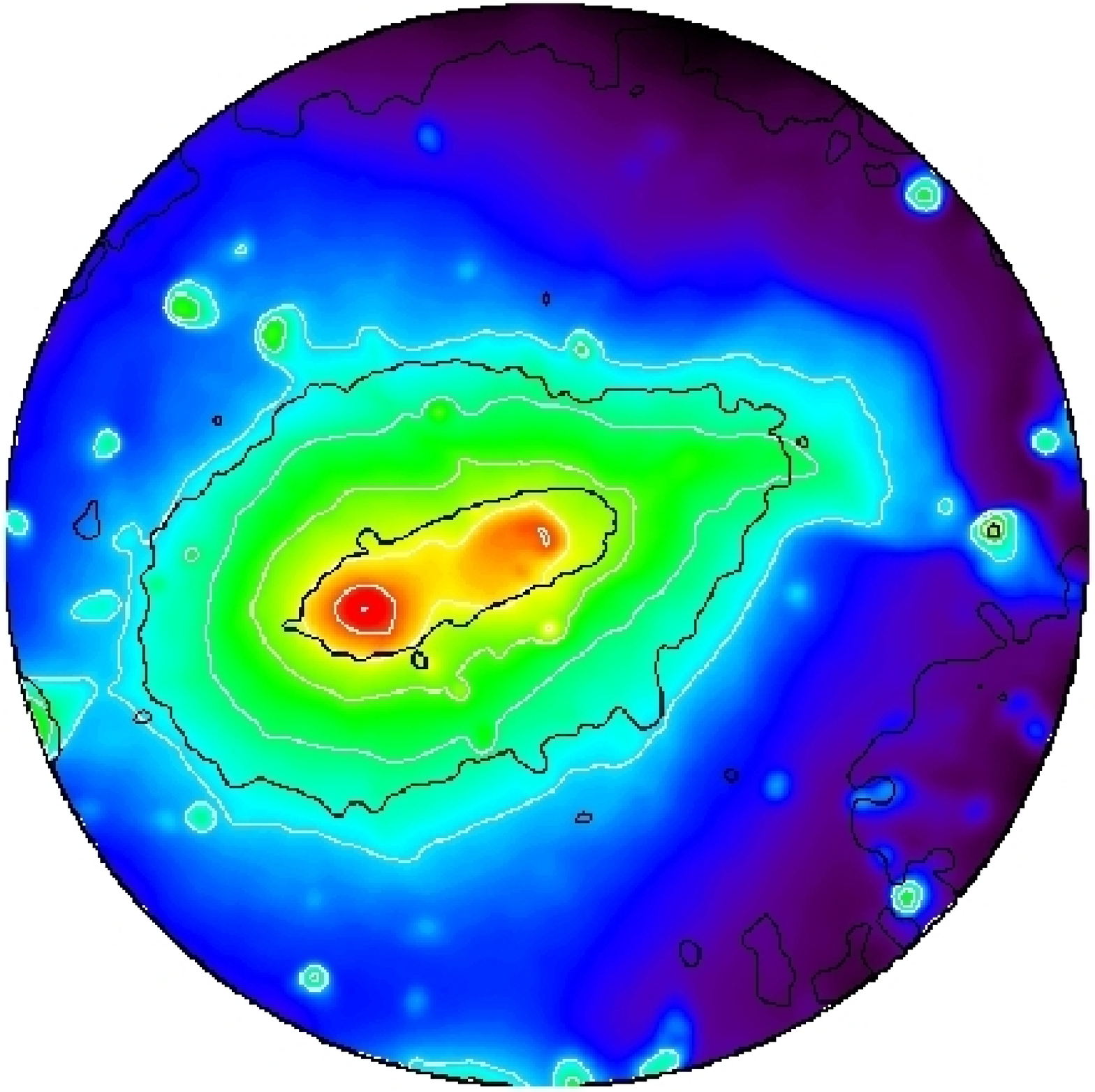}

\caption{Top panels: mass histories, $M_{500}$ versus redshift, for cluster A (left), B 
(middle) and C (right).  Middle panels: normalised RMS centroid displacement, a dimensionless estimator of a cluster's dynamical state (see text for details). The horizontal line indicates a value of 10 per cent (which we define as the threshold for a major merger) and filled squares in the top panels show outputs where the estimator exceeds this value. Bottom panels: examples of X-ray surface brightness maps for a cluster with 
low (0.01; left), intermediate (0.1; middle) and high (0.2) values of the RMS centroid shift, 
respectively. Here, contours illustrating equally spaced values of logarithmic surface brightness (white) and surface mass density (black)
are overlaid.  }
\label{mh:fig}
\end{figure*}

To define the properties of each cluster within the simulation data, 
DM haloes were first identified using a 
Friends of Friends (FoF) algorithm, where the position of the most bound 
DM particle was taken as the centre. The Spherical Overdensity 
(SO) algorithm was then used to grow a sphere around this centre and 
determine $r_{500}$, the radius containing a total mean density, 
$\bar{\rho}=500\rho_{\rm crit}(z)$, where 
$\rho_{\rm crit}(z)$ is the critical density for a closed universe at 
redshift, $z$. This radius was chosen as it approximately corresponds to 
the upper limit of the extent of X-ray observations. The three clusters are labelled A, B and C respectively. 

At the end of each simulation
($z=0$), the masses of clusters A, B and C within $r_{500}$ were $M_{500}=[6,3,6] \times 10^{14} h^{-1}{\rm M}_{\odot}$,
approximately corresponding to $[1.4,0.7,1.4]\times 10^6$ DM particles 
respectively. To determine
the evolution of each cluster with redshift, candidate progenitors were 
selected by finding all FoF groups at the previous output, whose 
centres lie within $0.5r_{500}$ of the present cluster's centre. We 
adopt a short (typically $b=0.05$, but decreasing to $b=0.025$ for 
problematic snapshots) FoF linking length to avoid the linking of two 
merging progenitors in close proximity, which can lead to fluctuations 
in the centre from output to output. We then select the most massive 
object as the progenitor, except when there are several candidates 
with similar mass, in which case we choose the one that is closest (this only occurs for cluster C). Our choice meant that where cluster C undergoes an almost
equal-mass merger at $z=0.5$, we did not end up
following the most massive object at higher redshift, but tests confirm
that this choice has no bearing on our main results.

Fig.~\ref{mh:fig} (top panels) shows how $M_{500}$ grows with time for
each of the three clusters. For convenience,
we have used redshift as the time axis; our study is limited to outputs between redshifts
0 and 1. By design, the mass histories vary significantly between the clusters: 
cluster A undergoes several major mergers (leading to abrupt jumps in mass) early on, 
then settles down at $z \sim 0.4$; cluster B accretes matter over the whole period; and
cluster C undergoes two massive mergers (at $z \sim 0.9$ and $z \sim 0.4$ respectively) 
with relatively quiet stages in between.

\subsection{Map making}\label{mapsec}

For our main analysis we constructed {\it maps} of surface mass density (dominated by the 
DM) and bolometric X-ray surface brightness for each cluster at each redshift. 
The former quantity is formally related to the volume density ($\rho$) as
\begin{equation}
\Sigma  = \int \, \rho \, dl,
\end{equation}
while the latter is related to both the electron density and temperature of the intracluster plasma
\begin{equation}
\Sigma_{\rm X}  = {1 \over 4\pi (1+z)^4} \, \int \, n_{\rm e}^2 \, \Lambda(T_{\rm e}) \, dl,
\end{equation}
although note that features are primarily due to fluctuations in the density.
For the analysis that follows, the explicit redshift dependence of the surface brightness is
ignored and we further assume the ideal conditions of infinite signal to noise (except for discreteness noise due to the finite number of particles employed).
The conversion from gas to electron density is performed assuming a fully ionised, $Z=0$
plasma (so $n_{\rm e} \sim 0.9 \rho_{\rm gas}/m_{\rm H}$) and the cooling rate is computed using the
tables calculated by \citet{sd_coolingtables} for $Z=0.3 Z_{\odot}$, the typical metallicity
of the intracluster medium (ICM).

The estimation procedure followed is similar to that employed by \citet{onura_mapmethod}. Briefly, a cuboid
is defined, centred on the cluster, with sides of proper length $2r_{500}$ in the X and Y directions and $8r_{500}$ in
the Z direction (to capture material associated with the cluster along the line-of-sight). Particles that reside
within this cuboid are then identified and projected in the Z direction on to a 2D array of $400 \times 400$ pixels.
The pixel size was chosen to sample length scales of at least the Plummer softening length (at $z=0$, 
$r_{500} \sim 1 \, h^{-1} {\rm Mpc}$), so that all {\it real} structures were capable of being resolved by the map.

The gas particles are not point-like, but spherical clouds of effective radius, $h$, and shape defined by the 
(spline) kernel used by the {\sc gadget2} SPH method \citep*[see][]{springeletal2001a}. Thus, all gas particles were smoothed
onto the array using the projected version of the kernel. To reduce the noise in the mass maps, dominated by DM
particles, densities and smoothing lengths were adopted in a similar way (defining $h$ such that each DM particle
enclosed an additional 31 neighbours) and smoothed using the same kernel as with the gas. 

The projected mass density at the centre of each pixel, ${\bf R}_0$, is therefore
\begin{equation}
\Sigma({\bf R}_0) = {1 \over A_{\rm pix}} \,  \sum_{i=1}^{N} \, m_i \, w(|{\bf R}_i-{\bf R}_0|,h_i),
\end{equation}
where $A_{\rm pix}$ is the pixel area, the sum runs over all $N$ particles within the cuboid region, $m_i$ is the mass of 
particle $i$ and $w$ is the projected SPH kernel, suitably normalised to conserve the quantity being smoothed. 

The (redshift zero) X-ray surface brightness is estimated using a similar equation
\begin{equation}
\Sigma_{\rm X}({\bf R}_0) = {0.9 \over 4\pi A_{\rm pix} m_{\rm H}} \, \sum_{i=1}^{N_{gas}} \, m_i \, n_{{\rm e},i} \, \Lambda(T_i) \, 
w(|{\bf R}_i-{\bf R}_0|,h_i),
\end{equation}
where the sum runs over all hot ($T_{i}>10^{6}$K) gas particles and we have assumed equivalence between the hot gas and electron temperatures.

The maps are re-centred for analysis, such that the new centre is set to the brightest pixel in the 
X-ray surface brightness map, as would generally be the case for observations. The allowed alteration is restricted to ensure that the centre does not `jump' to  a bright substructure (this is unrealistic, but possible because of our simple non-radiative model) as this would undermine 
the effort directed at following the assembling structure. 

\begin{figure*}
\centering
\includegraphics[width=0.3\textwidth]{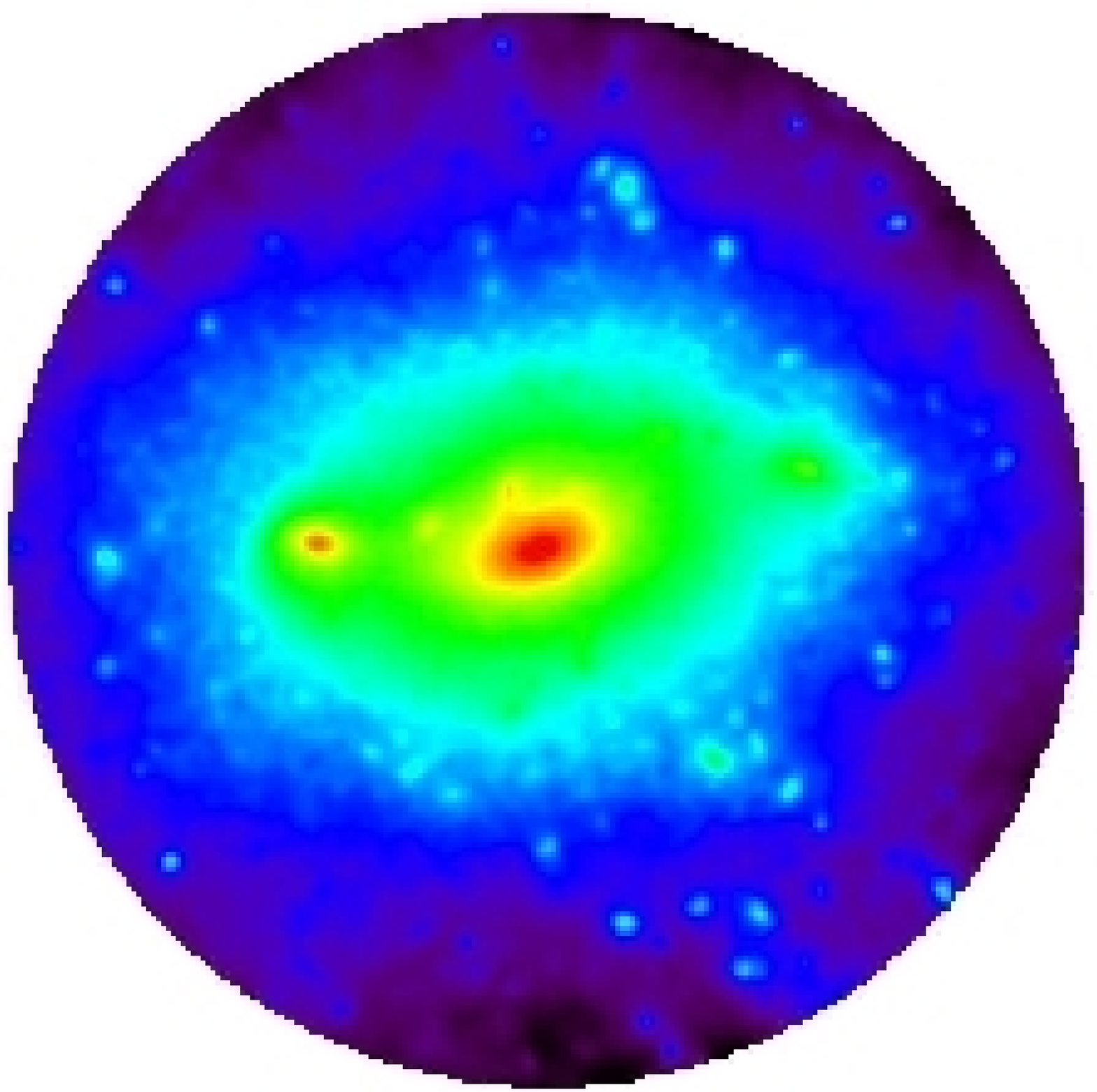}
\hspace{0.1\textwidth}
\includegraphics[width=0.3\textwidth]{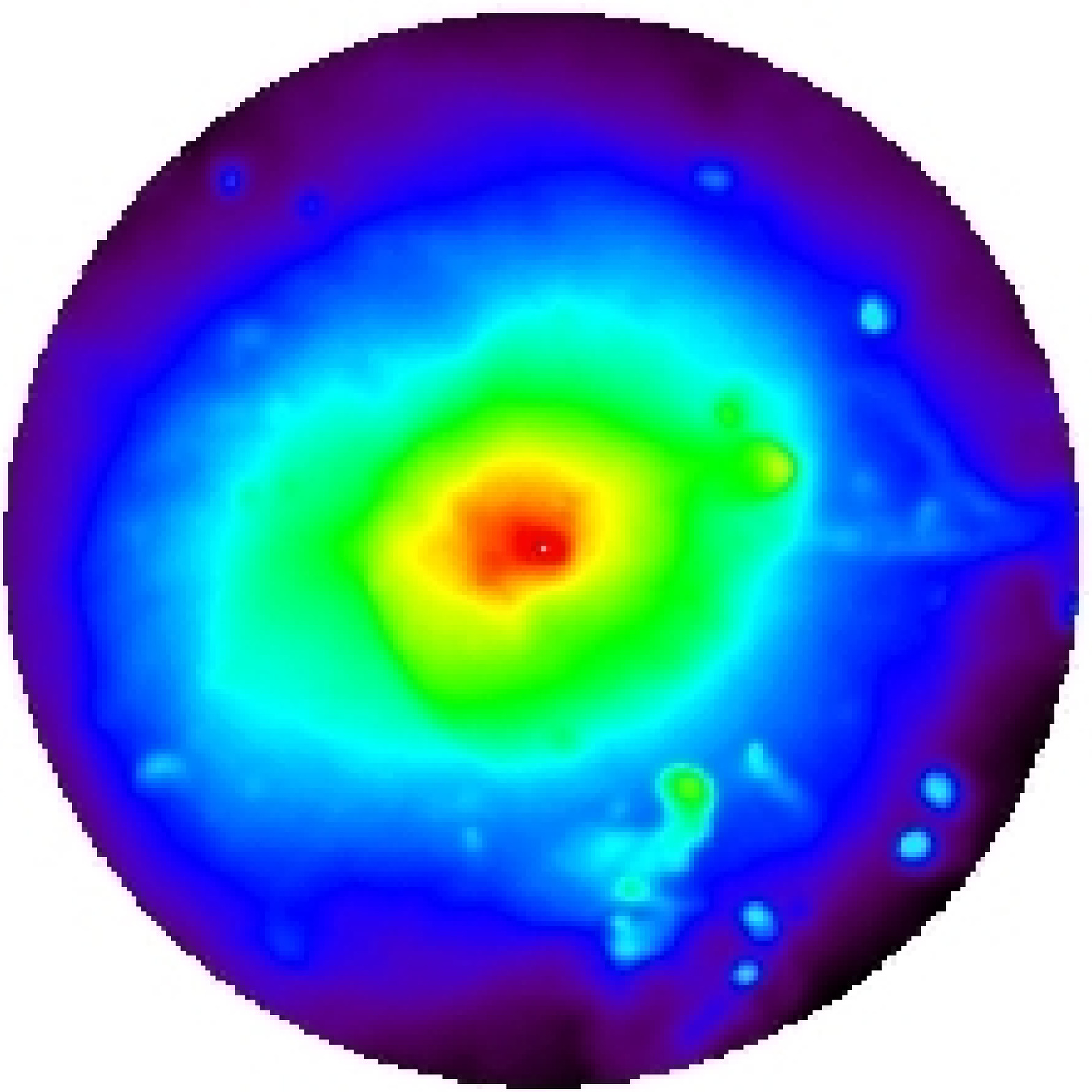}
\\
\includegraphics[width=0.3\textwidth]{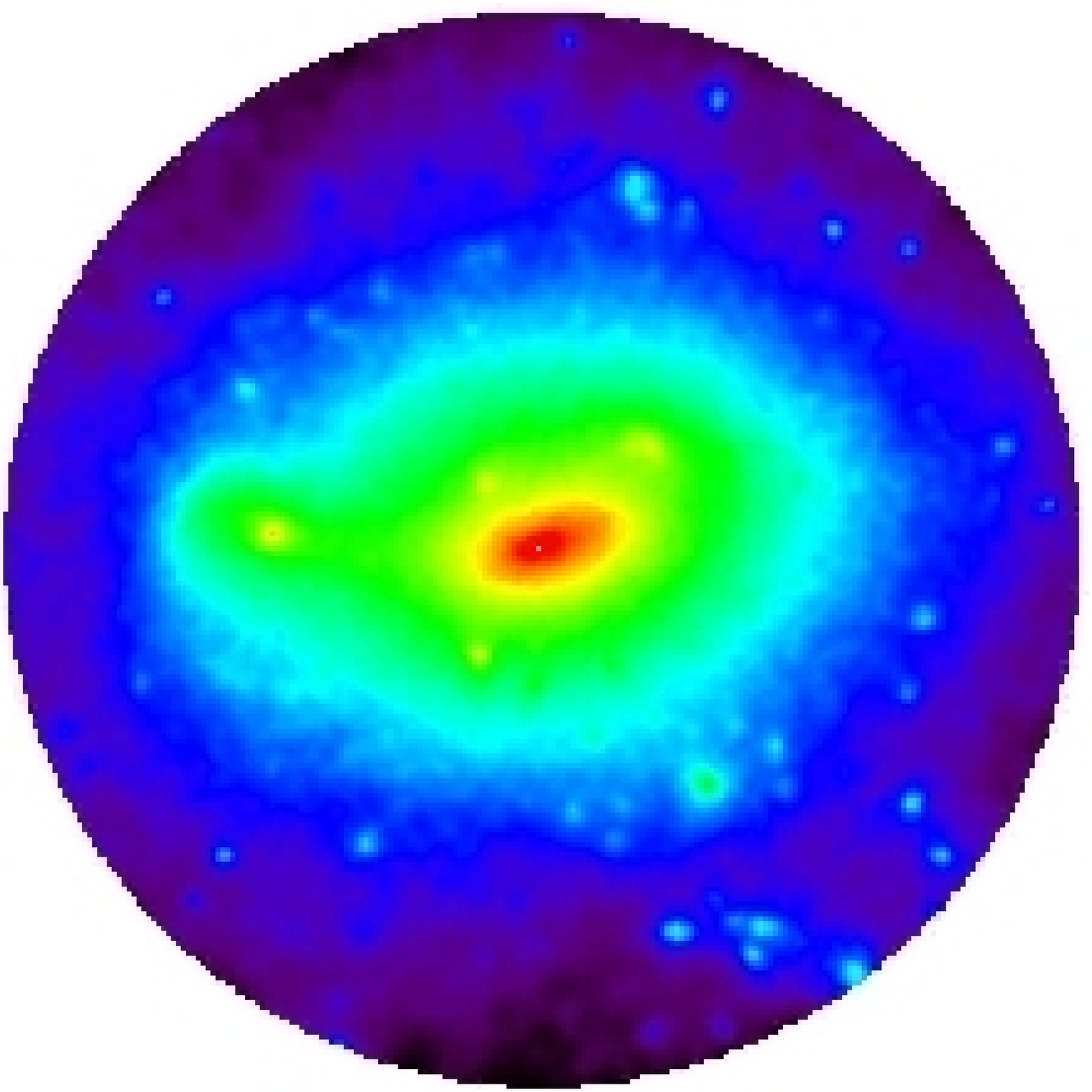}
\hspace{0.1\textwidth}
\includegraphics[width=0.3\textwidth]{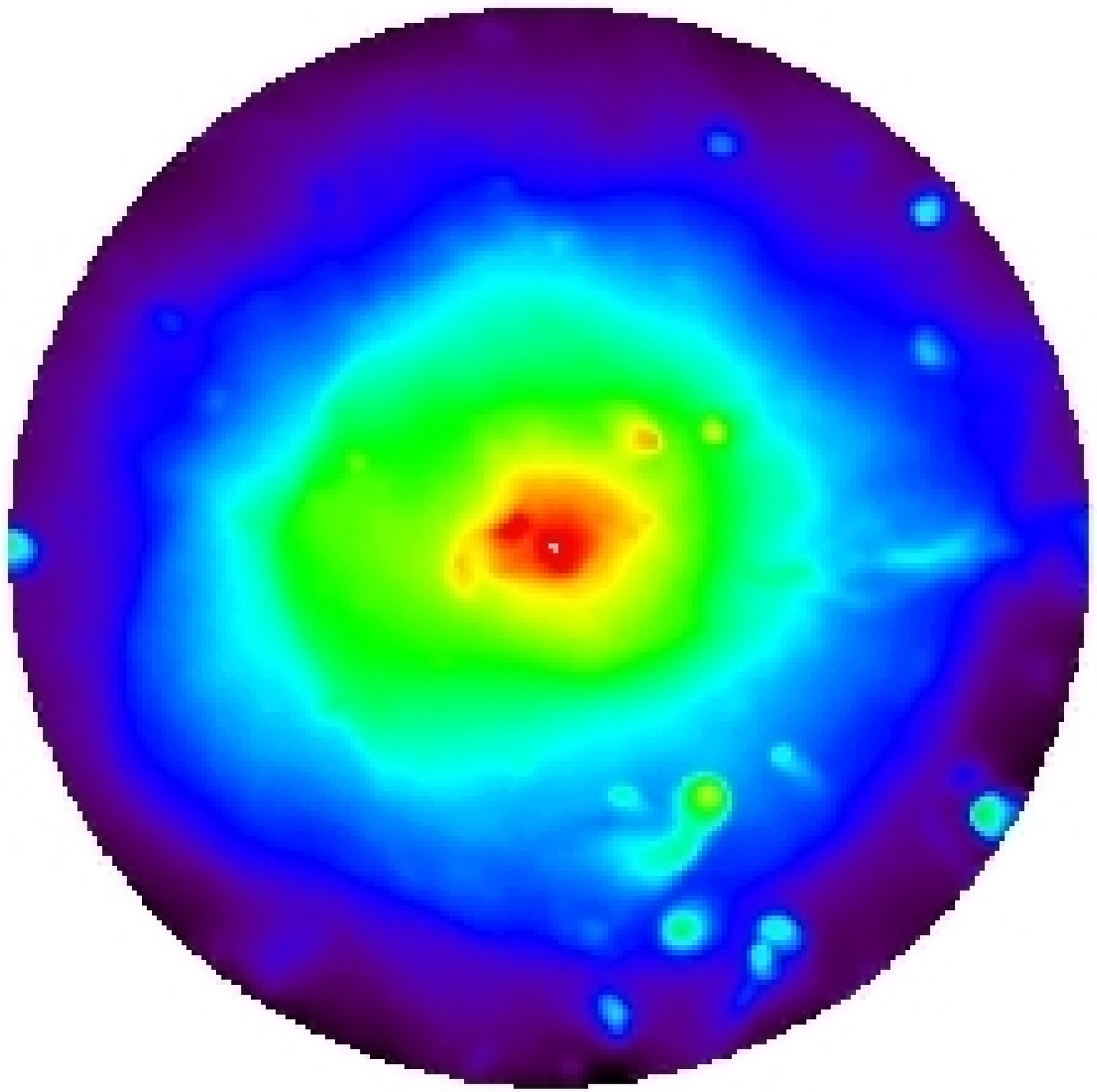}
\\
\includegraphics[width=0.3\textwidth]{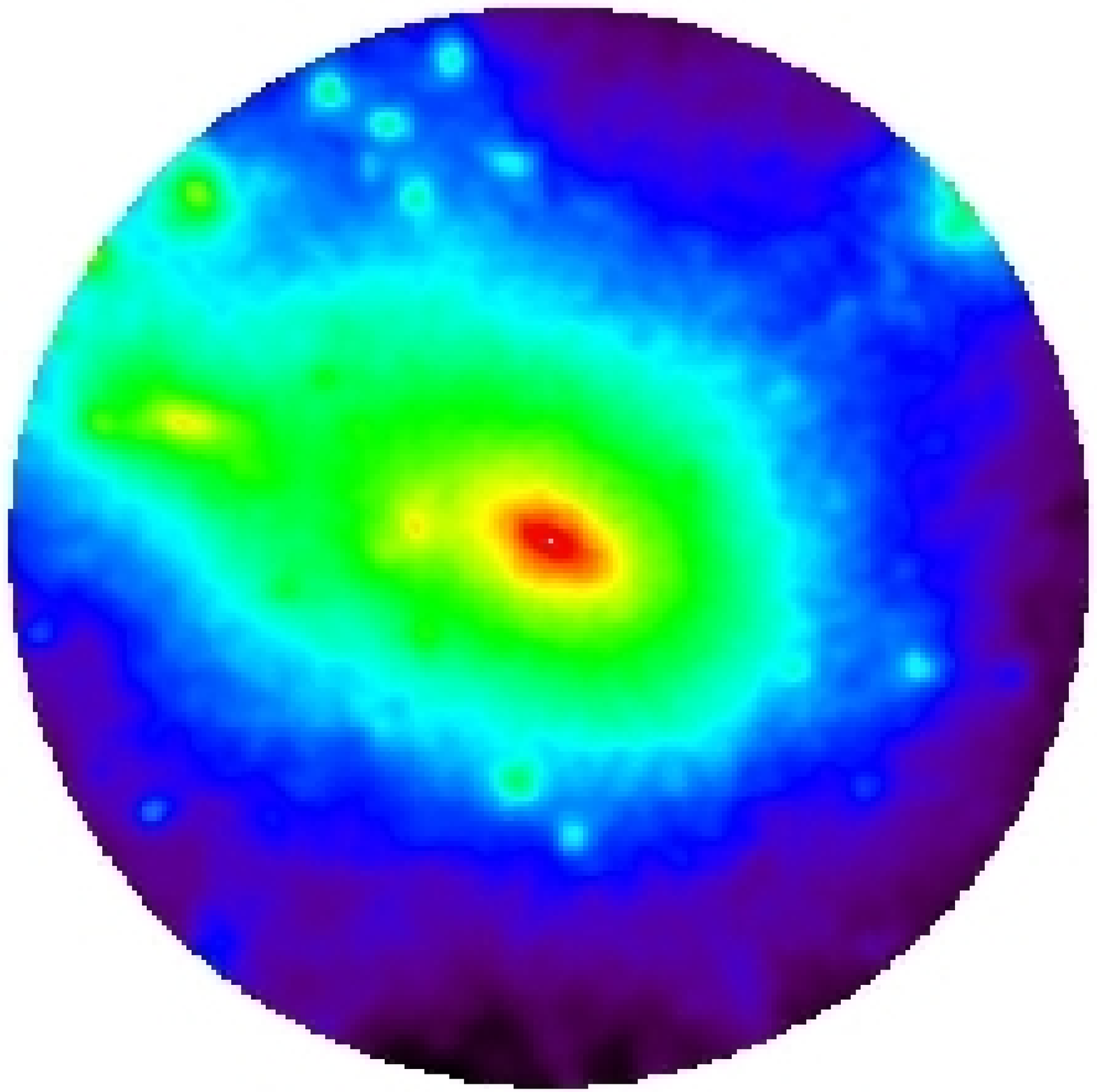}
\hspace{0.1\textwidth}
\includegraphics[width=0.3\textwidth]{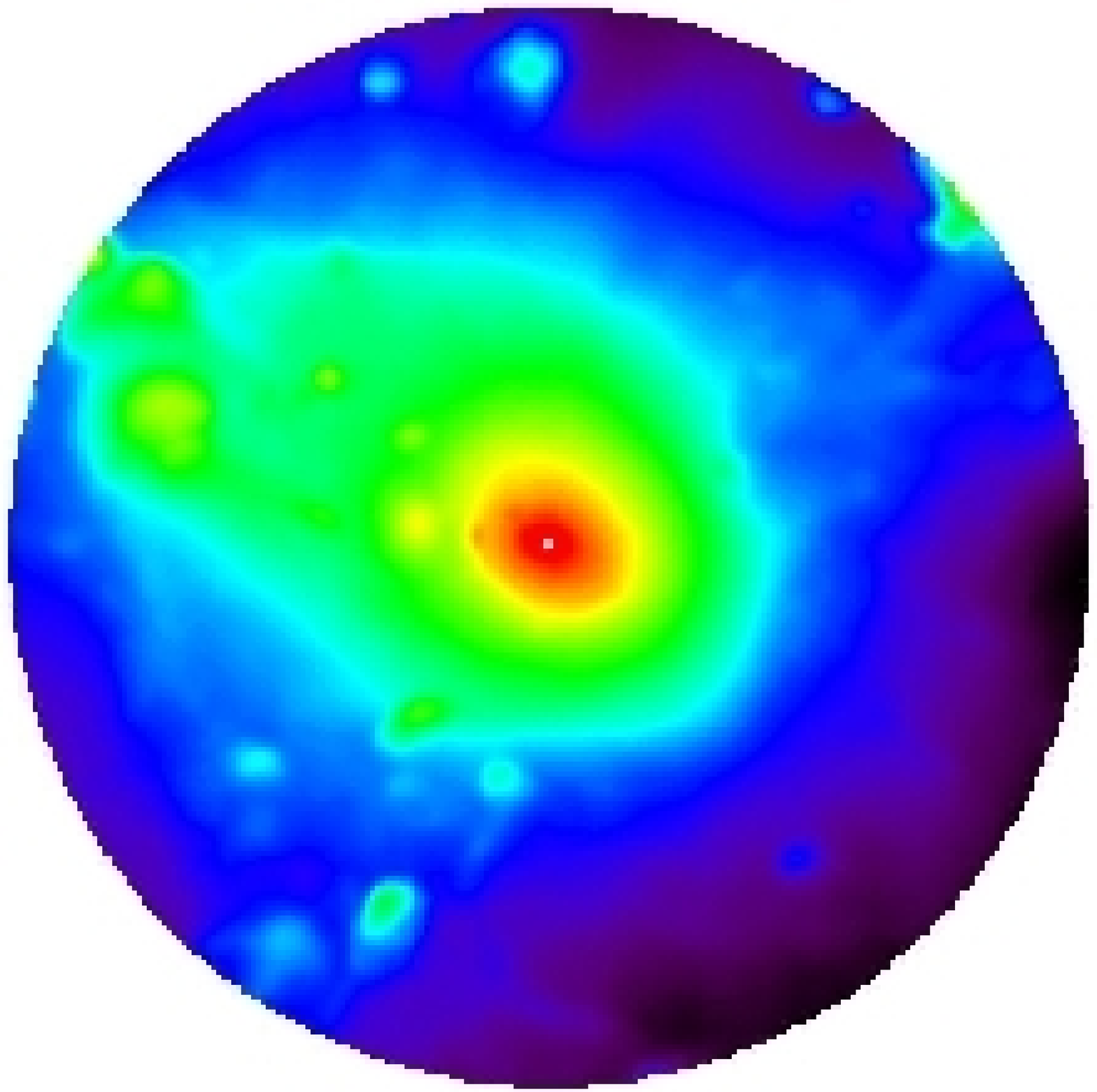}
\\
\includegraphics[width=0.3\textwidth]{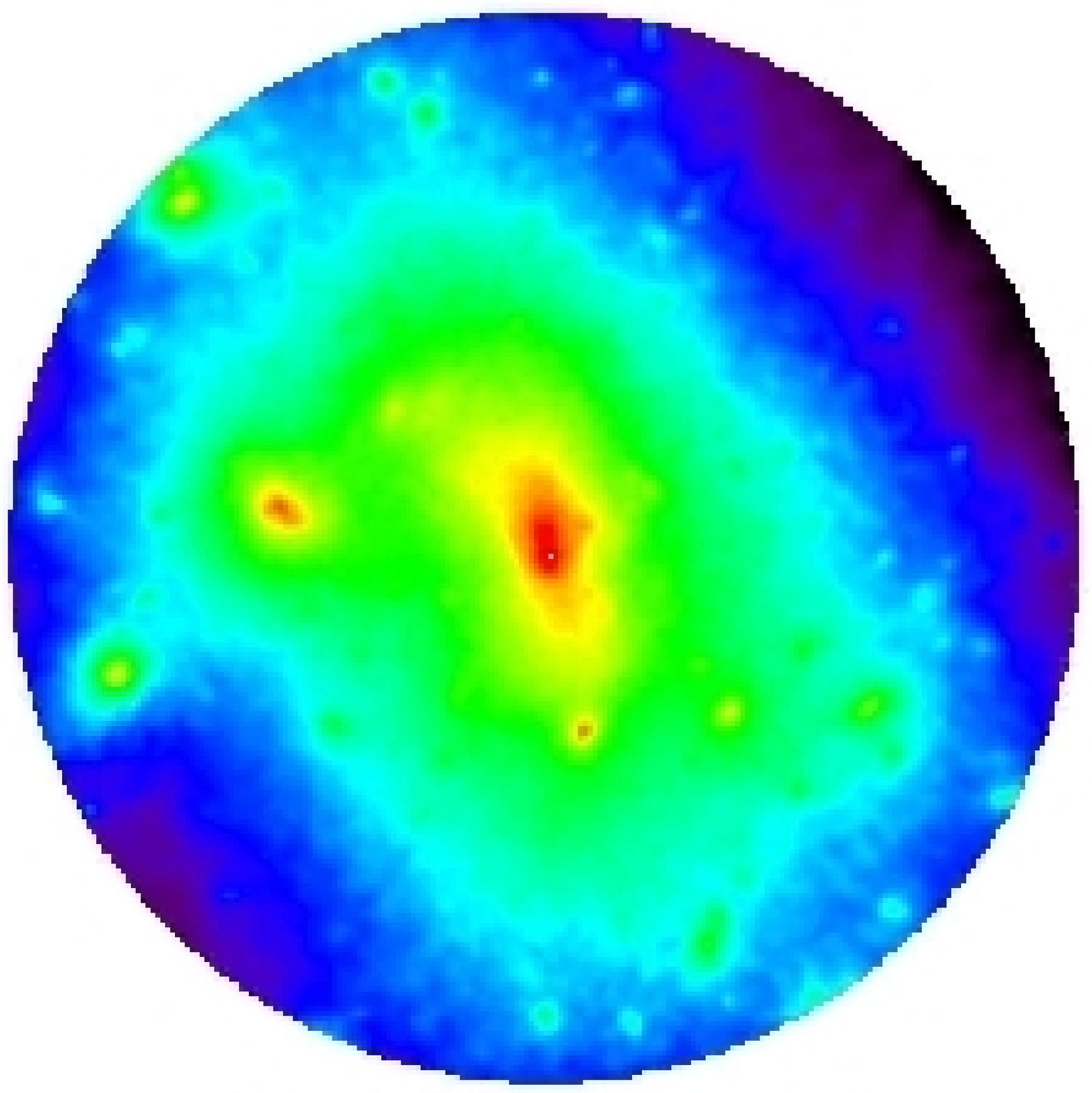}
\hspace{0.1\textwidth}
\includegraphics[width=0.3\textwidth]{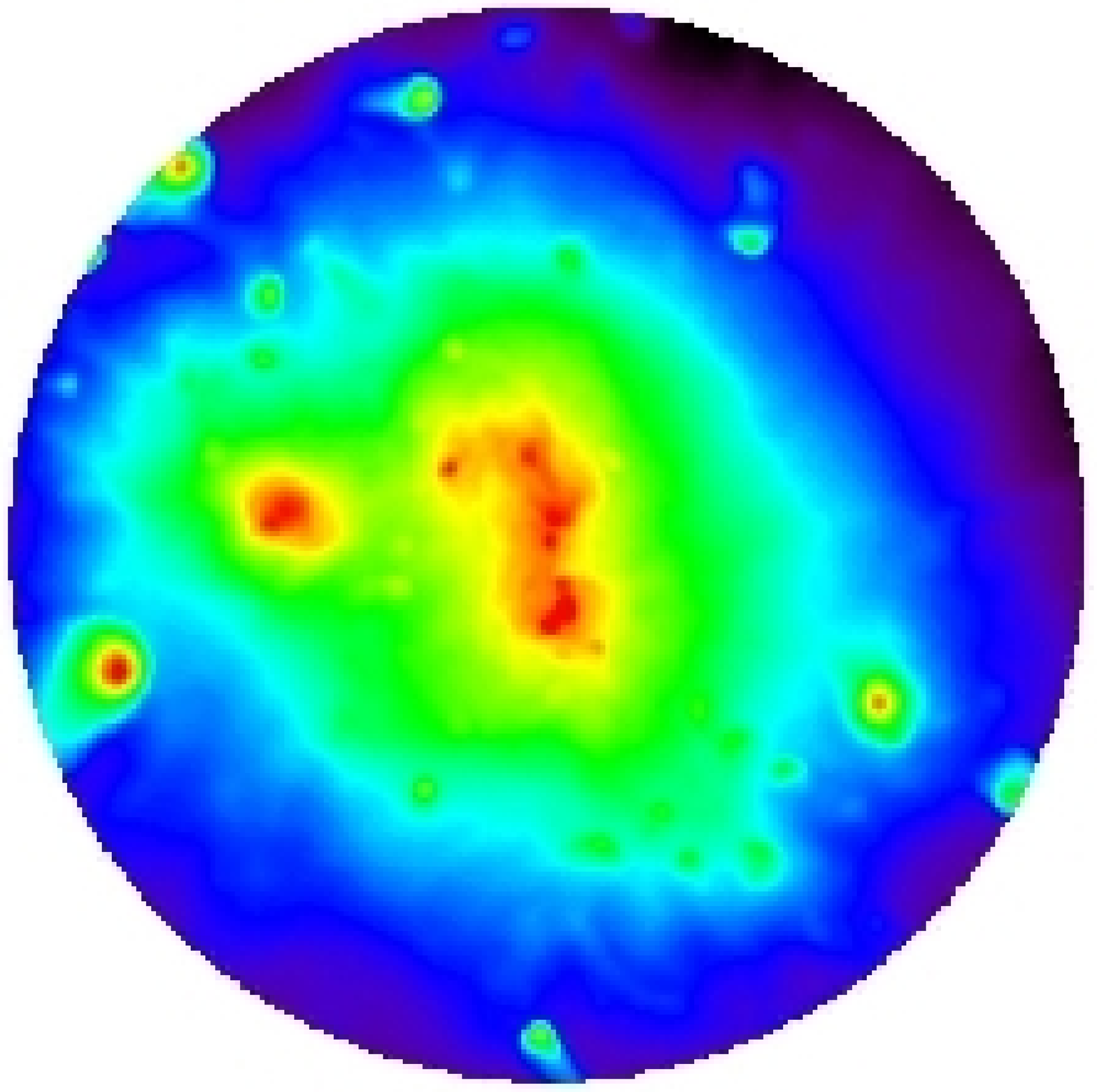}

\caption{Examples of cluster maps. Left column: logarithmic surface mass density maps for, top to 
bottom, cluster A cooling run, cluster A, B and C non-radiative runs at $z=0$. Right column: the same but for the logarithmic X-ray surface
brightness maps.}
\label{zeq0mapfig}
\end{figure*}

The bottom three rows of Fig.~\ref{zeq0mapfig} illustrate surface mass density (left column) and 
surface brightness (right column) maps for cluster A, B and C at $z=0$. Qualitatively, the strongest features 
are clearly present in both maps, but there are some differences, notably the lack of some
obvious subhaloes in the X-ray maps and extended features in the X-ray maps due to stripped
gas. It is also clear that the brightest X-ray substructures tend to be much rounder than
in the mass maps, as expected, since the gas traces the potential, which is smoother and more spherical than the density.

\subsection{Characterising dynamical state from the maps}\label{calcdynam}

\begin{figure*}
\centering
\includegraphics[width=0.33\textwidth]{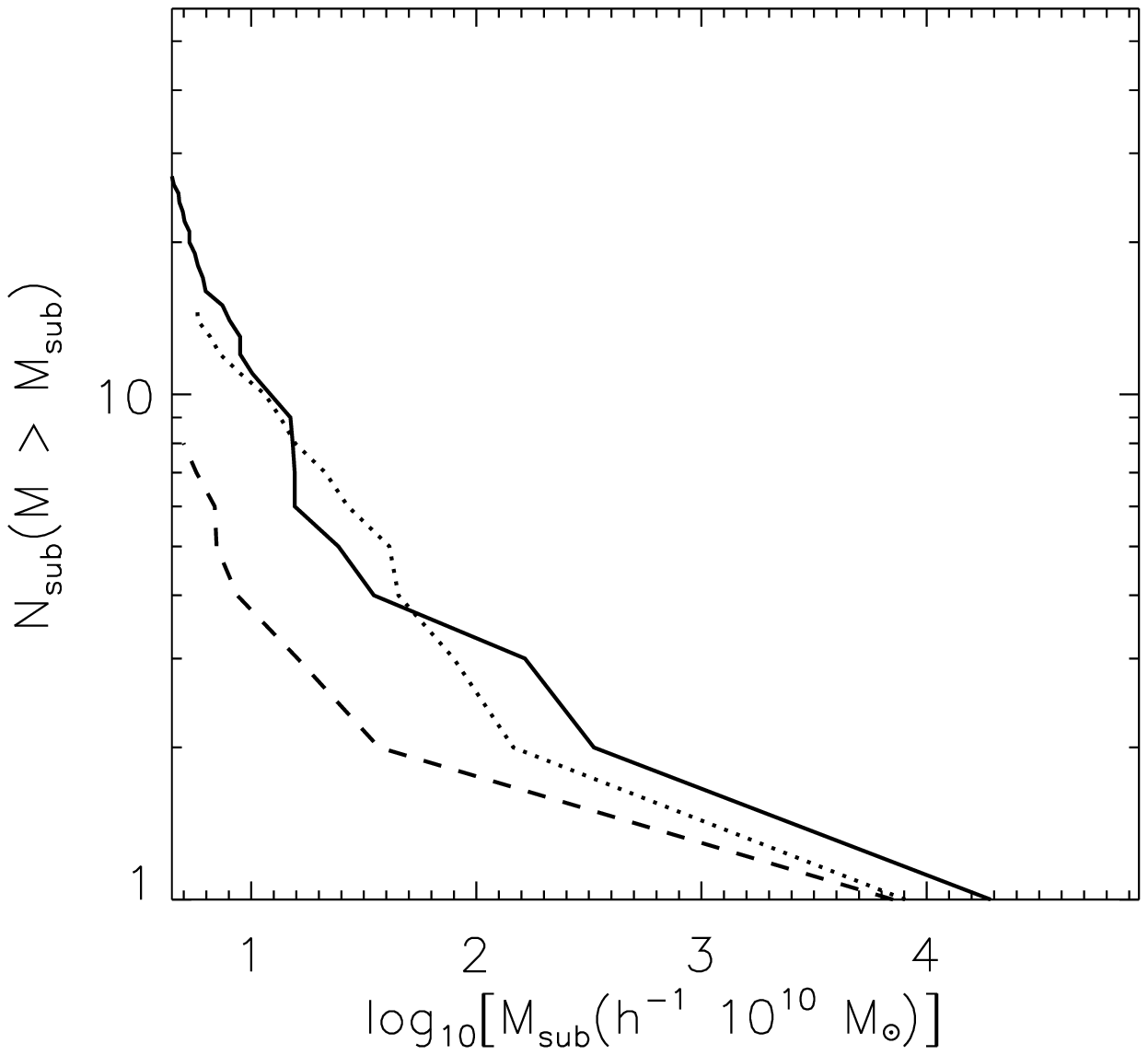}
\includegraphics[width=0.33\textwidth]{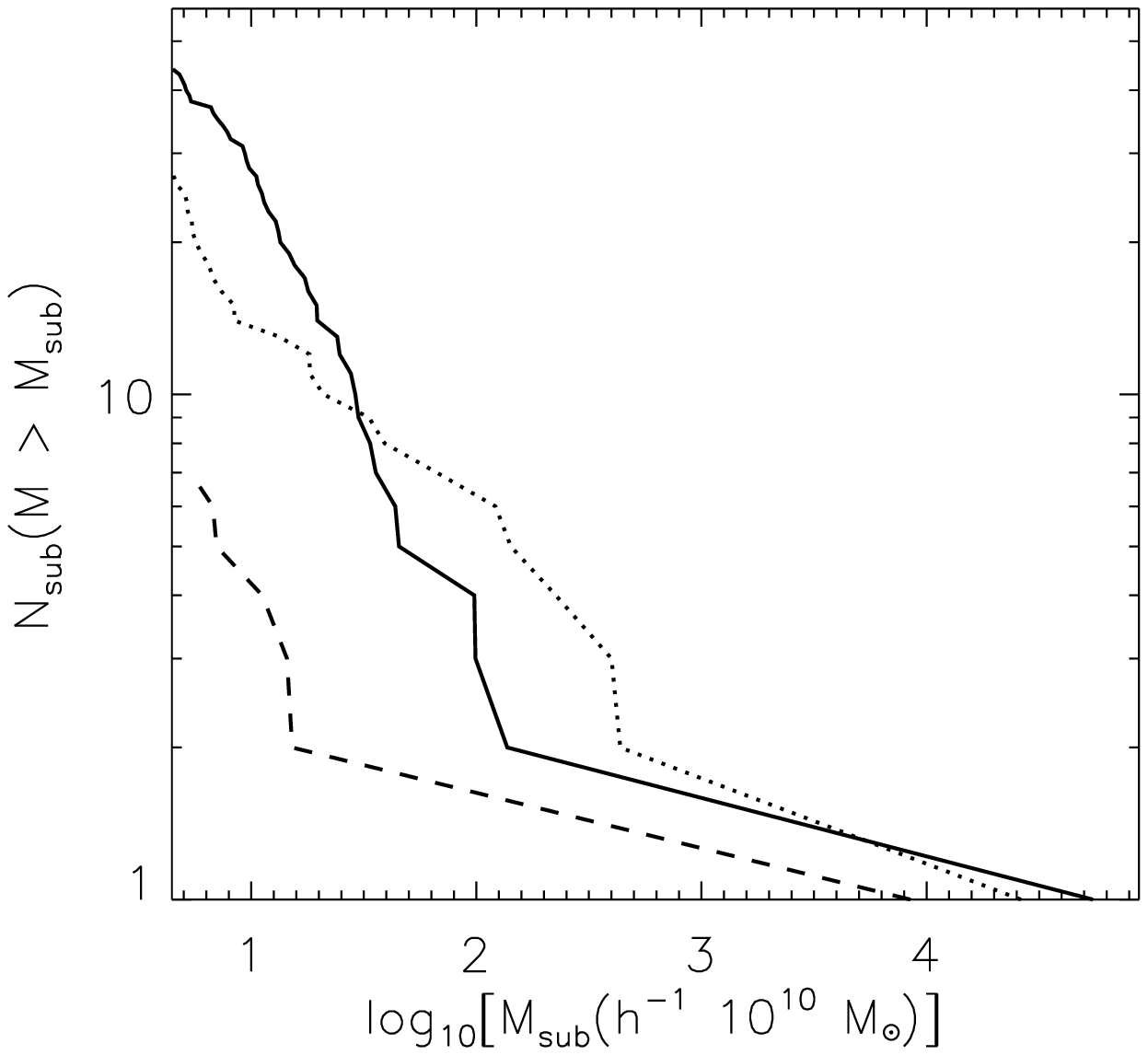} 
\includegraphics[width=0.33\textwidth]{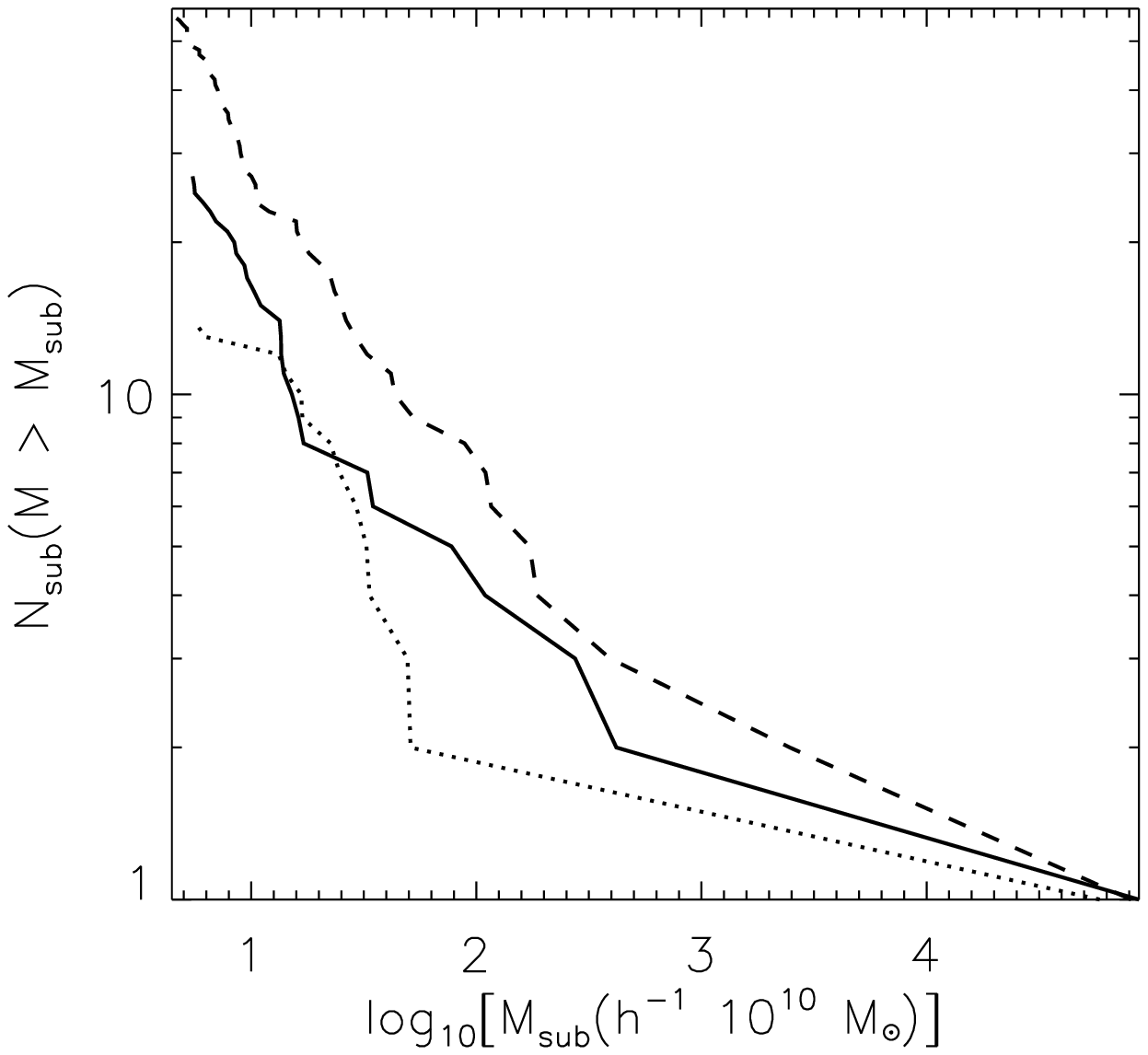}\\
\includegraphics[width=0.33\textwidth]{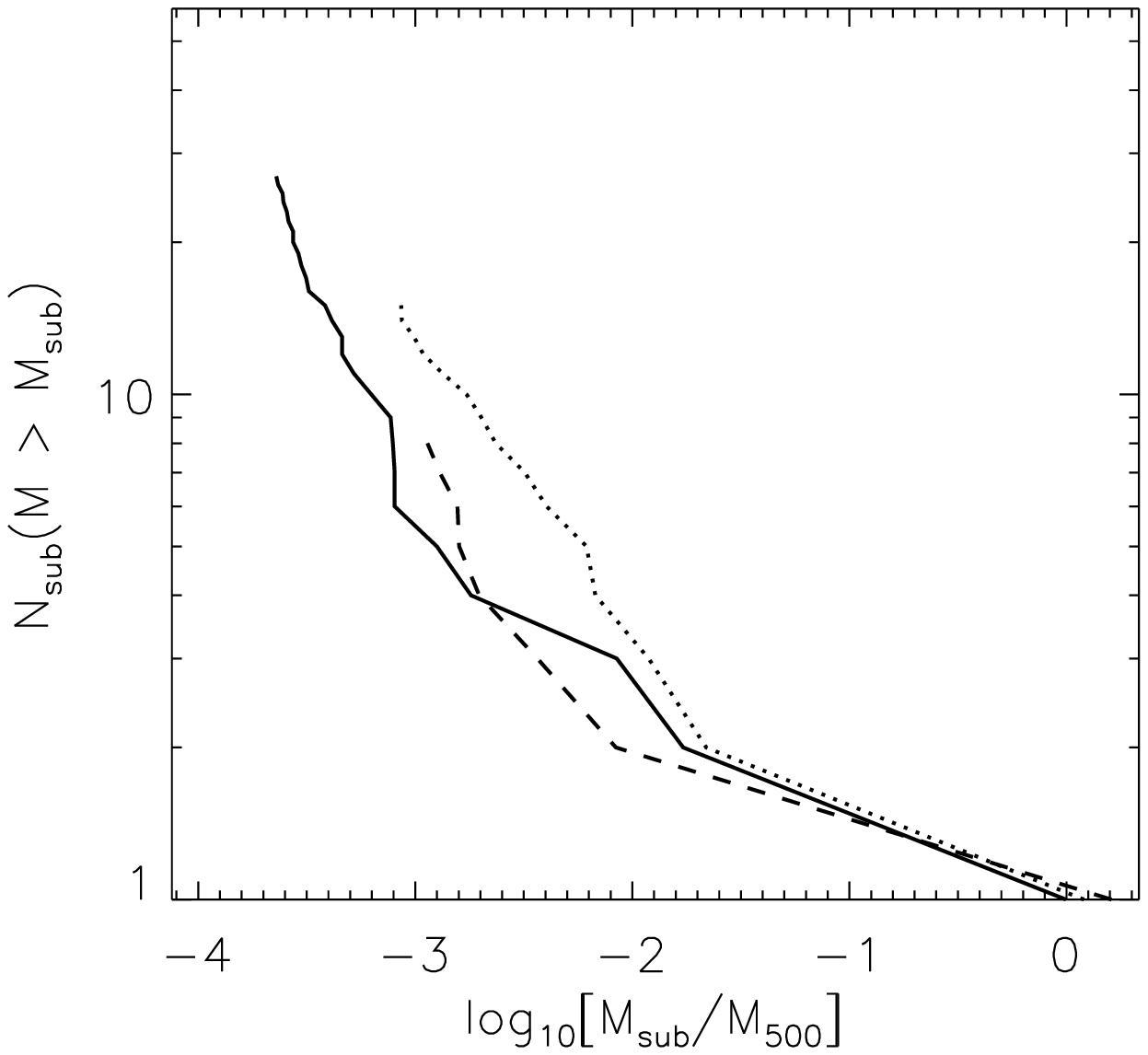}
\includegraphics[width=0.33\textwidth]{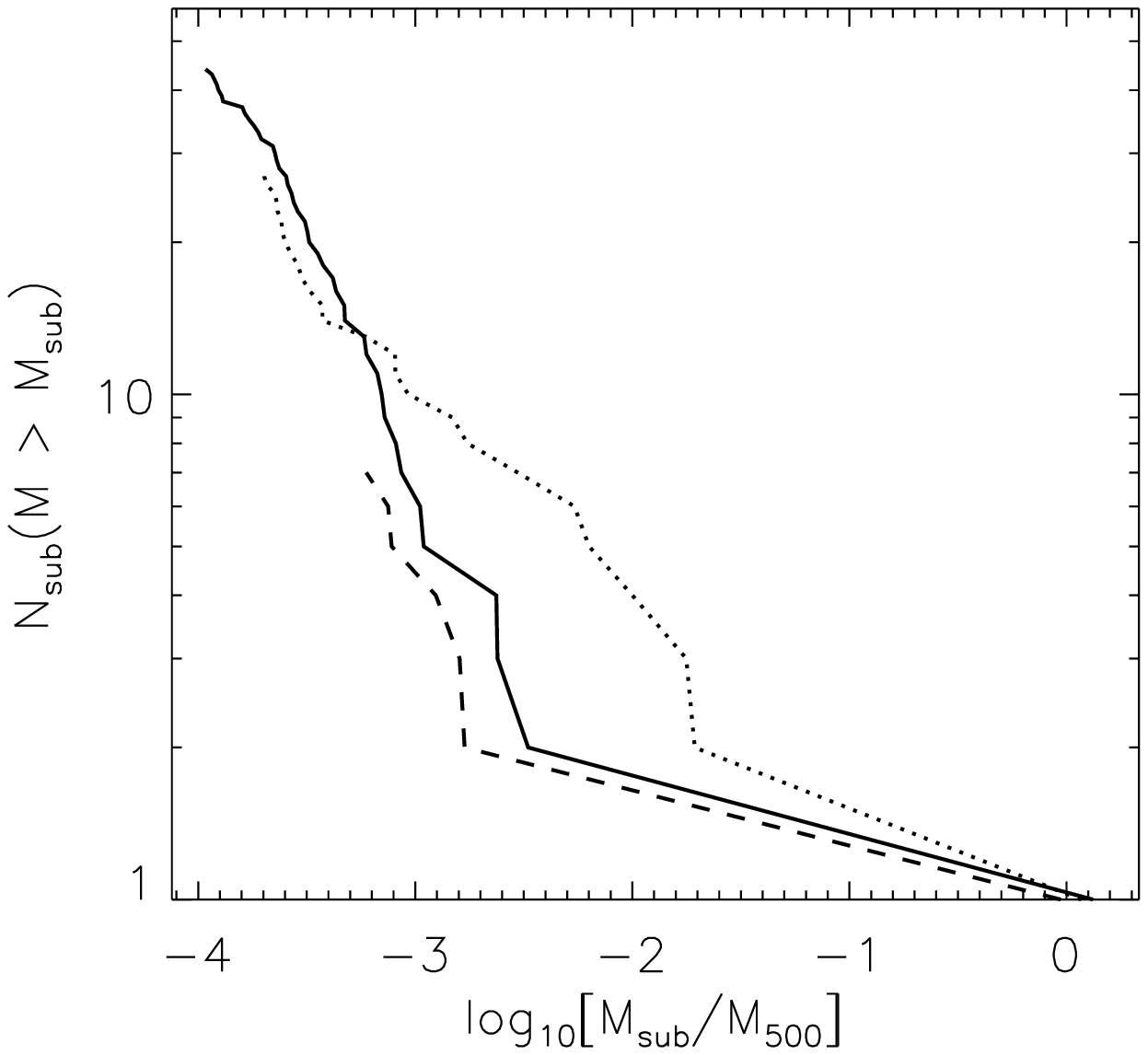} 
\includegraphics[width=0.33\textwidth]{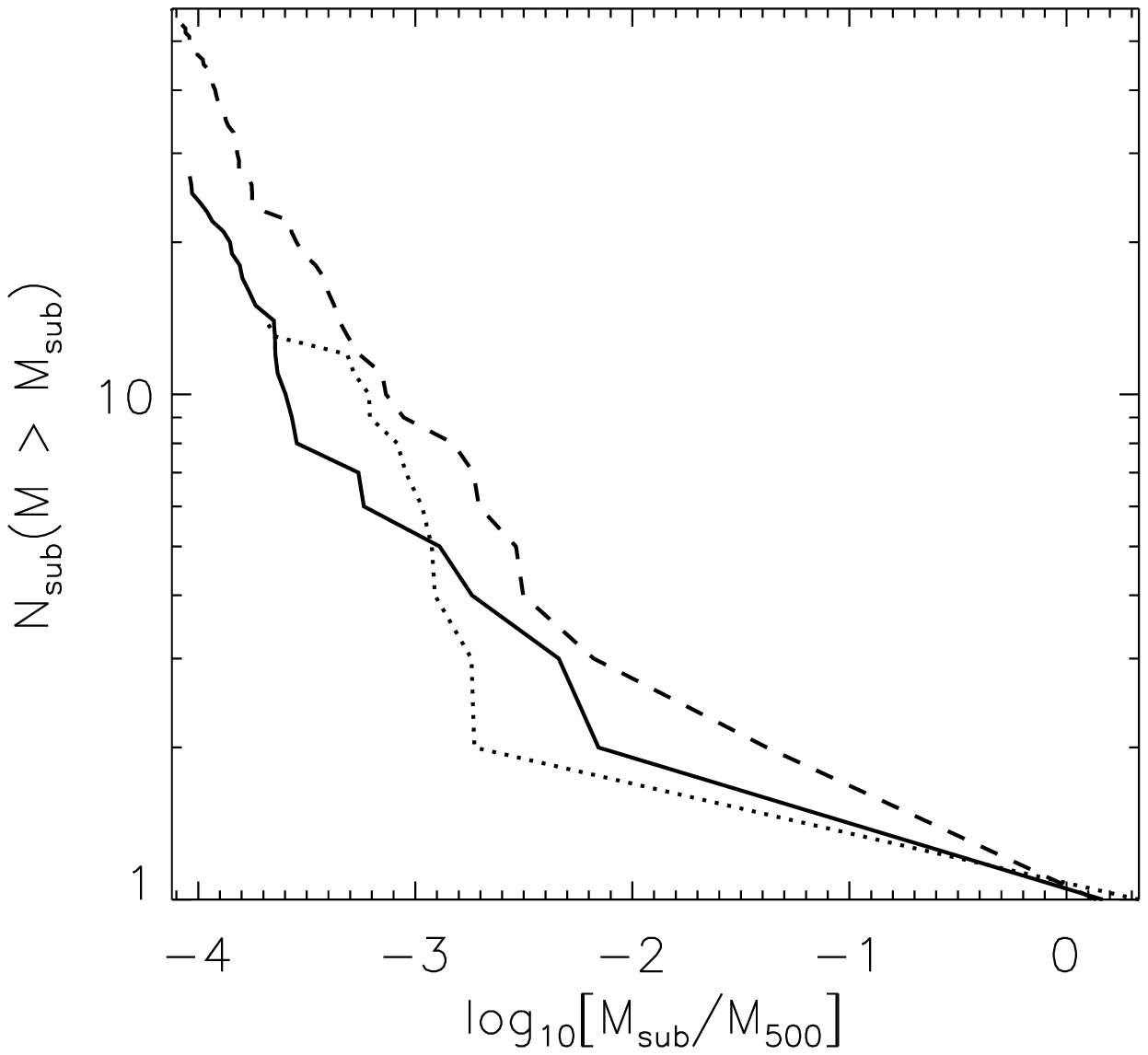}
\caption{Subhalo mass functions for cluster A (solid lines), B (dotted lines) and C (dashed lines), at redshifts, $z=1$ (left), $z=0.5$ (middle) and $z=0$ (right). First row: Cumulative subhalo (DM) mass functions. Second row: As for first row but with the subhalo DM masses scaled to $M_{500}$. Data are for subhaloes with their most bound particle within $r_{500}$.}
\label{mfclusfig}
\end{figure*}

Our first application of the maps is to estimate the dynamical state of the cluster from its visual
appearance. We employ the method of \citet{ohara:apj}, also found to provide a reliable indicator of 
dynamical state by \citet{clusmergers_poole06}, using idealised simulations of cluster mergers. For this method,
the displacement between the X-ray peak and centroid is calculated within circular apertures ranging 
from $r_{500}$ down to 0.05$r_{500}$ in radius, decreasing by 5 per cent each time, and then the RMS value of 
the displacement is computed, relative to $r_{500}$. We found this technique to be most effective when using
heavily smoothed maps to compute the centroid, so adopt the smoothing kernel used in our substructure detection 
algorithm (described in Section \ref{2dsec}), but with $\sigma = 0.1 r_{500}$. The position of the peak is always taken as the centre of the cluster,
as defined in the previous subsection. 

The variation of this RMS 2D statistic with redshift is shown in the middle panels of Fig.~\ref{mh:fig}. Values above 10 per cent (indicated with a horizontal line) are typically found when a cluster is undergoing a major merger \citep[see, for example,][]{clusmergers_poole06}. The redshifts at which this value is exceeded are also indicated with filled symbols in the mass histories (top panels). The bottom panels of the same figure give examples of clusters with low, intermediate
and high values of the RMS centroid shift, clearly showing an increase in dynamical activity.

\section[]{3D Subhalo detection}\label{detectsec}

Although the key objective of our analysis is to study the X-ray and mass 
maps of the clusters, we can draw important insight from an analysis of the 
full 3D data. In this section we identify 3D self-bound DM subhaloes in the map region (a cylinder of radius $r_{500}$ and length 8$r_{500}$, centred on the main cluster) and investigate their global properties. The information we 
glean from this analysis will help us to interpret our results in Section 
\ref{massgassec} by allowing us to distinguish the underlying physical 
mechanisms from any effects introduced by our method.

\subsection[]{Detection technique}

We use a version of {\sc subfind} \citep{springeletal2001b} to decompose FoF groups 
(identified for this purpose with $b=0.2$) into 3D self-bound subhaloes. 
The modified version, kindly provided by Volker Springel  \citep[see also][]{newsubfind}, identifies both gas and 
DM particles (and star particles when relevant) within each subhalo. A region larger than the final map region is analysed such that all subhaloes that contribute significantly to, but may not be fully within, the map region are included. 

We employ a threshold of 100 DM particles, corresponding to a mass,
$M \approx 4\times 10^{10}h^{-1}{\rm M}_{\odot}$, as our minimum resolution
limit for the subhalo catalogues. As we will show, this is significantly below our 2D completeness limit (determined in Section \ref{2d3dsec}). As our 3D subhalo catalogues will form the basis for comparison with 2D substructure, we consider not only subhaloes that lie entirely within the map region, but also those with at least $75$ per cent of their mass along the line of sight (as defined in Section \ref{simsec}), even if their centre co-ordinates are outside $r_{500}$ in projection. Note that, even if less than 100 per cent of the subhalo's particles are within the map region, the whole DM mass of the subhalo is still recorded.

The mass of each subhalo is computed using
only the DM particles, to reduce its dependence on the amount of gas
stripping that has occurred (the mass, $M_{\rm sub}$, therefore refers
to the DM mass of the subhalo). We take the centre of the subhalo to be
the position of the most bound particle, but also calculate a {\it projected}
centre, to facilitate matching with the substructures in the map. This was determined to be the position of the peak projected number of DM particles, i.e. the co-ordinates of the cell containing the most particles when each subhalo's DM particles are binned according to 
their X and Y co-ordinates (particles with Z co-ordinates outside the map region are excluded). 

\subsection{Properties of 3D subhaloes}\label{3dpropertiessec}

\begin{figure*}
\centering
\includegraphics[width=0.33\textwidth]{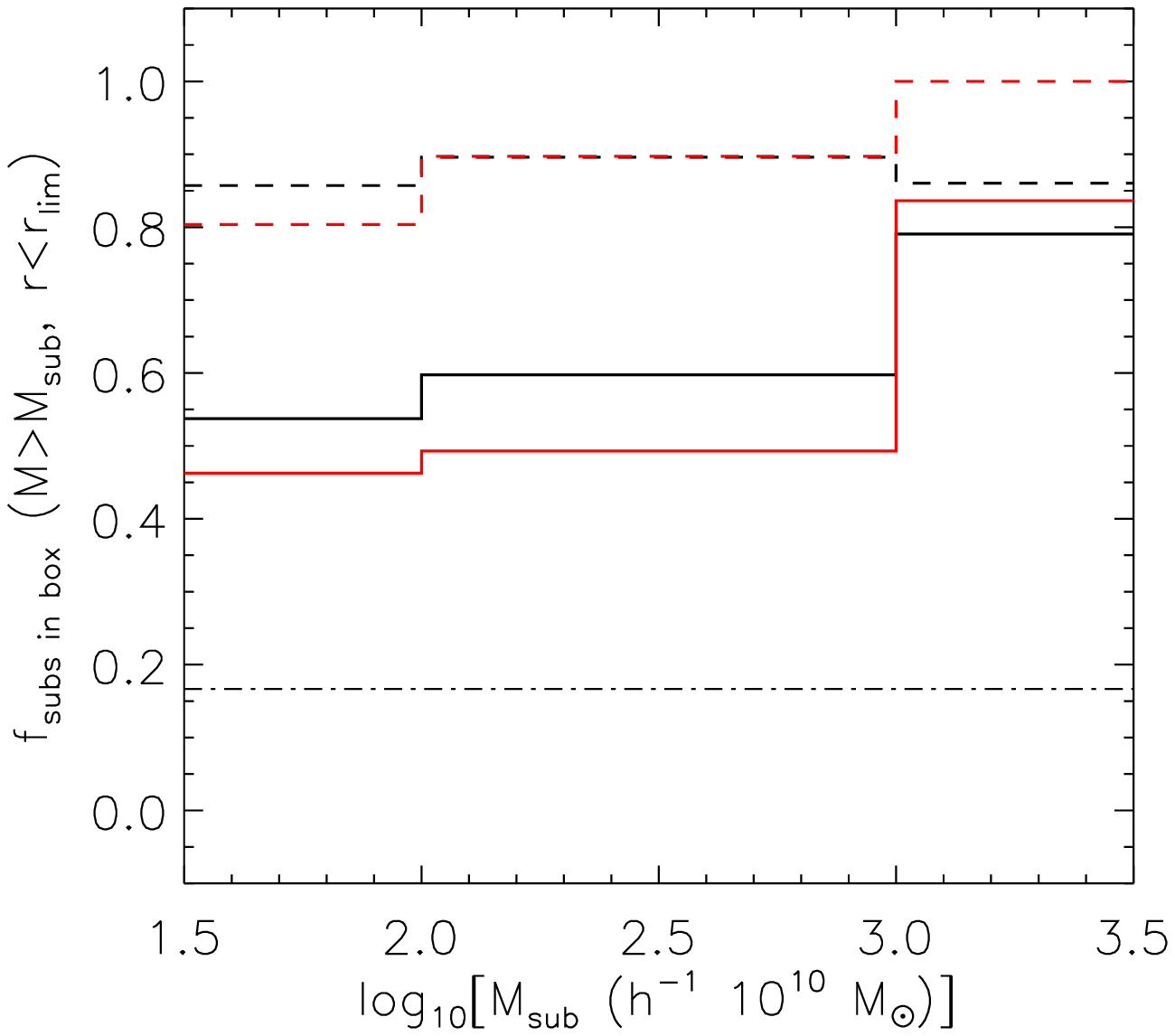}
\includegraphics[width=0.33\textwidth]{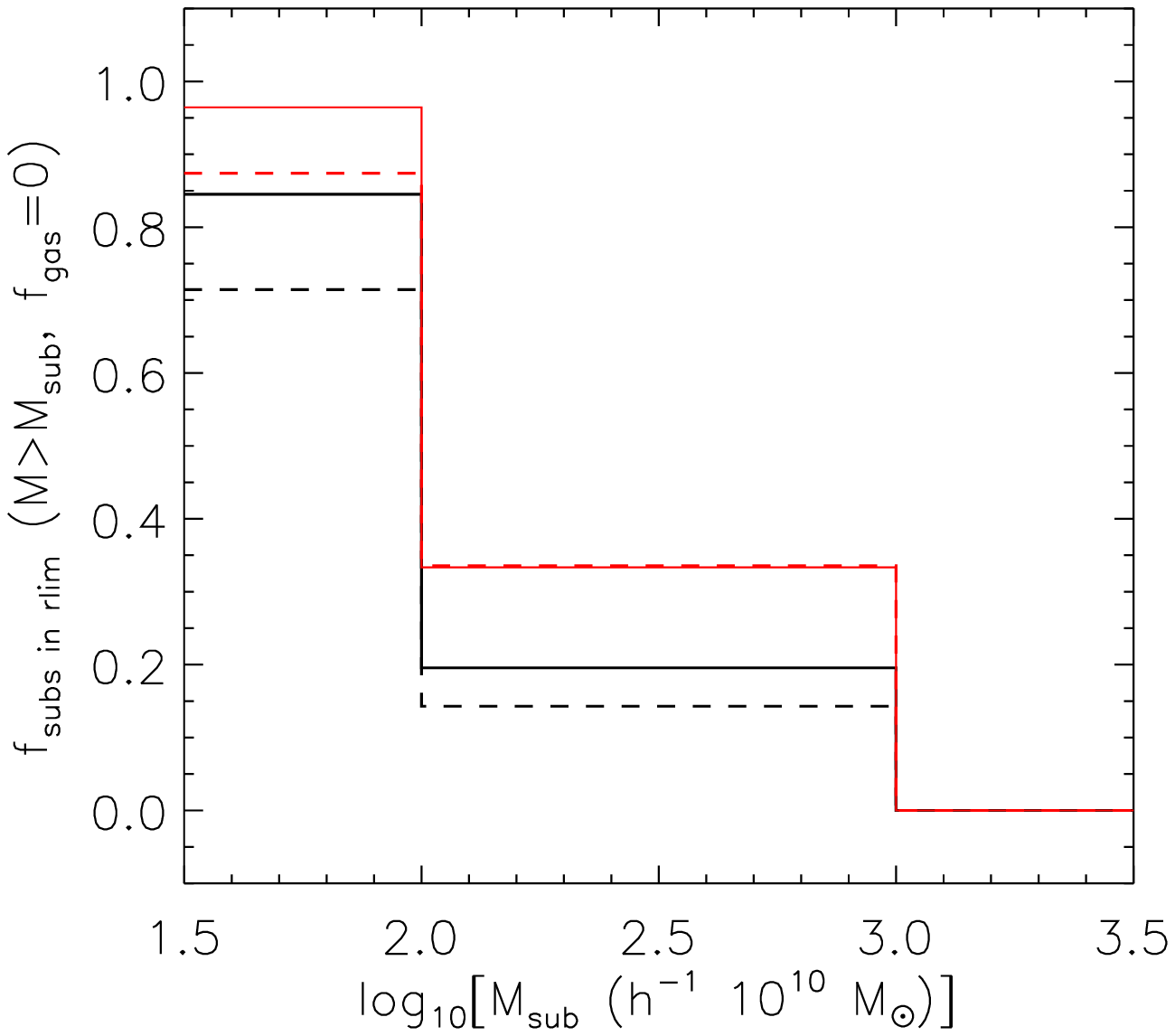}
\includegraphics[width=0.33\textwidth]{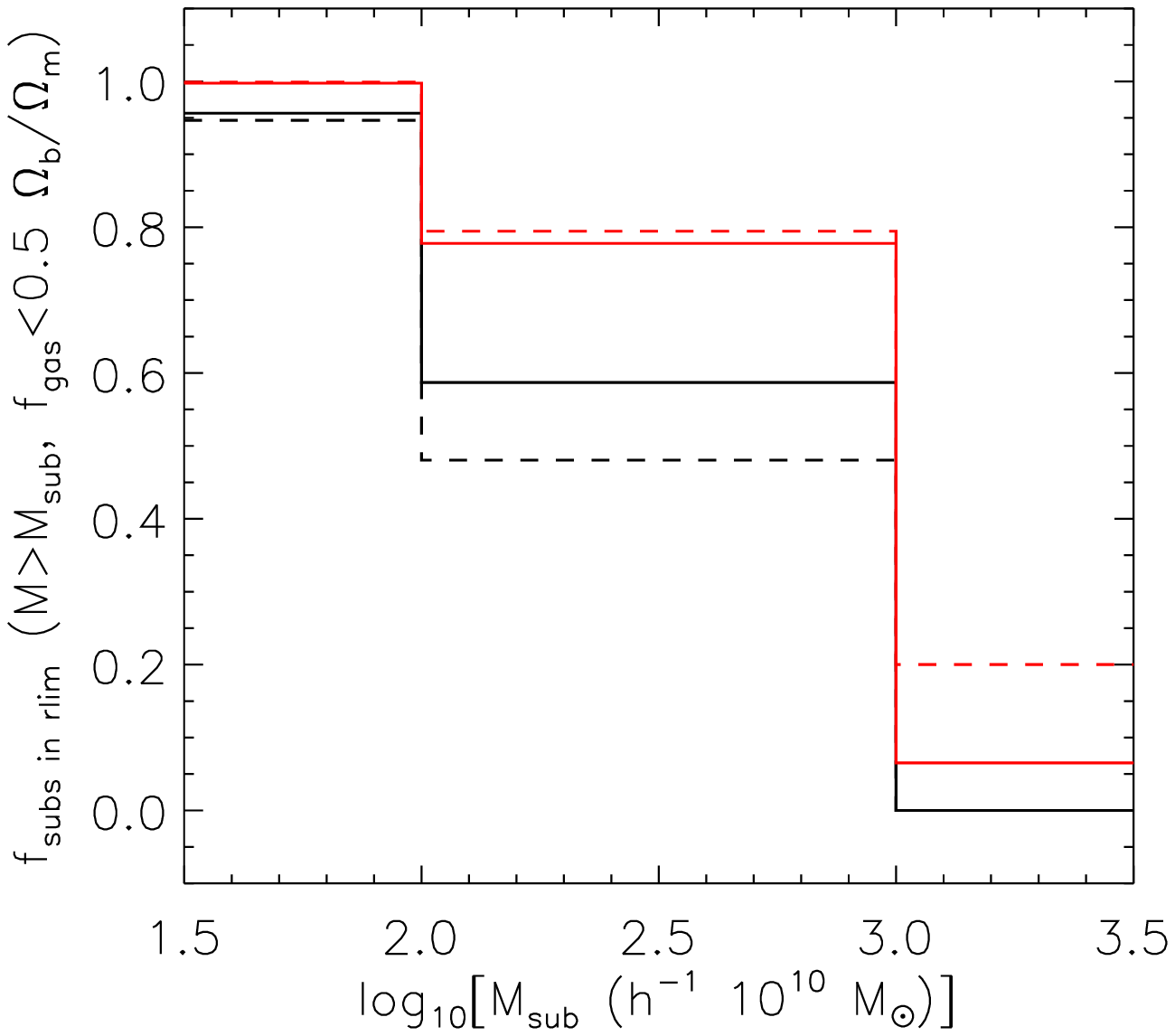}
\caption{
Left panel: fraction of resolved subhaloes within the full map region that lie inside 
$r_{500}$ (solid lines) and 2$r_{500}$ (dashed lines). Dot-dash line represents fraction within $r_{500}$ expected for a uniform distribution (from ratio of volumes). 
Middle panel: fraction of subhaloes within $r_{500}$ (solid) and 
the the full map region (dashed) that have no hot gas ($f_{\rm gas}=0$). Note that all lines are at zero in high mass bin.
Right panel: as in the middle panel but for subhaloes that have little hot gas
($f_{\rm gas} \leq 0.5 \, \Omega_{\rm b}/\Omega_{\rm m}$). Note that both black lines are at zero in high mass bin.
In all panels, results for the same subhalo DM mass 
($10^{11}<M_{\rm sub}/h^{-1}{\rm M}_{\odot}<10^{12}$; $10^{12}<M_{\rm sub}/h^{-1}{\rm M}_{\odot}<10^{13}$;  
$M_{\rm sub}>10^{13}h^{-1}{\rm M}_{\odot}$) and redshift (red lines: $0 \leq z \leq 0.2$; black lines: 
$0.5 \leq z <1.0$) intervals are shown. Note that data for all three clusters are stacked here and that the main clusters themselves are included in these data.}
\label{boxVr500}
\end{figure*}

Before we begin discussion of the results in this section, it is important to note that we always include the main cluster halo in the subhalo data. This is important to facilitate the comparison to 2D substructures detected in the maps later on, as the {\it cores} of the clusters (see the central mass density peaks clearly evident in the first column of Fig. \ref{zeq0mapfig}) are detected in 2D and these cores, therefore, are detected as 2D substructures in their own right.

In Fig.~\ref{mfclusfig} we show the cumulative subhalo DM mass function for subhaloes with their most bound particle inside 3D $r_{500}$, down to our imposed resolution limit of 100 DM particles. The results for cluster A (solid), B (dotted) and C (dashed) are shown individually for $z=1$ (left), 0.5 (middle) and 0 (right) in the first row.
Note that the main cluster itself is the most massive subhalo. The total number of resolved subhaloes, ranging from less than 10 to nearly 60, depends on cluster mass. For example, cluster C has significantly fewer subhaloes at $z=1$ and $z=0.5$ than the 
other two clusters, but has more at $z=0$. This increase reflects cluster C's major merger at $z \sim 0.4$, as seen in Fig.~\ref{mh:fig}. 
However, when the subhalo masses are scaled to the parent cluster mass, the scatter between clusters and redshifts is much smaller, as shown in the second row. 

We have also examined how the properties of subhaloes in the map region vary depending on whether or not they lie within $r_{500}$ in 3D, to assess the impact of subhaloes projected along the line of sight. It is particularly important that we examine the distribution of subhaloes, because of the unusual geometry we are using (a cylinder, rather than a sphere). In the left panel of Fig.~\ref{boxVr500}, we show the fraction of subhaloes (including the main cluster halo) in the entire map region that lie within $r_{500}$ (solid) and $2r_{500}$ (dashed) in 3D for the redshift intervals $0 \leq z \leq 0.2$ (red lines) and $0.5 \leq z <1.0$ (black lines). These redshift intervals were chosen to include an equal number of snapshots (11). Around half of the low mass subhaloes lie within $r_{500}$ which is significantly higher than for a uniform distribution, for which we would expect, $\frac{V_{\rm sphere}}{V_{\rm cylinder}} = \frac{4}{3}\pi r_{500}^{3} / 8\pi r_{500}^3 = 1/6$ (indicated with the dot-dash line). Nearly all subhaloes ($80-100$ per cent) lie within $2r_{500}$, suggesting that the contribution to the map from substructure outside the cluster's virial radius is small. The rise, compared to lower mass bins, in the fraction of subhaloes  with $M_{\rm sub}>10^{13}h^{-1}{\rm M}_{\odot}$ within $r_{500}$ is due to the presence of the cluster cores. The cluster cores dominate this bin (in number) and since the map region is centred on them, they are always within $r_{500}$ by design. At high redshift, the fraction of galaxy-sized ($M_{\rm sub}<10^{13} h^{-1}{\rm M}_{\odot}$) subhaloes within $r_{500}$ is approximately $10$ per cent higher than at low redshift. This is likely to be caused by the effects of tidal forces, stripping the DM as the subhalo orbits in the cluster potential. This effect may reduce the likelihood of finding subhaloes which are dark in X-rays in this mass range at low redshift, since they may move to lower DM mass bins (via tidal stripping) shortly after their hot gas is removed. 

Given the aims of this investigation, we want to try to place some limits on the fraction of DM substructures without X-ray emission we expect to find. In the middle and right panels of Fig.~\ref{boxVr500}, we now plot the fraction of subhaloes (within $r_{500}$ (solid) and the full map region (dashed)), with no hot (T$>10^{6}K$) gas ($f_{\rm gas}=0$) and little hot gas ($f_{\rm gas}\leq 0.5 \Omega_{\rm b}/\Omega_{\rm m}$), respectively. This somewhat arbitrary threshold, $f_{\rm gas}\leq 0.5 \Omega_{\rm b}/\Omega_{\rm m}$, was chosen simply to distinguish hot gas-poor subhaloes from hot gas-rich subhaloes. Note that, by definition, the rest of the subhaloes ($1-f_{\rm subs}$) fall into the latter category and have $f_{gas}>0.5  \Omega_{\rm b}/\Omega_{\rm m}$. The main trend apparent is that the fraction of empty or low-gas subhaloes is higher at lower mass, in agreement with \citet{2004MNRAS.350.1397T}, who find the survival time of hot gas in subhaloes is a strong increasing function of subhalo mass.  Without the added effects of radiative cooling and energy injection from galactic winds, for example, our results already predict that the vast majority of galaxy-sized ($10^{11}\le M_{\rm sub}/h^{-1}{\rm M}_{\odot}<10^{13}$) subhaloes are substantially depleted of hot gas, while the opposite is true on group (and cluster) scales. We also find that more subhaloes have no hot gas at low redshift than at high redshift, in agreement with \citet{newsubfind}. We note that the middle panel of Fig.~\ref{boxVr500} is insensitive to the temperature threshold, since the vast majority of subhaloes with no hot ($T>10^{6}$K) gas have no gas of any temperature.

The vast difference between the fraction of empty subhaloes in the lowest mass bin and at higher masses (e.g. $\simeq 80$ per cent of subhaloes with $10^{11}\le M_{\rm sub}/h^{-1}{\rm M}_{\odot}<10^{12}$  already gas-free at high redshift, yet still only $\simeq 30$ per cent with $10^{12}\le M_{\rm sub}/h^{-1}{\rm M}_{\odot}<10^{13}$  gas-free at low redshift) is qualitatively in agreement with \citet{2004MNRAS.350.1397T}, if we assume higher redshift to indicate less time since infall. They find complete removal of hot gas within $1$ gigayear (typically massive galaxies) to $3$ gigayears (typically groups) of entering the cluster's virial radius on average. Results from \citet{2008MNRAS.383..593M} are in general agreement, but indicate $\simeq 30$ per cent of hot gas in a halo (typically a massive galaxy) can survive much longer ($\simeq10$ gigayears); a result shown to improve colours of satellite galaxies in semi-analytic models \citep{fontetal_2008}. We find that the majority of subhaloes with $M>10^{12}h^{-1}{\rm M}_{\odot}$ always retain some hot gas and indeed at least $20$ per cent have $f_{\rm gas}> 0.5 \Omega_{\rm b}/\Omega_{\rm m}$. This shows our results are compatible with subhaloes retaining some of their original hot gas, although it seems in most cases the majority is removed. 

\citet{newsubfind} find that stripping is very efficient with $\simeq99$ per cent of all subhaloes in $r_{\rm vir}$ being gas-free at $z=0$. Note that this percentage will be dominated by their low mass subhaloes which are most numerous (and most gas-deficient) and so compares well with the percentage ($90$ per cent) that we find in our low mass bin. It remains to be seen how much gas has to be stripped before the likelihood of detecting the substructure in both the total mass and X-ray surface brightness maps is affected.

\section[]{2D Substructure detection}\label{2dsec}

A number of authors have used 2D weak lensing maps and X-ray images of clusters, both to compare the spatial distribution of hot gas and dark matter in these objects \citep[e.g.][]{bulletclowe,2007ApJ...668..806M,bullet2} and to help infer their dynamical state, by measuring the offset between the centres of these two components \citep{smithetal2005}. The scope of the information about the underlying 3D system which such 2D comparisons could potentially provide, has not yet been explored and the present study is the first attempt to do this. 

The key features of this piece of work  are a simple, yet effective, technique for identifying substructure in 2D maps of simulated clusters, in combination with an easy-to-use method for mapping 2D mass substructures to both 2D X-ray substructures and 3D subhaloes. First, we analyse our `perfect' observations (i.e. noise-free maps). This allows us to establish how many projected mass and X-ray substructures can, in principle, be uniquely identified despite projection effects and the intensity of the cluster background.  This approach also provides insight into the fate of a projected mass substructure's hot gas when an X-ray counterpart is not found; the maps (unlike 3D data) allow immediate visual follow up and reveal interesting features of the stripping process. We will explore how much of this is observable with current techniques in Section \ref{real}, by degrading the map resolution and adding noise to both map types.

\subsection[]{Detection technique}\label{ussec}

\begin{figure*}
\centering
\subfigure[Pre-smoothed mass map.]{
  \label{pstm}
  \includegraphics[width=0.23\textwidth]{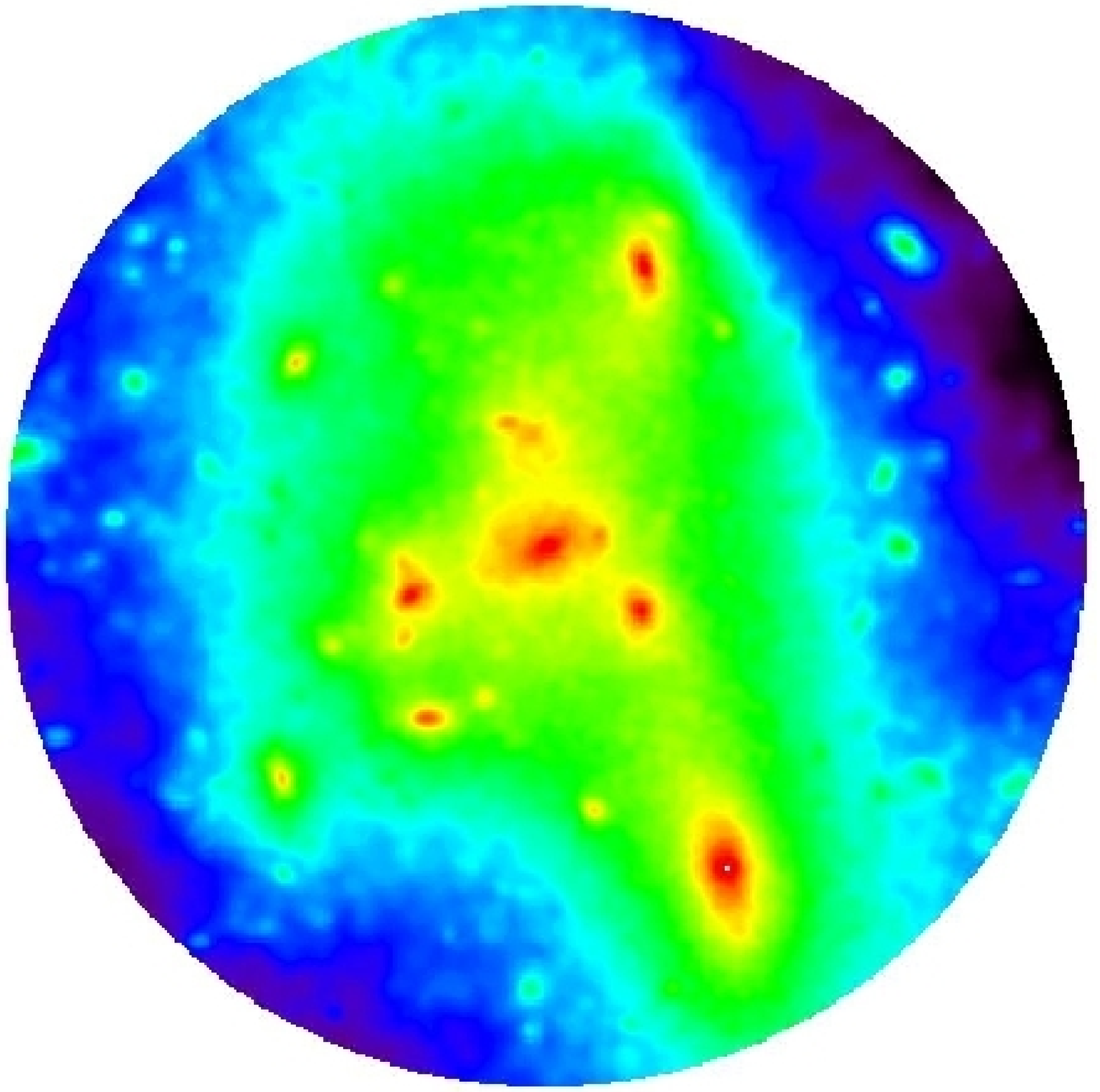}}
  \hspace{0.005\textwidth}
\subfigure[Unsharpmask for mass map.]{
  \label{smoothtm}
  \includegraphics[width=0.23\textwidth]{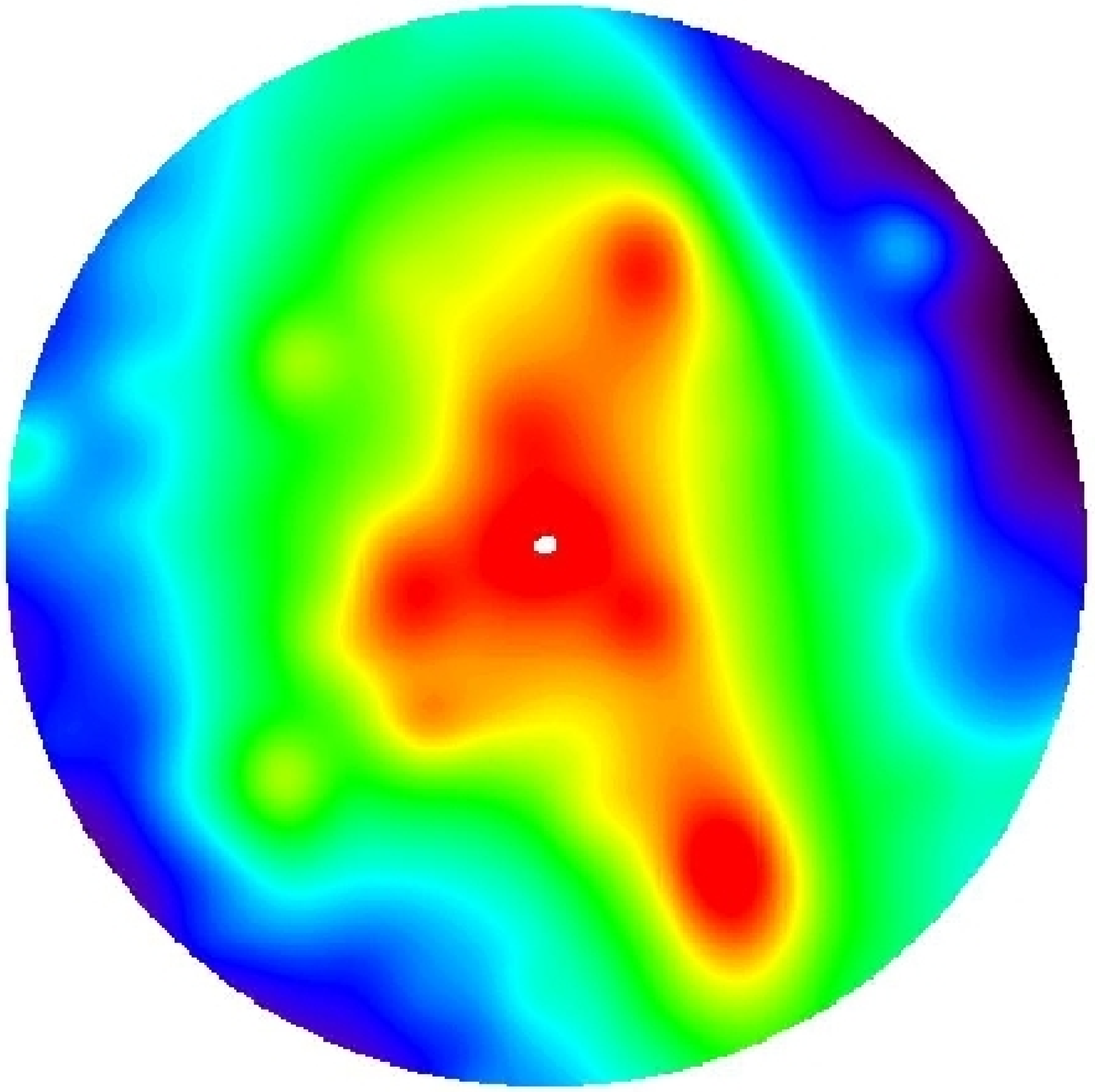}}
   \hspace{0.005\textwidth}
\subfigure[Unsharpmasked mass map after a cut has been made.]{
  \label{ustm}
  \includegraphics[width=0.23\textwidth]{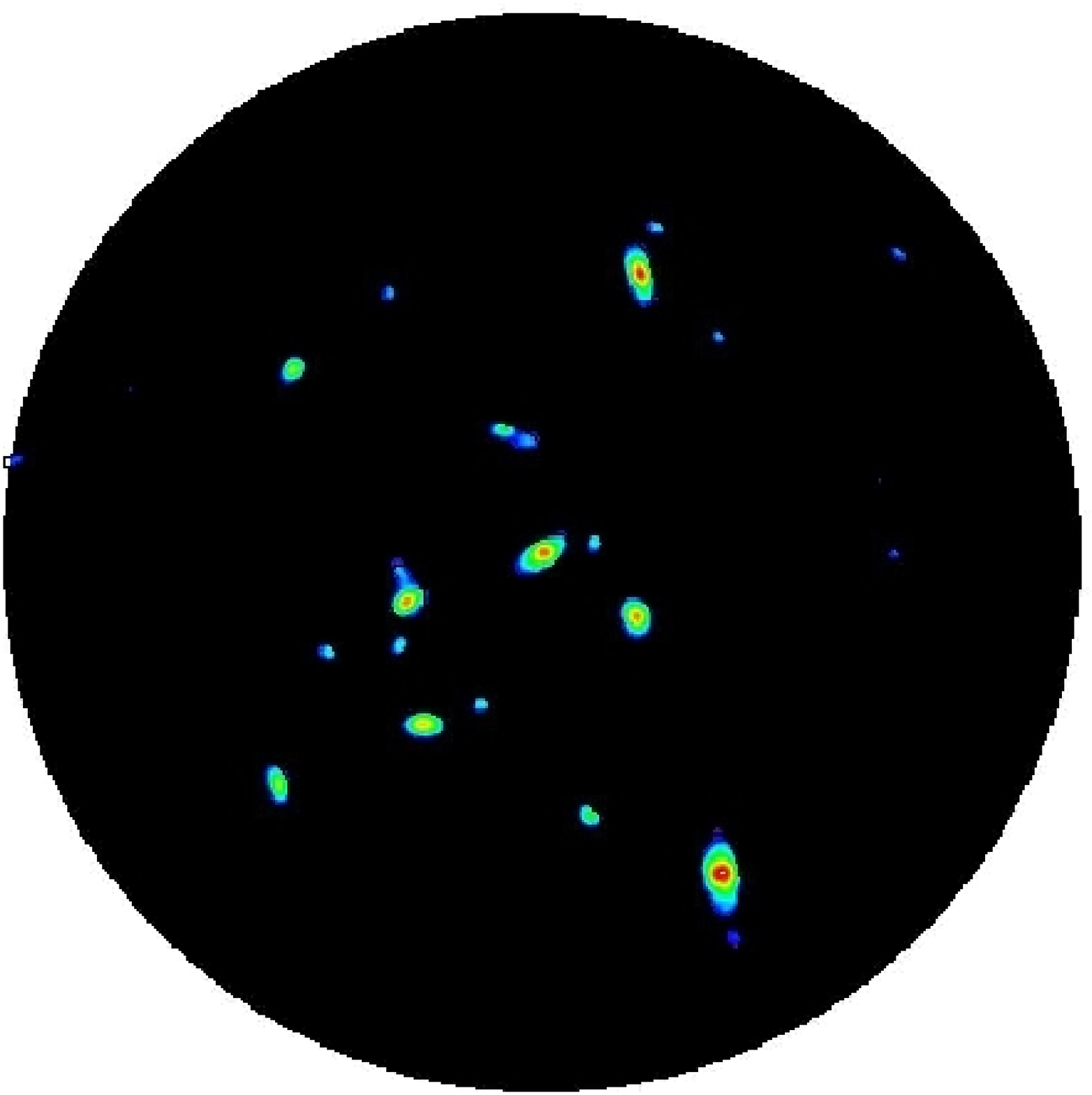}}
  \hspace{0.005\textwidth}
\subfigure[Substructures detected in mass map.]{
  \label{finaltm}
  \includegraphics[width=0.23\textwidth]{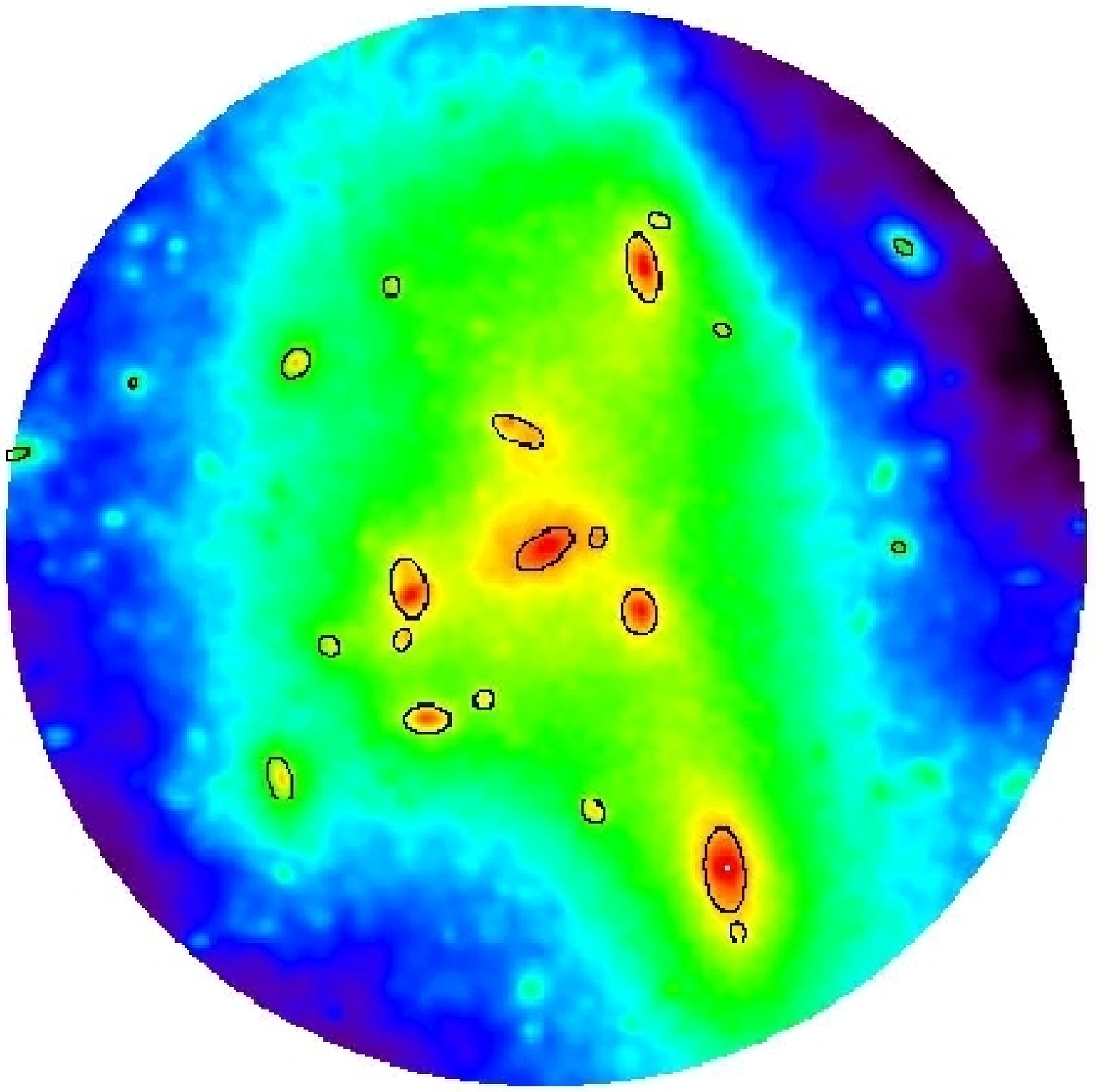}}\\
  
 \subfigure[Pre-smoothed X-ray map.]{
  \label{pssb}
  \includegraphics[width=0.23\textwidth]{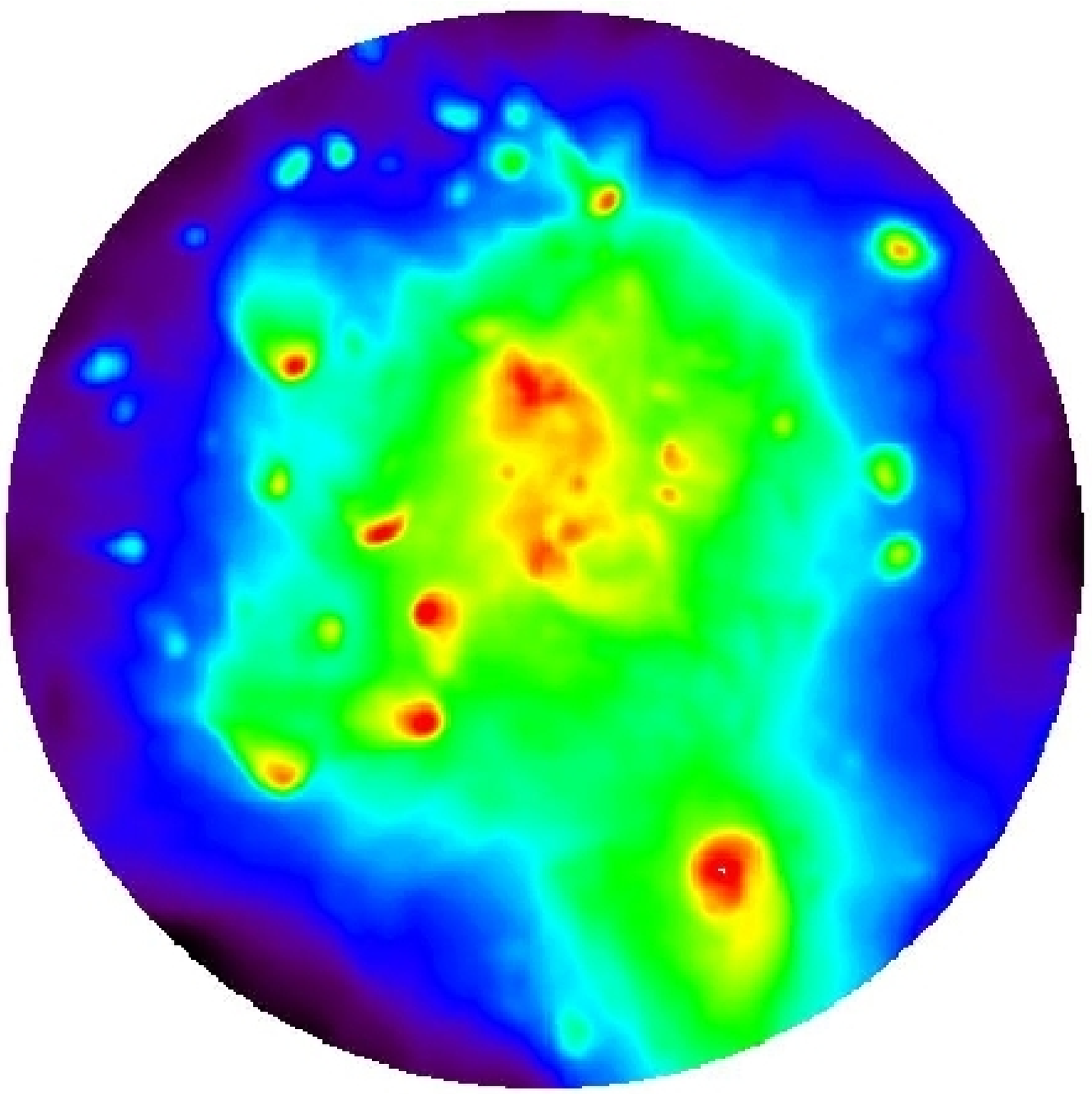}}
  \hspace{0.005\textwidth}
\subfigure[Unsharpmask for X-ray map.]{
  \label{smoothsb}
  \includegraphics[width=0.23\textwidth]{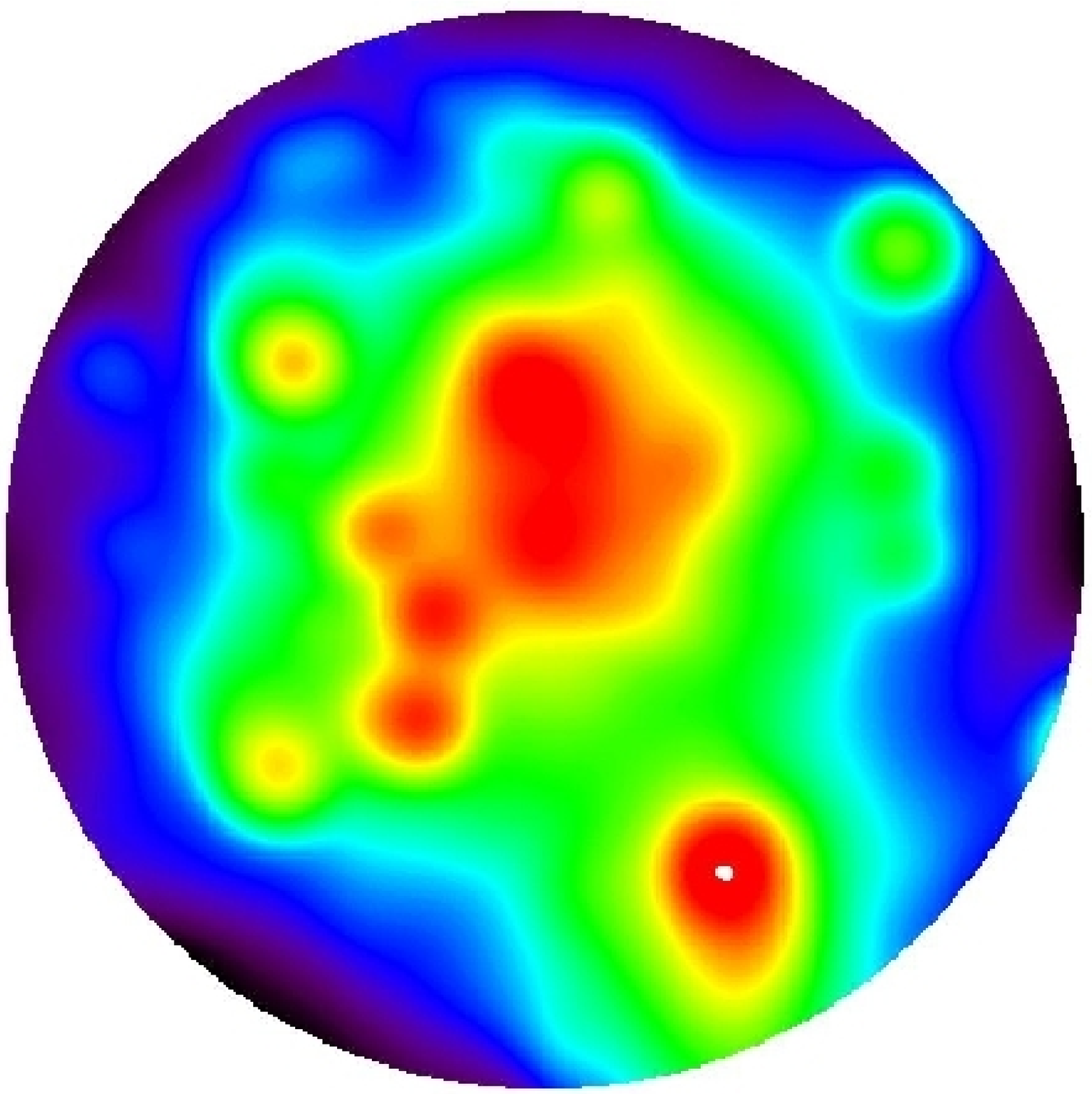}}
   \hspace{0.005\textwidth}
\subfigure[Unsharpmasked X-ray map after a cut has been made.]{
  \label{ussb}
  \includegraphics[width=0.23\textwidth]{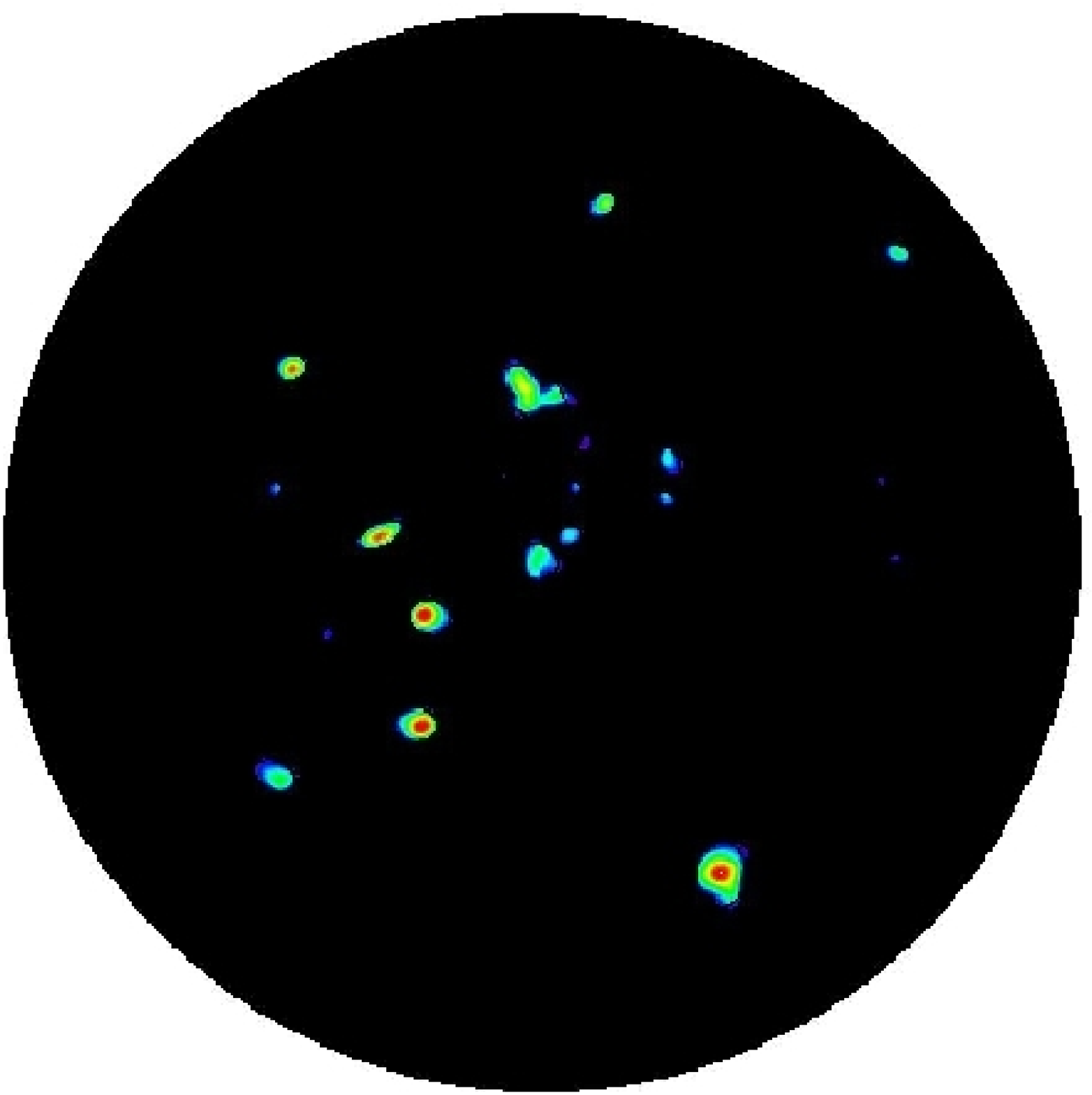}}
  \hspace{0.005\textwidth}
\subfigure[Substructures detected in X-ray map.]{
  \label{finalsb}
  \includegraphics[width=0.23\textwidth]{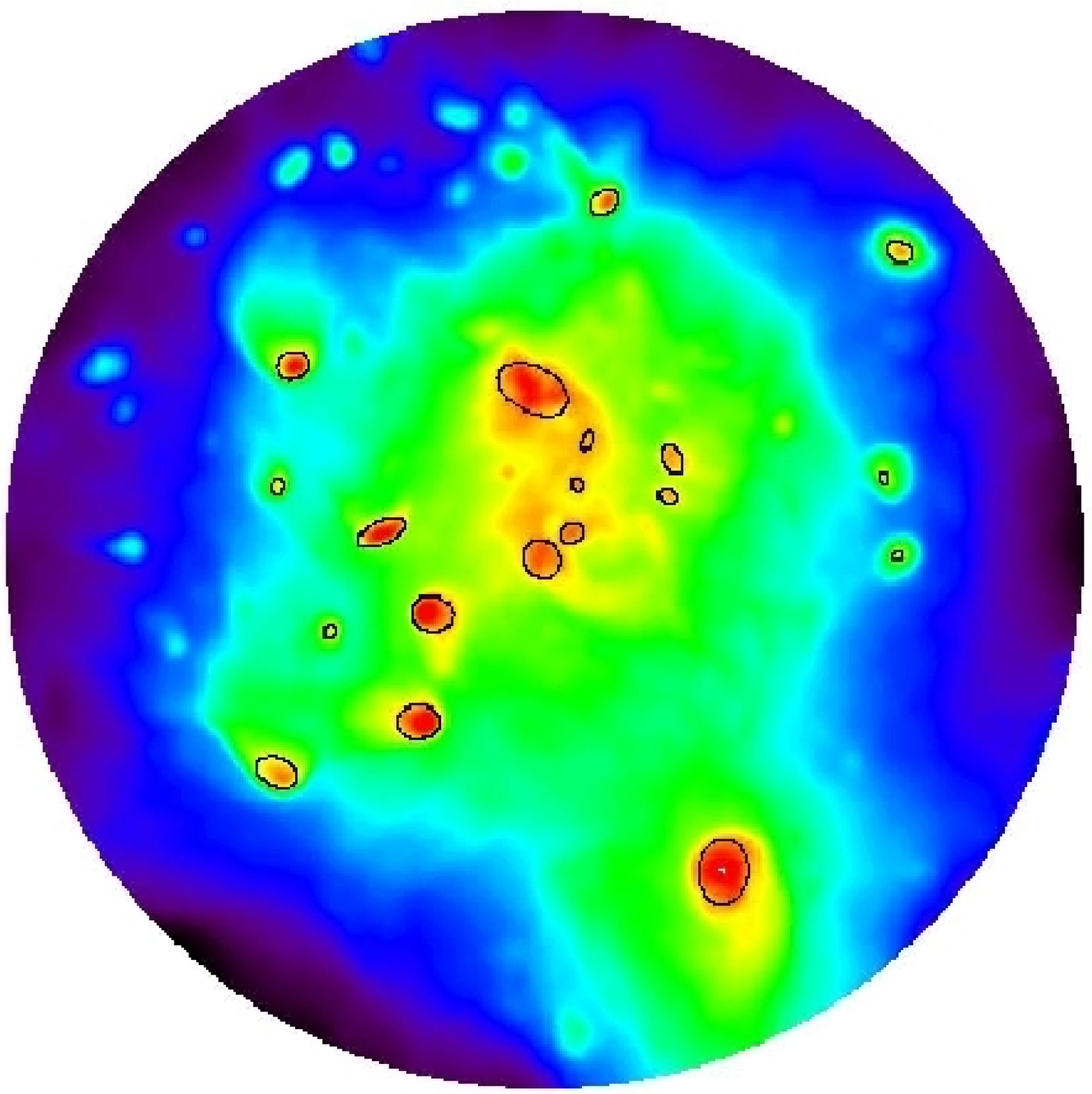}}

\caption{An example of our substructure detection procedure for cluster C at $z = 0.162$. The
top row corresponds to the surface mass density maps and the bottom row to the X-ray maps 
surface brightness maps. See text for further details.}
\label{maps:fig}
\end{figure*}

The first step towards detection is to enhance the substructure in the maps. For this purpose we use a method based on the unsharp-masking technique, in order to remove the cluster background. 
The unsharp-masking technique itself has already been used as a visual aid by highlighting small-scale structure in X-ray images of galaxy clusters, for example \citet{fabperseus03,fabcentaurus}. 
The main advantage is that it does not rely on the cluster being circularly symmetric, recovering the distribution of substructure well even in the most complex scenarios (i.e. multiple mergers, as is sometimes the case in our simulations, especially at high redshift).

The first stage of the procedure is to smooth the maps 
with a preliminary Gaussian filter. This could be used to emulate the point-spread-function
of a real instrument, but here we set the Full-Width-Half-Maximum (FWHM) to simply match the 
spatial resolution of the simulation, as our results are presented in the limit of no added 
noise (other than intrinsic discreteness noise due to the finite number of particles employed).
Our maps contain a fixed number of pixels (200) across $r_{500}$, corresponding to a length scale for each pixel
of around $5 h^{-1}{\rm kpc}$ at $z = 0$, which is the equivalent Plummer softening length of our simulation
(held fixed in proper units over the redshift range of interest here). 
The minimum length scale that should be trusted is around 3 times this, corresponding to the extent
at which the gravitational force law becomes perfectly Newtonian in the {\sc gadget2} code. Furthermore,
$r_{500}$ is smaller at higher redshift, so our pixel scale is also smaller. We therefore set 
${\rm FWHM}=15 h^{-1}{\rm kpc}$ (physical) for all maps studied in this paper. It should also be noted that 
the maps are generated to be larger in X and Y than required so that the larger maps can be analysed to 
avoid edge effects in the region of interest. Panels (a) and (e) in Fig. \ref{maps:fig} illustrate examples of these {\it pre-smoothed} maps.

The second stage is to convolve the pre-smoothed maps again with a broader Gaussian kernel, to 
create the unsharp-mask image, shown in Fig. \ref{maps:fig}, panels (b) and (f). Here we fix $\sigma_{2}$ to be 0.05 $r_{500}$ (corresponding to a FWHM ranging from approximately $35 h^{-1}{\rm kpc}$ to $120 h^{-1}{\rm kpc}$ over the redshift range), which was deemed to be the most effective value
from extensive testing. This twice-smoothed version of the map is then subtracted from the 
pre-smoothed map, leaving a map showing just the enhancements to the cluster background. 

\begin{figure}
  \includegraphics[width=0.5\textwidth]{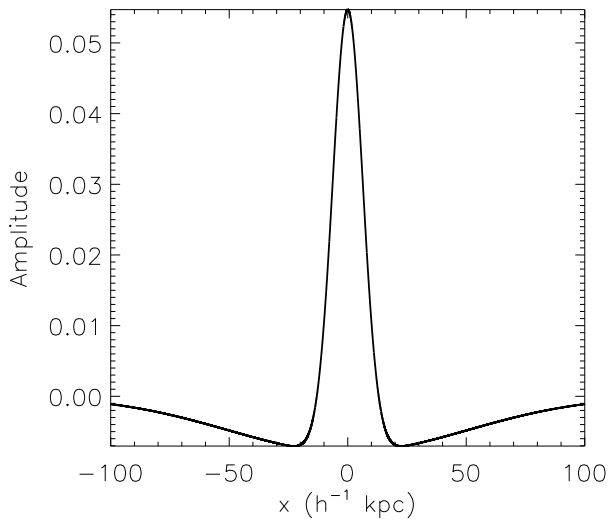}
 \caption{1D visualisation of the convolution kernel equivalent to the whole 
background-subtraction procedure. For this example, a typical value of 1  $h^{-1}$ Mpc is used for the cluster radius, giving $\sigma_{2}=50$ $h^{-1}$ kpc, $\sigma_{1}\sim 6$  $h^{-1}$ kpc (as defined in equation \ref{sig1eq}) and $r_{0}\sim 12$ $h^{-1}$ kpc (as defined in the 1D analogue of equation \ref{radkernel}). Note that this is for illustration only and as such the area under each Gaussian is normalised to 1 between $\pm\infty$ in 1D before the difference of the two is taken.}
\label{mexhatfig}
\end{figure}

Utilising the commutative, distributive and associative properties of convolution, it is possible to 
derive one function that, when convolved with the map image, produces the same result as the series of 
operations described above. The kernel used to generate the pre-smoothed map approximates the Gaussian function, which is given by
\begin{equation}
G_{prelim}=N_{1} e^{ -\frac{ x^{2}+y^{2} }{ 2\sigma_{1}^{2} } },
\end{equation}
where the normalisation,
\begin{equation}
N_{1}=\frac{1}{2\pi\sigma_{1}^{2}},
\end{equation}
and, in this case, 
\begin{equation}
\sigma_{1}=\frac{15 h^{-1}{\rm kpc}}{\sqrt{8\ln{2}}}
\label{sig1eq}
\end{equation}
which is set by the spatial resolution of the simulation. Similarly,
the combined operations of pre-smoothing and generating the unsharp-mask image is simply
\begin{equation}
G_{U.S.}=N_{1,2} e^ {-\frac{x^{2}+y^{2}} {2(\sigma_{1}^{2}+\sigma_{2}^{2}) } },
\end{equation}
where the normalisation is now
\begin{equation}
N_{1,2}=\frac{1}{2\pi(\sigma_{1}^{2}+\sigma_{2}^{2}) },
\end{equation}
and $\sigma_{2}$ is $0.05$$r_{500}$. The function
representing the entire procedure,
\begin{equation}
F=G_{prelim} - G_{U.S.},
\label{kernel}
\end{equation}
which is a close approximation to the Mexican-hat function or the matched filter defined by \citet{matchedfilter_babul}, is shown in 
Fig.~\ref{mexhatfig} . 

The size of substructures that are detected are dependent on the combination of the standard 
deviations of the Gaussians used to obtain the final image. We derive an expression that 
characterises the width of the kernel and therefore the scale of substructure to which our 
technique is sensitive. The characteristic width of the function in Fig.~\ref{mexhatfig} 
can be determined by calculating the radius at which the amplitude of the function is zero. 
The radius of the zero-points, $r_{0}$, is given by,
\begin{equation}
r_{0}=\sqrt{2} [\frac{\sigma_{1}^{2}+\sigma_{2}^{2}}{\sigma_{2}^{2}/\sigma_{1}^{2}} 
\ln[\frac{\sigma_{1}^{2}+\sigma_{2}^{2}}{\sigma_{1}^{2}}] ]^{\frac{1}{2}}.
\label{radkernel}
\end{equation}
Since $\sigma_{2}$ is expressed in units of $r_{500}$, it has a slight redshift dependency (e.g. for cluster B, $r_{500}=0.34$ $h^{-1} {\rm Mpc}$ at $z=1$ and 0.78 $ h^{-1} {\rm Mpc}$ at $z=0$), 
meaning that more extended substructures will be detected at lower redshifts. However, the 
increase in the value of the kernel width is only of the order of 20 per cent of its maximum value over the range of 
redshifts studied, $0 \leq z < 1$ (e.g. $r_{0}=0.014$ $h^{-1} {\rm Mpc}$ for $z=1$ and 0.018 $h^{-1} {\rm Mpc}$ for $z=0$, averaged over the 3 clusters). We are limited to detect only 2D mass substructures of the order of the size of the kernel and these 2D mass substructures will, of course, be associated with a 3D subhalo mass. Since the typical size of a 3D subhalo above a given mass is larger at lower redshift, due to the decrease in the critical density as the universe expands, the trend for the kernel to also be larger at lower redshift actually reduces the redshift-dependence of the minimum 3D subhalo mass which we can detect in 2D.

\begin{figure*}
\centering
\includegraphics[width=0.33\textwidth]{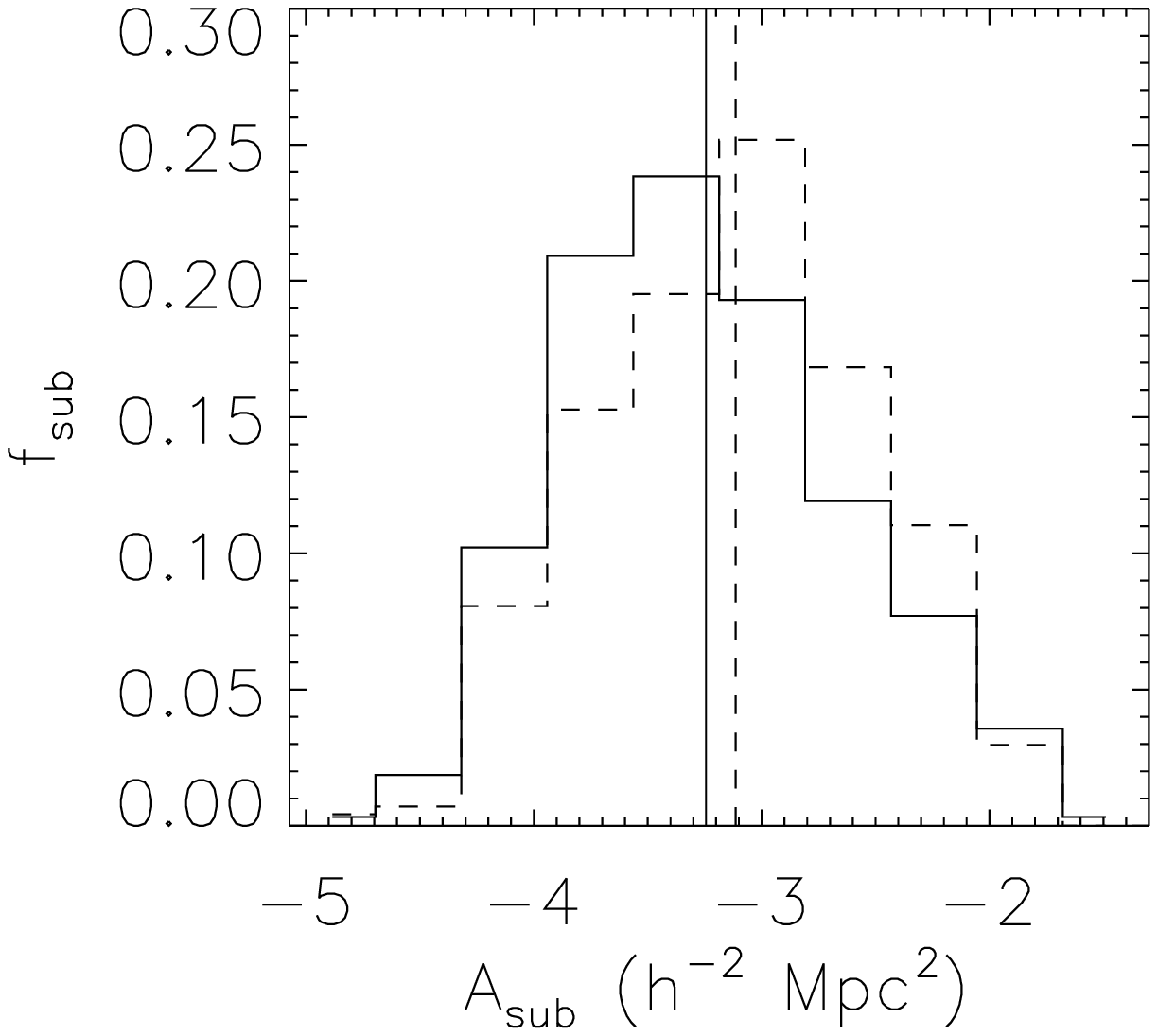}
\includegraphics[width=0.33\textwidth]{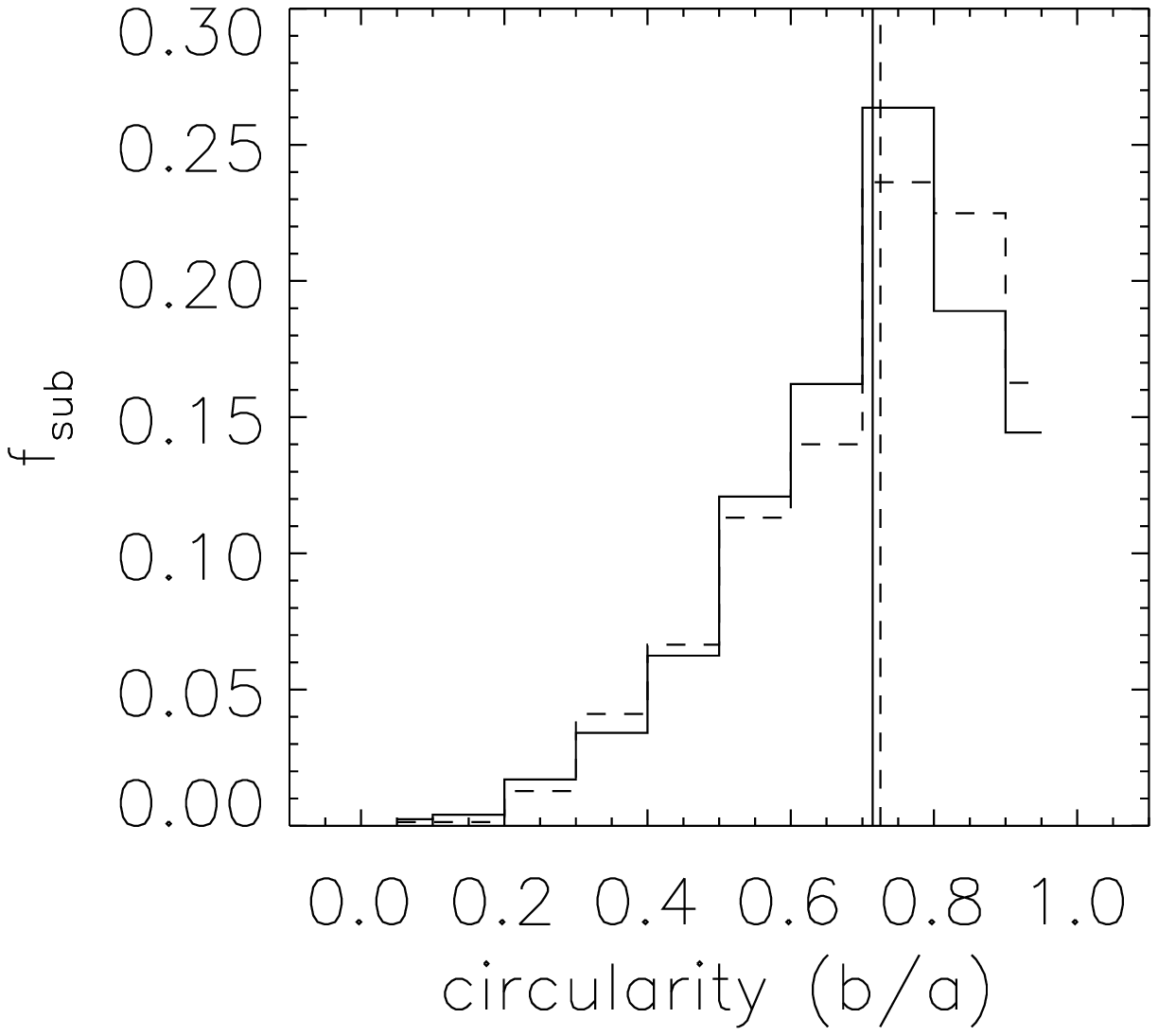}
\includegraphics[width=0.33\textwidth]{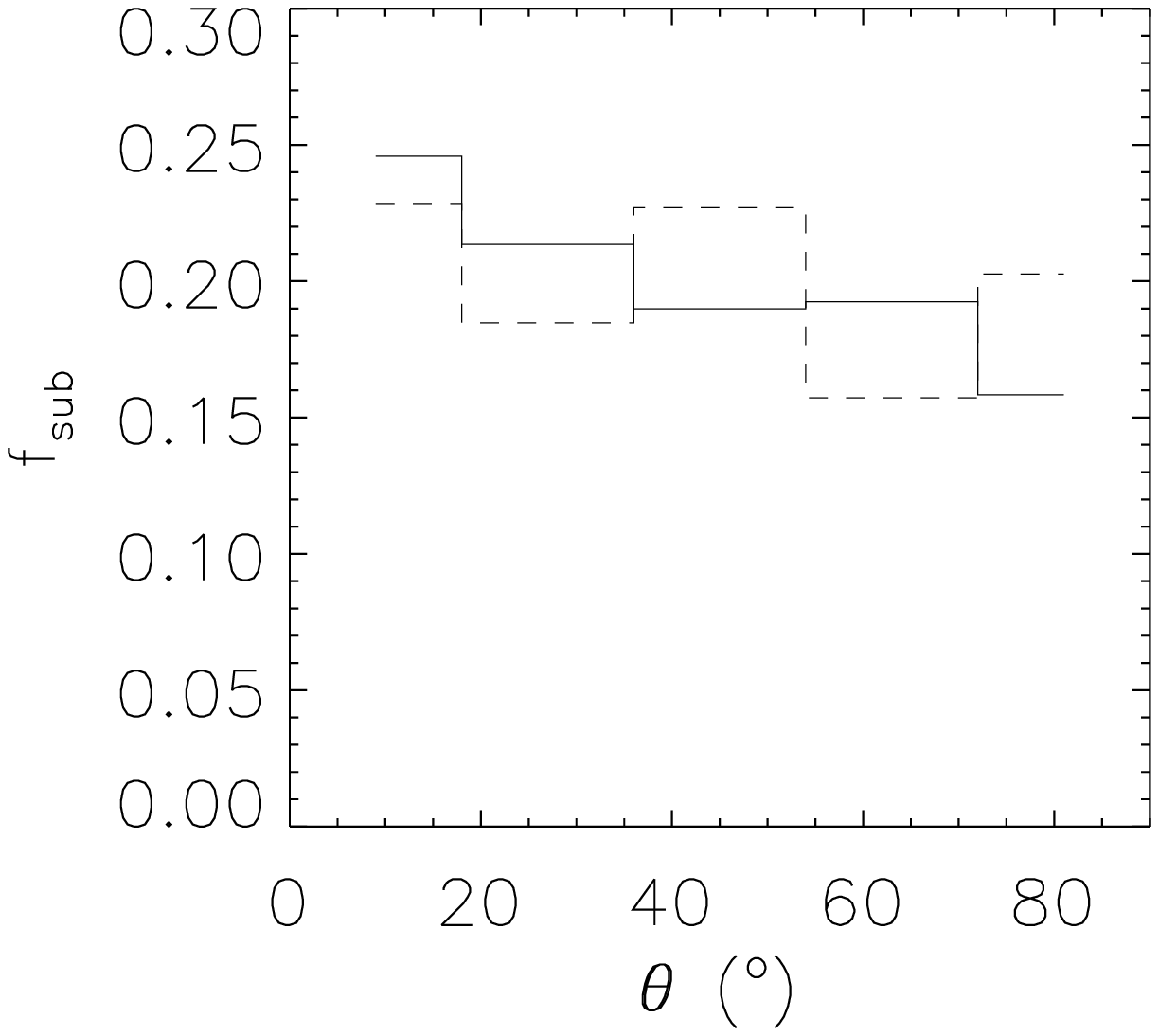}
\caption{Left: Area distribution of substructures. Middle: Distribution of circularities for substructures. Right: Angle between semi-major axis vector of substructure and radial vector from substructure centre to cluster centre. Surface mass density (solid) and X-ray surface brightness (dashed) $3\sigma$ catalogues. Vertical lines indicate the mean values in each case.
}
\label{ellipsefig}
\end{figure*}

In order to pick out the true substructures from other fluctuations, any pixels with values less than  
$\mu + X\sigma$ (where $X$ is an integer, representing our detection threshold,  
and $\mu$ and $\sigma$ are the mean and standard deviations 
of the residual substructure maps) are discarded. Examples of the resulting maps at this stage of the procedure are shown in  Fig. \ref{maps:fig}, panels (c) and (g). Substructures are then defined similarly to the FoF technique but in 2D,
grouping together neighbouring pixels with values greater than the background level. Ellipses are fitted to these pixel groups by finding the eigenvectors (corresponding to the direction of the semi-major and semi-minor axes) and the eigenvalues (whose square roots correspond to the magnitude of the semi-axes) of the moment of inertia tensor. This allows us to determine the extent, orientation and circularity of the 2D substructures (see below).

We investigate three values of $X$ for the projected total mass map: 1, 3 and 5, and evaluate which 
is most successful when the comparison with the 3D subhalo data is made in Section \ref{2d3dsec}.
It was found that the X-ray surface brightness maps respond slightly better to our technique due to the 
fact the gas distribution is far smoother (because it traces the gravitational potential) and contains fewer small-scale fluctuations, 
meaning that less stringent cuts are required in order to achieve the same results (upon visual 
inspection). Therefore, the selection of the parameters used to define the catalogue of X-ray 
substructures is undertaken separately to that for the mass substructures. We found that $X=5$ is too stringent for the X-ray maps, removing substructures that are clearly 
visible by eye, whereas $X=1$ and $X=3$
produce reasonable results for both the X-ray and the mass maps.

\subsection[]{Properties of 2D substructures}\label{2dpropertiessec}

The total number of substructures detected in the 1, 3 and 5$\sigma$ total mass catalogues (which consist of 90 maps, i.e. 30 per cluster) is 3224, 1233 and 680 respectively. It is clear from these numbers that, as would be expected, the higher the value of {\it X}, the lower the number of detections. There are also considerably fewer X-ray substructures than total mass substructures for the same $X$ value, with 1169 in the $1\sigma$ X-ray catalogue and 707 in the $3\sigma$ X-ray catalogue. This can in part be attributed to the smoother distribution of the hot gas, which responds differently to the unsharp-masking procedure. However, it is also apparent (on visual inspection) that there is simply less inherent substructure in the X-ray maps, particularly on small scales.

First, we examine the distribution of total mass (solid) and X-ray surface brightness (dashed) substructure areas, $A_{\rm sub}$, in the left-hand panel of Fig. \ref{ellipsefig}. $A_{\rm sub}$ is defined as the number of pixels attributed to the 2D substructure in the unsharp-masked image multiplied by the physical area of the individual pixels. The distributions are very similar, except the X-ray surface brightness distribution peaks at a slightly higher value of $A_{\rm sub}$, suggesting  the X-ray substructures are typically more extended (this is confirmed by visual inspection). There is little dependence on choice of catalogue here. 

We examine the shape of the substructures by plotting the distribution of circularities in the middle panel of Fig.~\ref{ellipsefig} (line styles as before). Here, we define circularity, $c=(b/a)$ where $a$ is the major axis and $b$ the minor axis of the ellipse. This distribution is very stable to choice of catalogue, suggesting that the morphology of the objects we detect 
changes little between catalogues. The distribution of mass substructures 
peaks at $c \sim 0.75$, due to the triaxial
nature of the DM substructure; the gas is slightly rounder and peaks at $c \sim  0.8$. 
\citet{knebe_projectsub} obtain a similar result for their projected sphericity of DM subhaloes, computed from the particles directly. This indicates that our detection method recovers the true 2D shape of the substructures successfully.

The right-hand panel of Fig.~\ref{ellipsefig} shows the distribution of the radial alignment of the 2D substructures with respect to the cluster centre. This is computed by first calculating the angle of inclination between the cluster centre (in 2D) and the centre of the ellipse representing a substructure. The alignment, $\theta$, is then found by subtracting from this the angle of inclination of the semi-major axis of the ellipse. The range of $\theta$ can be reduced to $0-90^{\circ}$ by treating opposite directions of the semi-major axis vector as equivalent. A mild tendency towards alignment is exhibited by the total mass substructures (solid), however the X-ray substructures (dashed) show no preferred direction. \citet{knebe_projectsub} perform a similar calculation for their projected DM subhaloes and found a much stronger tendency for alignment than we see here, when they considered all particles associated with the subhalo. However, they investigate the effect of varying the percentage of particles they analyse by limiting the alignment measurement to the inner regions of the subhaloes. The trend for alignment shown in their results weakens as smaller percentages of particles are considered and comes into agreement with observations when analysing the inner $10-20$ per cent. Our result is also in much better agreement with theirs for this region. This reflects the fact  that our 2D detection technique finds only the cores of the original 3D subhaloes, which is demonstrated by the small scale of the detected 2D substructures. Therefore, we are effectively performing our alignment and circularity analysis on only the innermost particles and so find best agreement with \citet{knebe_projectsub} when they similarly restrict their analysis.

\section[]{Comparison of 2D mass substructures to 3D subhaloes}\label{2d3dsec}

In this section, by comparing the 2D total mass substructures (described in Section 
\ref{2dsec}) with 3D self-bound DM subhaloes (described in Section 
\ref{detectsec}), we assess the reliability of our 2D detection method and 
infer the 3D properties (e.g. subhalo mass) of our 2D substructures. We examine the completeness (with respect 
to 3D) of our 2D catalogues, as well as the number of individually resolved high-mass objects 
they contain, in order to select one total mass catalogue that is most suited for the analysis 
in later sections. Our catalogues contain substructures identified in all three clusters and
at all redshifts ($0 \leq z < 1$).

Ideally, we want to be 
$100$ per cent complete down to at least $M_{\rm sub} \sim 10^{13} h^{-1}{\rm M}_{\odot}$, as this is the typical 
mass scale of substructures detected in current observations of the unusual systems discussed in Section \ref{intro}. However, high completeness at lower masses would be desirable as smaller subhaloes are 
the more likely ones to be found stripped of their gas \citep{2004MNRAS.350.1397T}. An additional criteria we wish to place 
on any detections is that, ideally, they are individually resolved (i.e. not confused with another 
subhalo that is nearby in projection). We also look at the purity of our 2D substructure catalogues by
assessing the fraction of 2D substructures which we fail to associate with 3D subhaloes.
 
 \subsection{Completeness}

\begin{figure}
\centering
  \includegraphics[width=0.5\textwidth]{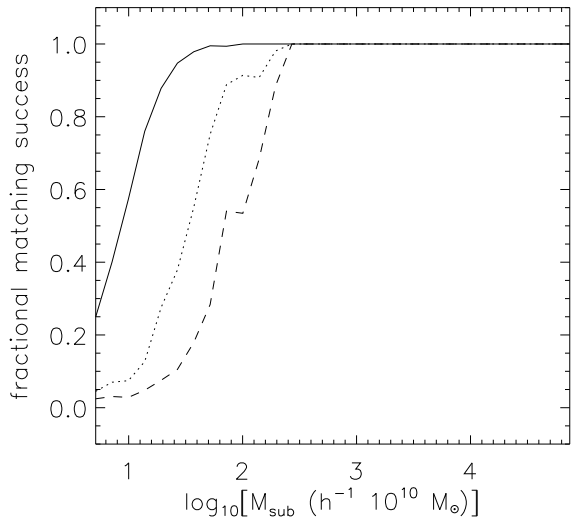}
\caption{Fractional matching success {\it per mass bin} of 3D subhaloes to 2D mass substructures as a function of 
subhalo DM mass for $1\sigma$  2D catalogue (solid line), $3\sigma$ 2D catalogue (dotted line) and $
5\sigma$ 2D catalogue (dashed line). Bins are equally spaced in $\log(M_{\rm sub})$.}
 \label{successmass}
\end{figure}

Firstly, we determine the completeness of each of our three
2D mass substructure catalogues ($X=1,3,5$). This is done by starting with the 3D subhaloes and looking for 2D counterparts in the mass maps, then examining the resulting
{\it matching success} (i.e.  the fraction of 3D subhaloes for which a 2D counterpart is found) per 3D subhalo mass bin.
The criterion for matching the 3D subhaloes to the 2D mass substructures is that the centre of the 
3D subhalo must lie within the ellipse that characterises the 2D substructure 
(with a $\simeq 20$ per cent margin for error, which was determined by experiment). 
As discussed in Section \ref{detectsec}, the default 3D centre is taken to be
the (projected) position of the most-bound particle in the subhalo, as 
identified by {\sc subfind}. This is a robust choice, comparing very well with
the peak surface density in the maps in the vast majority of cases. However,
during a complicated merger, we found that the most-bound particle can 
occasionally lie outside the cluster core (see Section 7.2), in which 
case we apply the position of the peak projected DM particle density of the 
cluster instead.

Multiple 3D subhaloes can be matched to the same substructure in the mass map; we refer to this as a {\it multiple match}. However, 2D substructures cannot 
share a 3D subhalo as our criterion means each subhalo is only ever matched to one 2D substructure. It should also be noted that we start with a limited 3D catalogue containing 
only those subhaloes whose centres are within the projected $r_{500}$ and match to the complete 
catalogue of 2D substructures, which extends slightly beyond the projected $r_{500}$ (i.e. outside the map). This simply prevents the failure to match a genuine 2D-3D pair when one substructure's centre lies slightly outside this boundary. 

Fig.~\ref{successmass} illustrates the completeness of our 2D catalogues as a function of subhalo mass (note that the main clusters are included in these data).
Specifically, it shows the fraction of all subhaloes in the map
region that are detected, including those subhaloes associated with the
same 2D substructure, due to source confusion or genuine projection effects
(detailed below). In all three catalogues we clearly associate 2D substructures with all 3D subhaloes that have 
$M_{\rm sub}>10^{13} h^{-1}{\rm M}_{\odot}$. The $1\sigma$, $3\sigma$ and $5\sigma$ catalogues are $90$ per cent 
complete per mass bin down to $3\times 10^{11}$,  $10^{12}$ and  $3\times 10^{12}  h^{-1}{\rm M}_{\odot}$ respectively.

The {\it cut-off} in completeness, below which our ability to retrieve 3D subhaloes from the projected data decreases sharply with mass, is a result of several limiting factors: the map resolution (effectively set by the pre-smoothing kernel size, $\sigma_{1}$), the choice of $\sigma_{2}$ and simply the intensity of the cluster background. Low mass subhaloes have poor contrast against the background since they add little mass in addition to the total mass along the line of sight and so are difficult to distinguish. The mass at which this cut-off occurs is most sensitive to $\sigma_{1}$. As we demonstrate in Section \ref{real}, when we increase $\sigma_{1}$ by a factor of around 10 (more typical of the resolution of weak lensing mass reconstructions), the $90$ per cent (per mass bin) completeness limit for the $3\sigma$ catalogue becomes $\sim 10^{13}  h^{-1}{\rm M}_{\odot}$ (see Fig.~\ref{successmass_obs}).

\subsection{Projection and Confusion}

\begin{figure}
\centering
\includegraphics[width=0.5\textwidth]{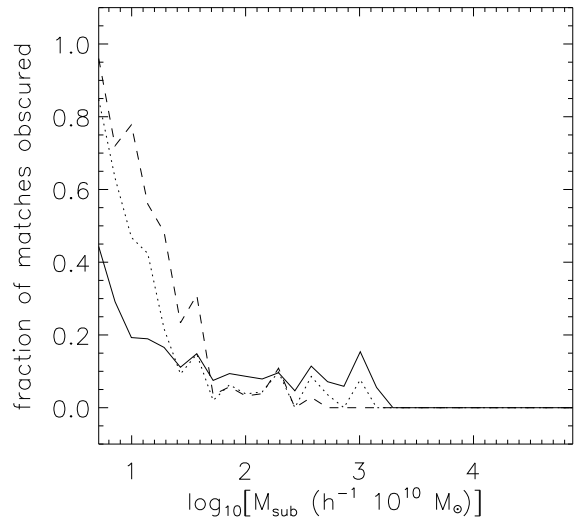}
\caption{Fraction of detected subhaloes {\it per mass bin} which are obscured (see text for definitions) as a function of 
subhalo DM mass for $1\sigma$  2D catalogue (solid line), $3\sigma$ 2D catalogue (dotted line) and $
5\sigma$ 2D catalogue (dashed line). Bins are equally spaced in $\log(M_{\rm sub})$.}
\label{confusionfig}
\end{figure}

Visual inspection of the projected mass maps reveals that two peaks that are very nearby can be detected as one 2D substructure (i.e. confused), if the lower density pixels between them are not removed when pixels $< \mu+X\sigma$ are discarded. This is not a concern if the mass ratio of the 3D subhaloes that have given rise to the 2D peaks is high (as the inclusion of the less massive object has little effect), or if 
they are both low mass subhaloes ($\sim 10^{11} h^{-1}{\rm M}_{\odot}$), below the mass range 
we are interested in. However if, for example, a subhalo with a mass, $M_{\rm sub}\sim10^{13} h^{-1} {\rm M}_{\odot}$ and the main cluster core give rise to two adjacent peaks in the map which are confused as one 2D substructure, we limit our opportunities to study the properties of the subhalo in detail. This is particularly important since subhaloes with $M_{\rm sub} >10^{13} h^{-1} {\rm M}_{\odot}$ are relatively rare. A related effect is that of projection, where two subhaloes that are aligned along the line of sight give rise to only one peak in the projected mass map. 

Here we do not distinguish explicitly between projection and confusion. Instead, we define a detected subhalo as {\it obscured} if it is part of a multiple match and is not the most massive subhalo involved; we would not consider such a subhalo to be individually resolved. Fig.~\ref{confusionfig} shows the fraction of detected subhaloes per mass bin which are obscured. As expected the obscured fraction at the high-mass end is lowest for the $5\sigma$ catalogue and highest for the $1\sigma$ catalogue, since the former is the most stringent when removing low density pixels between adjacent substructures, allowing them to be individually resolved. The trend reverses at low mass, however, because the removal of low density pixels also erases
small 2D substructures. Since this is more effective with a larger value of sigma, the fraction of obscured substructures (detected only because of their association with larger substructures) increases. For the $3\sigma$ catalogue, around 70 per cent at $10^{11} h^{-1}{\rm M}_{\odot}$, 5 per cent in the $10^{12}-10^{13} h^{-1}{\rm M}_{\odot}$ mass range and zero at the high-mass end, are obscured. On inspection of the maps, it is apparent that the obscured fraction at mass scales of $\sim 10^{13}  h^{-1}{\rm M}_{\odot}$ typically occurs in the final stages of a merger and results from confusion when the two objects coalesce. 

We adopt the $3\sigma$ catalogue from now on as it offers a small reduction in the obscured fraction at high masses while maintaining good completeness above $M_{\rm sub}=10^{12} h^{-1} {\rm M}_{\odot}$, detecting $98$ per cent of all subhaloes above this mass.

\subsection{Purity}

\begin{figure}
\centering
\includegraphics[width=0.5\textwidth]{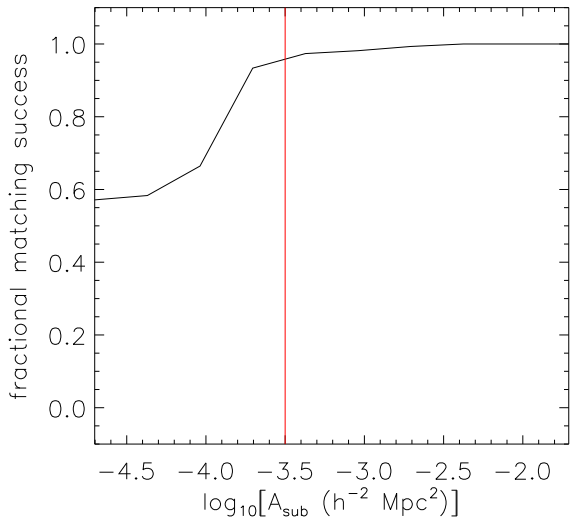}
\caption{Purity of our sample, i.e. the fraction {\it per area bin} of 2D mass substructures in our $3\sigma$
catalogue matched to 3D subhaloes as a function of the physical area of the substructure, $A_{\rm sub}$.
The vertical line marks $A_{\rm sub}=3\times10^{-4}h^{-2}{\rm Mpc}^2$, above which our sample can be
taken to be pure. Bins are equally spaced in $\log(A_{\rm sub})$.}
\label{successarea}
\end{figure}

We now consider the purity of our $3\sigma$ catalogue, by undertaking the matching procedure in reverse, i.e. starting with the 2D mass substructures and trying to identify 3D subhalo counterparts for these. The {\it matching success} (i.e. the fraction of 2D mass substructures that are successfully matched to a 3D subhalo, in this case) then provides a measure of the purity. This is important as it tells us in which regions of parameter space the raw 2D data could potentially be used directly, without reference to the 3D data for calibration. Fig.~\ref{successarea} shows the 
fractional matching success of 2D mass substructures to 3D subhaloes versus the characteristic physical 
area of the 2D substructure, $A_{\rm sub}$.  

We achieve very high {\it purity} down to $A_{\rm sub}\simeq 3\times10^{-4} h^{-2}{\rm Mpc}^2$, close
to the approximate projected area of our combined kernel, $F$, ($\pi r_{0}^2 \sim 10^{-3} h^{-2}{\rm Mpc}^2$; see equations \ref{kernel} and \ref{radkernel} for definitions of $F$ and $x_{0}$). 
Reasons 
for not finding a 3D subhalo to match every 2D substructure are three-fold. Firstly, we have detected a 
substructure associated with a 3D subhalo with less than 100 DM particles (i.e. our minimum 
allowed subhalo size). Secondly, the substructure detection is `false' i.e. we have detected a transient 
enhancement which does not constitute a self-bound subhalo. Or finally, there is an associated subhalo but 
matching has failed (matching becomes increasingly difficult as substructures become smaller).

\subsection{Correlation between mass and area}
\begin{figure}
\centering
\includegraphics[width=0.5\textwidth]{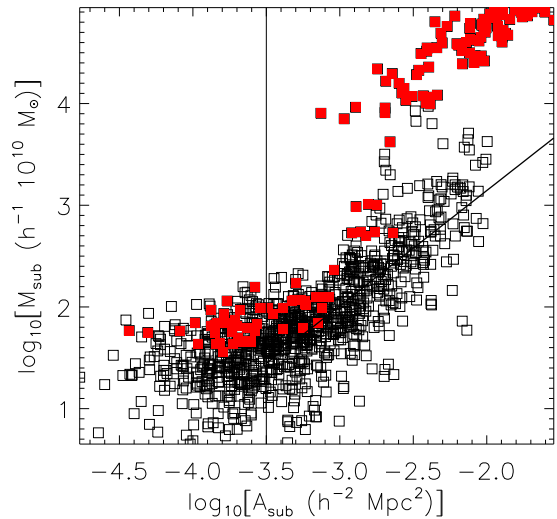}
\caption{Correlation between physical area, $A_{\rm sub}$, of 2D total mass substructure ($3\sigma$ catalogue) in unsharp-masked image 
and DM mass of 3D subhalo, $M_{\rm sub}$, to which it is matched. Filled squares show {\sc subfind} background 
haloes (see text for details). Vertical line shows purity threshold for 2D mass catalogue and the other is best-fitting line given by equation \ref{mafit}.}
\label{massarea}
\end{figure}

In Fig.~\ref{massarea}, we take all 2D total mass substructures in the $3\sigma$ catalogue which have been matched to 3D subhaloes and examine the correlation between the area of the 2D substructure, $A_{\rm sub}$ and the DM mass of its 3D subhalo counterpart, $M_{\rm sub}$. The figure contains data for all three clusters at all redshifts in the range $0\le z<1$. The filled squares indicate objects which are {\sc subfind} {\it background haloes} as opposed to subhaloes.  Background haloes consist of the most massive subhalo found in each FoF group plus any additional group particles that are gravitationally bound to it and which do not already belong to a subhalo. Effectively, the background halo is the parent halo in which the other subhaloes reside. The background haloes grouped in the top right are the cluster cores themselves, forming a separate population because the 2D detection corresponds to the core only, whereas the mass is that of the entire cluster. The background haloes at lower masses are smaller parent haloes which lie in front of or behind the main clusters. This figure shows that when we are above the completeness 
limit in terms of associated 3D subhalo DM mass ($M_{\rm sub}= 10^{12} h^{-1} {\rm M}_{\odot}$), most 2D substructures are also in the region where we know our 2D catalogue is pure 
(i.e. $A_{\rm sub}> 3\times10^{-4} h^{-2} {\rm Mpc}^2$); the converse does not hold, however.

For the rest of the substructures, that are not background haloes, we demonstrate a strong correlation between the 3D DM subhalo mass ($M_{\rm sub}$) and the 2D area ($A_{\rm sub}$). A least squares fit to this correlation yields,

\begin{equation}
\log{\frac{M_{\rm sub}}{10^{10}h^{-1}{\rm M}_{\odot}}}=(1.13\pm0.04)\log{\frac{A_{\rm sub}}{h^{-2}{\rm Mpc}^{2}}}+(5.4\pm0.1)
\label{mafit}
\end{equation}

where all points with $A_{\rm sub}>3\times10^{-4} h^{-2} {\rm Mpc}^2$ were considered. Using this correlation we can select a new threshold of $A_{\rm sub}=10^{-3} h^{-2} {\rm Mpc}^2$ (corresponding to a mass of $10^{12} h^{-1} {\rm M}_{\odot}$) producing a sample of substructures with both high purity and high completeness (we refer to such catalogues as {\it  pure}).
 
We can also estimate the intrinsic scatter in this relation using,

\begin{equation}
\sigma_{\log{(M_{\rm sub})}}=\sqrt{\frac{1}{N}\sum_{i}(\log{(M_{\rm i})}-\log{(M_{\rm fit})})^{2}}
\end{equation}

where $M_{\rm i}$ is the mass value of each data point and $\log{(M_{\rm fit})}$ is the value computed using equation \ref{mafit} for the corresponding area. We find $\sigma_{\log{(M_{\rm sub})}} = 0.35$ which suggests that the typical uncertainty in the DM mass of a subhalo is around a factor of $2$. For comparison, the fit was also made using the discarded $1\sigma$ and $5\sigma$ 2D mass catalogues instead and the intrinsic scatter in the resulting relations was very similar (0.30 and 0.37 respectively) suggesting the quality of the fit is independent of catalogue choice. Furthermore,  for given value of $A_{\rm sub}$, the maximum variation in the estimated value of $M_{\rm sub}$ when comparing all three 2D catalogues with each other is approximately a factor of $3$, comparable to the error from intrinsic scatter. We also note that the intrinsic scatter is greater at the higher redshift, for example it is 0.31 when fitting only to data for $0 \leq z \leq 0.2$ and 0.41 for $0.5 \leq z <1.0$. Such a correlation, though calibration-dependent, is potentially useful for providing a quick, rough estimate of subhalo DM mass determined from the observed area of a substructure in a weak-lensing map (assuming the substructure is resolved). 

\subsection{Summary}

We have matched our 2D total mass substructure catalogues to self-bound 3D subhaloes and have identified the $3\sigma$ catalogue as the most suitable for the analysis in future sections. This catalogue is at least 90 per cent complete in all subhalo mass bins above $10^{12} h^{-1} {\rm M}_{\odot}$ and pure above a projected area of $3\times10^{-4} h^{-2}{\rm Mpc}^2$, which is close to the resolution limit of our kernel. We also note a strong correlation between the 
(observable) area of the 2D substructure and the DM mass of the 3D subhalo. Using this we derive an area threshold, $10^{-3} h^{-2}{\rm Mpc}^2$, above which our substructure catalogues have both high purity and completeness. Projection and confusion effects above the completeness limit are minimal. 

\section[]{Comparison of substructure in the hot gas and dark matter components}\label{massgassec}

We now address the main aim of the paper; comparing the substructure in the X-ray surface brightness and total mass maps. Again, we apply a simple matching technique, this time to our pairs of maps and then 
attempt to  explore the underlying physical mechanisms which govern the resulting matching success, whilst 
also trying to constrain any potential biases our method may have introduced.

The catalogues of 2D mass substructures and 2D X-ray substructures are compared for each snapshot. The criterion for a match is that there is some overlap of the region enclosed by the ellipse that characterises the mass substructure and the region enclosed by the ellipse that characterises the X-ray substructure. In order to keep our method simple, we allow both 2D total mass and 2D X-ray surface brightness substructures to be matched to more than one substructure of the other type, rather than using additional matching criteria to prevent this. We use the term {\it single match} to refer to a unique pairing of one mass substructure with one X-ray substructure and the term {\it multiple match}  for a mass substructure which has been matched to more than one X-ray substructure or vice versa.

\subsection{Direct matching}

An important feature of this work is the use of 2D data (maps) so, with this approach in mind, we first undertake the matching with no reference to the 3D subhalo data. This will allow us to confirm how reliable a picture the 2D data alone can provide as we can later compare our results to those which have been calibrated against the 3D subhalo information.

As in Section \ref{2d3dsec} we undertake the matching procedure in two ways; starting with the 2D total mass substructures and seeking an X-ray counterpart for each and then starting with the 2D X-ray substructures and seeking a mass counterpart for each. Table \ref{catcomparetab} summarises the results of these matching processes, where the data for 2D mass substructures comes from the former and that for 2D X-ray substructures from the latter. Here we use the subscripts TM and SB to signify substructures in the total mass and X-ray surface brightness maps, respectively.

Firstly, it is encouraging that the fraction of substructures which are matched to more than one object ($f_{\rm multimatched}$) is very low ($2 - 10$ per cent) regardless of choice of 2D X-ray catalogue or whether only the pure sample of 2D mass substructures is used. High numbers of single matches are preferred as this suggests the effect of confusion is limited and that the number of false matches is low. 

The ratios of X-ray to total mass substructures ($\frac{N_{\rm SB}}{N_{\rm TM}}$) show that when using the full 2D mass catalogue, there is always a dearth of X-ray substructures and so, regardless of the criterion employed, there will be unmatched 2D mass substructures. However, when using the pure 2D mass catalogue, there is a factor of $\sim2 - 3$ more X-ray substructures. The fraction of total mass substructures that are matched ($f_{\rm matched}$) roughly doubles when moving to the pure sample, suggesting that the majority of mass substructures discarded to obtain purity did not have an X-ray counterpart. This could be interpreted in one of two ways; the discarded 2D mass `substructures' were false detections and so one would not expect to find any corresponding substructure in the X-ray emitting gas, or they were real but may have corresponded to low mass 3D subhaloes which are less likely to have retained their hot gas. In fact, around $75$ per cent of 2D mass substructures below the purity threshold were matched to a 3D subhalo, suggesting it is the latter effect that dominates. Interestingly, when moving from the $1\sigma$ to the $3\sigma$ X-ray catalogue the fraction of mass substructures matched decreases, but the same quantity for the X-ray increases. Here, the $3\sigma$ X-ray catalogue is more pure as a greater fraction of its substructures can be matched to 2D mass substructures, however the $1\sigma$ X-ray catalogue is more complete since a greater absolute number of its substructures are matched to 2D mass. A similar trade-off between purity and completeness was seen when matching 2D mass substructures to 3D mass subhaloes in Section \ref{2d3dsec}. The added complication here, of course, is that unlike the 2D mass substructures and 3D subhaloes, we cannot assume a 1:1 correspondence between the 2D mass and 2D X-ray substructures (in fact the deviation from this is the motivation for this work), so a completeness limit cannot really be established. 

Even more surprising than the large fraction of 2D mass substructures with no X-ray counterpart, is that a 2D total mass counterpart cannot be found for a high percentage of the 2D X-ray substructures. Even when considering the $3\sigma$ X-ray catalogue, which picks out only the most defined 2D substructures in the hot gas, and matching this with the full $3\sigma$ 2D mass catalogue, 40 per cent of the X-ray substructures still go unmatched. Investigating the properties of the matched and unmatched substructures should provide insight into this result.

\begin{center}
\begin{table}
\caption{General results of direct matching between the 2D total mass and 2D X-ray surface brightness substructure catalogues.Top rows: matching between all substructures in the X-ray catalogue, to all those in the total mass catalogue. Bottom rows: direct matching of all substructures in the X-ray catalogue to a pure ($A_{\rm sub,TM}> 10^{-3} h^{-2} {\rm Mpc}^{2}$) total mass catalogue. Columns: Multiple of $\sigma$ used in indicated catalogue ($X$), Ratio of number of substructures ($\frac{N_{\rm SB}}{N_{\rm TM}}$), fraction of total substructures in the catalogue indicated that are successfully matched ($f_{\rm matched}$), fraction of all matched substructures in the catalogue indicated that are matched more than once ($f_{\rm multimatched}$). }
\label{catcomparetab}
\begin{tabular}{p{0.035\textwidth} p{0.035\textwidth} p{0.035\textwidth} p{0.035\textwidth} p{0.035\textwidth} p{0.035\textwidth} p{0.035\textwidth}}
\hline
\textbf{$X_{\rm TM}$} &\textbf{$X_{\rm SB}$} & \textbf{$\frac{N_{\rm SB}}{N_{\rm TM}}$} &\multicolumn{2}{l}{\textbf{$\bf{f_{\rm matched}}$}} &\multicolumn{2}{l}{\textbf{$\bf{f_{\rm multimatched}}$}} \\
\\
 & & & \textbf{TM} & \textbf{SB} & \textbf{TM} &\textbf{SB}\\
 \hline
 \multicolumn{7}{l}{\it Full 2D total mass substructure catalogue}\\
3 & 1 & 0.95 & 0.43 & 0.45 & 0.06& 0.07\\
 3& 3 & 0.57 & 0.32&  0.59 & 0.08 & 0.03\\
\hline
\multicolumn{7}{l}{\it Pure 2D total mass substructure catalogue}\\
3 & 1 & 2.83  & 0.77 & 0.30 & 0.10& 0.04\\
 3& 3 & 1.71 & 0.67&  0.43 & 0.10 & 0.01\\
\hline
\end{tabular}
\end{table}
\end{center}

\begin{figure}
\centering
  \includegraphics[width=0.48\textwidth]{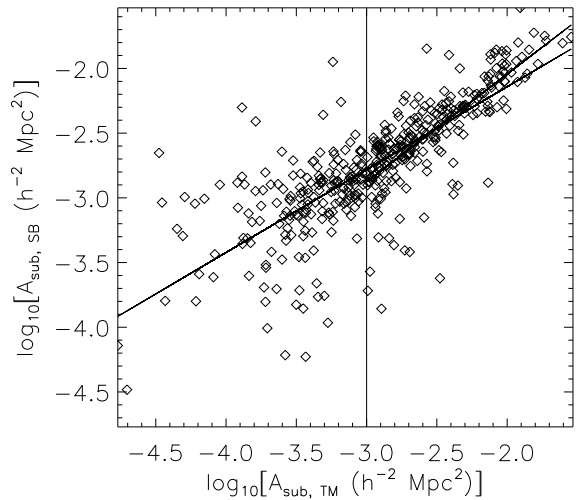}
\caption{Correlation between areas of matched pairs of 2D total mass, $A_{\rm sub,TM}$, and 2D X-ray, $A_{\rm sub,SB}$, substructures. Data for single matches only (see text for definition) for matching between $3\sigma$ total mass and $1\sigma$ X-ray catalogues (data set using $3\sigma$ X-ray catalogue has been omitted for clarity but shows a similar distribution). Solid lines are a least squares fit to all the data (extended line) and just that above the purity threshold ($A_{\rm sub,TM}\sim10^{-3}h^{-2}{\rm Mpc}^{2}$; indicated with a vertical line) for the 2D mass catalogue; The latter is given by equation \ref{aafit}. The complete data set has a correlation coefficient of 0.78.}
\label{aa_ee}
\end{figure}

Focussing first on those substructures that are matched, we compare the properties of the 2D matched pairs. Fig. \ref{aa_ee} demonstrates the tight correlation between the area of singly-matched X-ray ($1\sigma$ catalogue) and total mass ($3\sigma$ catalogue) substructures. The best-fitting line for matched pairs from this combination of catalogues, where the 2D mass map substructure is in the pure region ($A_{\rm sub,TM}>10^{-3}h^{-2}{\rm Mpc}^{2}$) is given by,

\begin{equation}
\log{\frac{A_{\rm sub,SB}}{h^{-2}{\rm Mpc}^{2}}}=(0.83\pm0.04)\log{\frac{A_{\rm sub,TM}}{h^{-2}{\rm Mpc}^{2}}}-(0.4\pm0.1)
\label{aafit}
\end{equation}

Including all multiple matches as well increases the scatter, but the relationship is still clearly evident. Note that the X-ray substructures are slightly larger than the total mass substructures; this is partly due to the use of the $1\sigma$ catalogue here, but also due to the more extended nature of the hot gas (for comparison, the gradient when using the $3\sigma$ X-ray catalogue is $0.91\pm0.06$, still less than 1). The outliers above the line can mostly be attributed to small 2D mass substructures being matched to highly elliptical 2D X-ray substructures, which are usually features near the centre of the main cluster corresponding to subhaloes actively undergoing stripping. In many cases it is impossible to tell whether the match is valid or not, however the scatter occurs below the threshold area which demarks where our catalogue is pure ($A_{\rm sub,TM}\sim10^{-3}h^{-2}{\rm Mpc}^{2}$). With this in mind it is unsurprising that some of the scatter here also results from spurious detections, i.e. mass substructures that are later found not to correspond to a subhalo. Scatter below the line seems to arise from two situations 1) a small gas feature is detected that overlaps with a large mass substructure which has been stripped of its gas, i.e. the two are in chance alignment, 2) the match appears genuine yet the gas substructure is small, suggesting the outer regions of gas have already been stripped.

\begin{figure}
\centering
  \includegraphics[width=0.48\textwidth]{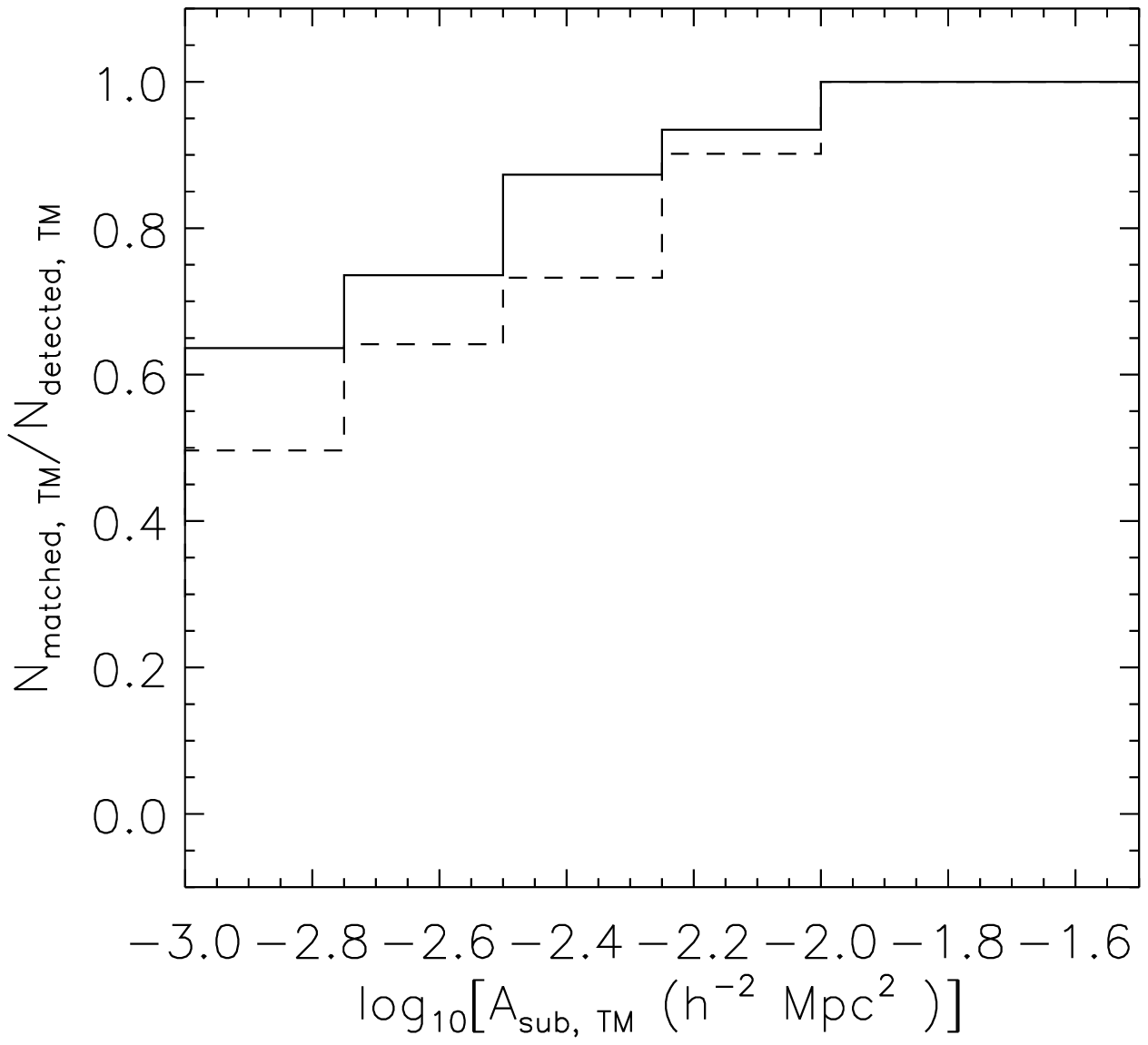}
   \label{fractm_direct}
  \hspace{0.1in}
  \includegraphics[width=0.48\textwidth]{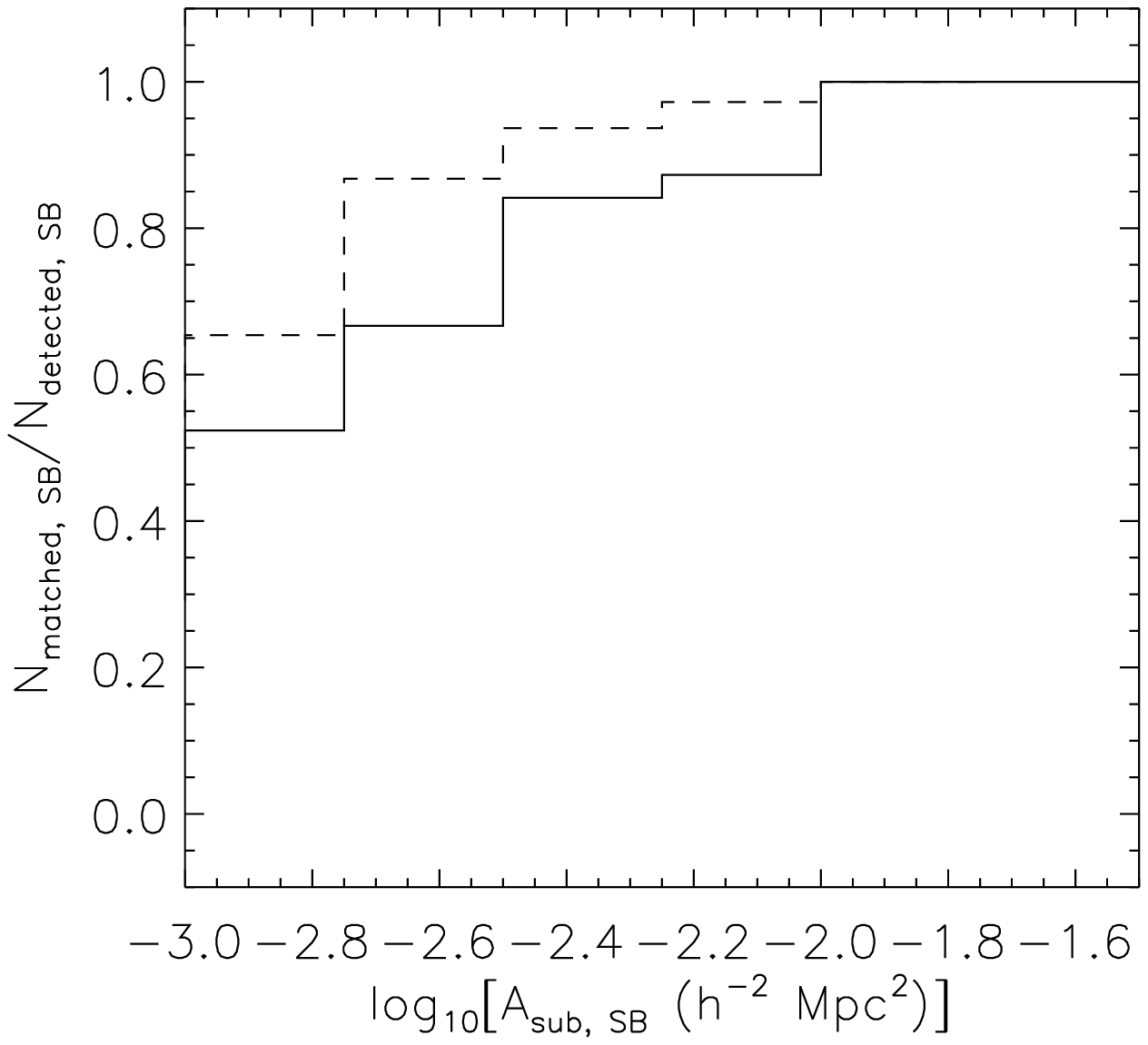}
    \label{fracsb_direct}
 
\caption{Matching success {\it per area bin} of $3\sigma$ 2D mass substructure catalogue to $1\sigma$ (solid) and $3\sigma$ (dashed) 2D X-ray substructure catalogues versus area, for substructures above the purity threshold ($A_{\rm sub}=10^{-3}h^{-2}{\rm Mpc}^2$). Bins are equally spaced in $\log(A_{\rm sub})$. Top: Fractional matching success when starting with the 2D total mass substructures and seeking a 2D X-ray counterpart for each. Bottom: Fractional matching success when starting with the 2D X-ray surface brightness substructures and seeking a 2D mass counterpart for each.}
\label{successvarea_gas}
\end{figure}

Fig. \ref{successvarea_gas} shows the fraction of all 2D total mass substructures matched to X-ray (top) and the fraction of all 2D X-ray substructures matched to total mass (bottom) as a function of substructure area. Above $10^{-3}$ $h^{-2}$ ${\rm Mpc}^{2}$ the matching success is $\geq50$ per cent per bin for both catalogue types, however this still suggests a very high number of  substructures do not have counterparts in the other map. There are also many large unmatched substructures, for example around 10 per cent of 2D mass substructures in the $-2.25\leq \log{(A_{\rm sub}/h^{-2}{\rm Mpc}^{2})}\leq -2.0$ range. Using equation~\ref{mafit} we can infer that this corresponds to a mass of approximately $10^{13} h^{-1} {\rm M}_{\odot}$, suggesting these correspond to fairly massive 3D subhaloes. Increasing the radius of the unsharp-masking kernel used to detect the X-ray substructures, to twice that of the fiducial kernel (i.e. $0.1r_{500}$), yields a similar matching success. This indicates that the results of the substructure comparison that are shown here, do not depend significantly on this aspect of the substructure detection procedure.

It is interesting that there is also a significant number of unmatched X-ray substructures, even at large areas. One would initially assume that once hot gas is separated from its DM subhalo it would disperse and so not be detected as a stand-alone substructure. Large unmatched 2D X-ray substructures were followed up individually by visual inspection of the maps and it appears that there are three main categories. These are: 1) clearly defined substructures which are so displaced that they cannot visually be associated with one particular DM substructure (although there are typically candidates in the vicinity), 2) clearly defined substructures that are slightly offset from a nearby dark mater substructure, and 3) detections of gas `features' in the vicinity of the core during merger events - these 'substructures' cannot be directly associated with a DM substructure and it is not necessarily appropriate to do so. 

Scenario 1 incorporates the most clear-cut examples of 2D X-ray substructures which are indisputably unmatched, whereas scenario 2 also includes those whose definition as matched or unmatched is somewhat subjective, as it is clear which mass substructure they belong to even though they are spatially distinct from it (for our purposes we call these unmatched). An example of the displacement of the X-ray component of a substructure can be seen in Section \ref{casestudies} (see Case Study 2, Fig.~\ref{gasdisplaced}). This example is a simple one, however, because there are often numerous hot gas-deficient mass substructures nearby to confuse matters and make determining which one the X-ray substructure originated from impossible (here we also have the time sequence to help us with this).

 Scenario 3 is sensitive to the choice of X-ray catalogue so, treating this type of detection as unwanted, we can conclude that the $1\sigma$ catalogue suffers from more false detections (by our definition), which goes part way to explain its lower overall matching success. Scenario 2 can also be sensitive to the catalogue choice, as if the displacement between substructures is small, then the increase in area of a $1\sigma$ X-ray detection can be enough to meet the overlap criterion in cases where it wasn't met for the $3\sigma$ X-ray detection. For this reason we continue to show the main results for matching to both the $1\sigma$ and $3\sigma$ X-ray catalogues, although it should be noted this effect doesn't have a big impact on the matching success. Furthermore, in Section~\ref{paramsec}, where we simplify the discussion by showing results for one only X-ray catalogue, it is the $1\sigma$ that is chosen (despite its more numerous spurious detections) since it provides the most conservative estimate of the number of 2D mass substructures for which an X-ray counterpart cannot be found.

\subsection{Matching 2D X-ray substructures to 2D mass substructures with a 3D counterpart}\label{submatch}

Subhalo mass is expected to be a crucial factor when looking for mismatches between DM and hot gas, as gas stripping procedures have more effect on low-mass subhaloes \citep{2004MNRAS.350.1397T}. Here, the 3D subhalo data becomes an invaluable tool, not only because it effectively calibrates our 2D total mass substructure catalogues, but because it allows us to probe the effect of subhalo mass on the success of the matching to the 2D X-ray substructures. We repeat the matching procedure outlined above, but this time only use 2D mass substructures which have successfully been associated with a 3D subhalo in Section \ref{2d3dsec}. It should be noted that, in this section, {\it 2D mass substructure catalogue} now refers to the calibrated version of the original catalogue, containing only those 2D mass substructures which have a 3D subhalo counterpart. In the case of 2D mass substructures that have been associated with more than one subhalo (see Section \ref{2d3dsec}), $M_{\rm sub}$ refers to the combined DM mass of these subhaloes.

Table \ref{catcomparetab3d} shows the overall statistics for matching both the full and pure (i.e. $A_{\rm sub,TM}>10^{-3}$ $h^{-2}{\rm Mpc}^{2}$) catalogues of 2D mass substructures with a 3D subhalo counterpart to the 2D X-ray substructure catalogues. It is worth highlighting here that the results in Table \ref{catcomparetab} and \ref{catcomparetab3d} are almost identical for the pure 2D mass substructure catalogue because, by definition, the vast majority of 2D mass substructures in the pure sample have a 3D counterpart. Note that the ratios of X-ray substructures to total mass substructures in the first row (full 2D mass catalogue) are larger than the equivalent ratios in Table \ref{catcomparetab}, which does not include any reference  to the 3D subhalo data. This difference can be directly attributed the fact that a 3D subhalo counterpart could not be identified for around $10$ per cent of the original 2D mass substructures (in the $3\sigma$ catalogue) and so $N_{\rm TM,Table \ref{catcomparetab3d}}/N_{\rm TM,Table \ref{catcomparetab}} \simeq 0.9$, whereas $N_{\rm SB}$ remains the same. 

There is a slight improvement in the fraction of 2D mass substructures matched to X-ray after calibrating the mass substructures against the 3D subhalo data (i.e. $f_{\rm matched,Table \ref{catcomparetab3d}}>f_{\rm matched,Table \ref{catcomparetab}}$), primarily because this process will have removed any spurious detections from our 2D mass catalogues. These are unlikely to be matched to an X-ray substructure, simply because they are due to discreteness noise in the total mass map or an artefact of the unsharp-masking procedure and, as such, we would not expect these to be correlated with features in the X-ray surface brightness map. The removal of these unmatched mass substructures results in a boost to the overall matching success and slightly reduces the fraction of 2D mass substructures matched more than once ($f_{\rm multimatched}$) to $3\sigma$ X-ray substructures. Despite this effect, however, the overall matching success still remains surprisingly low, with a maximum value of $45$ per cent, achieved when matching to the $1\sigma$ X-ray catalogue. 

This overall statistic is dominated by substructures with low associated 3D subhalo masses as these are far more numerous. From Fig.~\ref{mfclusfig} we can estimate that there are approximately $3$ times more subhaloes with $10^{11}<M_{\rm sub}/h^{-1}{\rm M}_{\odot}<10^{12}$ than subhaloes with $M_{\rm sub}/h^{-1}{\rm M}_{\odot}>10^{12}$ (for subhaloes with their most bound particle within $r_{500}$ only). From the middle panel of Fig.~\ref{boxVr500}, it is apparent that $85-95$ per cent of 3D subhaloes in this mass range (and with most bound particle within $r_{500}$) have no hot gas. With these two results in mind, it is not surprising that the total percentage of 2D mass substructures with 3D subhalo counterparts which also have X-ray counterparts is biased so low. This effect can be further demonstrated by considering the overall matching success to X-ray for the pure 2D mass substructure catalogue (second row, Table \ref{catcomparetab3d}). This is significantly higher, with a maximum value of $77$ per cent (again for the $1\sigma$ X-ray catalogue). Here, on removing substructures with $A_{\rm sub,TM}<10^{-3}$ $h^{-2}{\rm Mpc}^{2}$ to achieve a pure sample we have, by virtue of the $M_{\rm sub}-A_{\rm sub}$ correlation (Equation \ref{mafit}), removed substructures with an associated value of $M_{\rm sub}<10^{12} h^{-1} {\rm M}_{\odot}$.

\begin{figure}
\centering
  \includegraphics[width=0.48\textwidth]{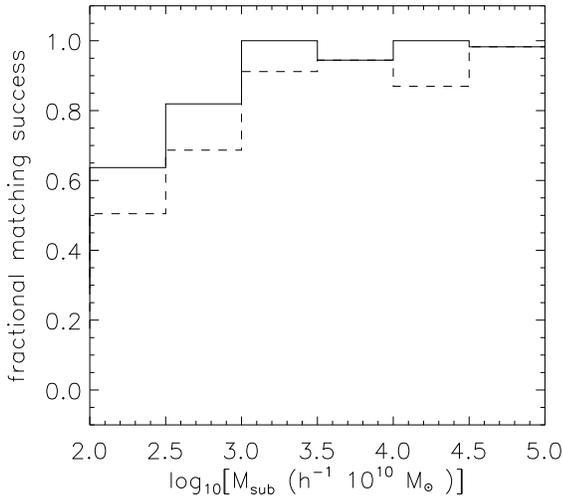}
 \caption{Fractional matching success {\it per mass bin} of $3\sigma$ 2D mass substructure catalogue to $1\sigma$ (solid) and $3\sigma$ (dashed) 2D X-ray substructure catalogues versus subhalo mass. Data are for 2D mass substructures for which a 3D subhalo counterpart was found. Bins are equally space in $\log(M_{\rm sub}) $, the subhalo DM mass.}
 
\label{successvmass_gas_3d}
\end{figure}

It is clear from the overall matching statistics that the matching success depends heavily on the 3D subhalo mass, so we now examine this dependency in more detail. Fig. \ref{successvmass_gas_3d} shows the success in matching 2D total mass substructures to 2D X-ray substructures as a function of the associated subhalo mass.  It should be remembered that $M_{\rm sub}$ is the subhalo DM mass, not total mass, and therefore is independent of the amount of gas removal a subhalo may have undergone, other than the secondary effect of the remaining DM being more prone to tidal stripping.  We have opted to include only those mass substructures matched to subhaloes above $M_{\rm sub}=10^{12} h^{-1} {\rm M}_{\odot}$ (the completeness limit). However, we note that the alternative choice of a sample with $A_{\rm sub,TM}>10^{-3}$ $h^{-2}{\rm Mpc}^{2}$ (the purity limit) made little difference to this figure for the mass range shown.

\begin{center}
\begin{table}
\caption{General results of matching between the 2D total mass substructures (which have been successfully matched to a 3D subhalo) and 2D X-ray surface brightness substructure catalogues.Top rows: matching between all substructures in the total mass  catalogue, to all those in the X-ray catalogue. Bottom rows: matching between a pure ($A_{\rm sub,TM}> 10^{-3} h^{-2} {\rm Mpc}^{2}$) total mass catalogue and all substructures in the X-ray catalogues. Columns: Multiple of $\sigma$ used in indicated catalogue ($X$), Ratio of number of substructures ($\frac{N_{\rm SB}}{N_{\rm TM}}$), fraction of total substructures in the catalogue indicated that are successfully matched ($f_{\rm matched}$), fraction of matched substructures in the catalogue indicated that are matched more than once ($f_{\rm multimatched}$). }
\label{catcomparetab3d}
\begin{tabular}{p{0.06\textwidth} p{0.06\textwidth} p{0.06\textwidth} p{0.06\textwidth} p{0.09\textwidth}}
\hline
\textbf{$X_{\rm TM}$} &\textbf{$X_{\rm SB}$} & \textbf{$\frac{N_{\rm SB}}{N_{\rm TM}}$} &\textbf{$\bf{f_{\rm matched}}$}& \textbf{$\bf{f_{\rm multimatched}}$} \\
\hline
 \multicolumn{5}{l}{\it Full 2D total mass substructure catalogue}\\
3 & 1 & 1.03 & 0.45 & 0.06 \\
 3& 3 & 0.62 & 0.34 & 0.07 \\
\hline
\multicolumn{5}{l}{\it Pure 2D total mass substructure catalogue}\\
3 & 1 & 2.84  & 0.77 & 0.10 \\
 3& 3 & 1.72 & 0.67 & 0.10  \\
\hline
\end{tabular}
\end{table}
\end{center}

For the $1\sigma$ X-ray catalogue, the matching success per mass bin rises gradually with DM subhalo mass:  it is $>95$ per cent for cluster cores ($M_{\rm sub}\gtrsim10^{14} h^{-1} {\rm M}_{\odot}$), $\simeq 95$ per cent for groups ($M_{\rm sub} \sim 10^{13} h^{-1} {\rm M}_{\odot}$) and $\simeq 65$ per cent for galaxies ($M_{\rm sub} \sim 10^{12} h^{-1} {\rm M}_{\odot}$). A similar trend is seen for the $3\sigma$ catalogue, except that the success within a given mass bin is around 10-15 per cent lower. 

It is expected, based on the trend for $f_{\rm gas}$ decreasing with subhalo mass already demonstrated, there would be a {\it cut-off mass} below which the fraction of substructures with an X-ray component would fall off sharply from $100$ per cent (similar to that seen in Fig \ref{successmass} when matching 2D substructures to 3D subhaloes). This feature is present (at $M_{\rm sub} \sim 10^{13} h^{-1} {\rm M}_{\odot}$), however the fall off is much more gradual and surprisingly the success rate is slightly below $100$ per cent even above this value. The position of the cut-off mass and the rate of fall-off thereafter, is independent of X-ray catalogue choice and there is a deviation in the predicted success of only $\sim 10$ per cent. 

\begin{figure}
\centering
  \includegraphics[width=0.5\textwidth]{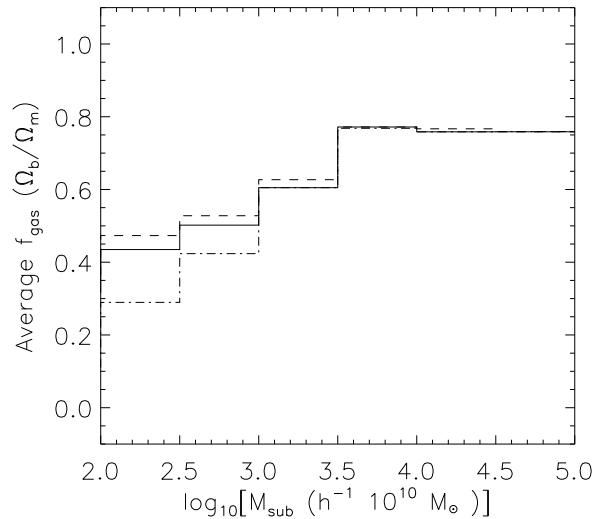} 
\caption{Average hot gas fraction {\it per mass bin} of most massive 3D subhalo matched to a 2D mass substructure in the 2D catalogue (dot-dash). Gas fractions are in units of $\Omega_{\rm b}/\Omega_{\rm m}$. Other lines show the same quantity, but only for those 2D mass substructures that are also successfully matched to a 2D X-ray substructure. Bin values and other line-styles are as for Fig.~\ref{successvmass_gas_3d}.}
\label{avgasfracVmass}
\end{figure}

In order to link these results to the composition of the underlying 3D subhaloes, we examine the hot (${\rm T}>10^{6}{\rm K}$) gas fractions, $f_{\rm gas}$, of the subhaloes that are matched to the 2D mass substructures. In cases where more than one subhalo is associated with the same 2D mass substructure, we calculate $f_{\rm gas}$ for the most massive subhalo (but note that the exclusion of 2D substructures matched to more than one subhalo from the following analysis makes little difference to the results). Fig. \ref{avgasfracVmass} shows the average hot gas fraction, $f_{\rm gas}$, per mass bin for different samples of 2D mass substructures (binning is identical to that in Fig. \ref{successvmass_gas_3d}). Interestingly, the average $f_{\rm gas}$ for all 2D mass substructures (dot-dash) shows the same trend with subhalo mass as the matching success, suggesting that the two are closely linked, as it would have been reasonable to assume.

The solid and dashed lines show the average $f_{\rm gas}$ value for 2D mass substructures which are matched to a substructure in the $1\sigma$ or $3\sigma$ X-ray catalogue, respectively. These values are slightly higher than those for all the substructures (with and without X-ray counterparts) suggesting that only low mass 3D subhaloes with a higher than average hot gas fraction will be successfully detected in 2D and matched in the X-ray maps. We also examined the minimum $f_{\rm gas}$  of a total mass substructure {\it with} a 2D X-ray counterpart and the maximum $f_{\rm gas}$ of one {\it without} a 2D X-ray counterpart, per mass bin, for both X-ray catalogues. These quantities effectively give the hot gas fraction thresholds that define the X-ray substructure catalogues. The minimum $f_{\rm gas}$ was catalogue independent, suggesting the lower detection limit is dominated by another factor, however, the maximum $f_{\rm gas}$  was found to be significantly higher for the $3\sigma$ catalogue at low masses (compared to the $1\sigma$ catalogue and the average value), occasionally by a factor of around $2$. Since the average $f_{\rm gas}$ of a mass substructure successfully matched to an X-ray substructure is catalogue independent, yet the maximum $f_{\rm gas}$  can be much higher for the $3\sigma$ catalogue, this suggests that that the difference in numbers of substructures contained in each catalogue is primarily due to hot gas distribution rather than mass. The objects picked up in the $1\sigma$ catalogue but not in the $3\sigma$ catalogue have $f_{\rm gas}$  values higher than the average detected, i.e. if $f_{\rm gas}$  is the controlling factor, they should appear in both catalogues. However, if a subhalo had a significant hot gas fraction, but this had been displaced from its centre or the dense core of the hot gas had been disrupted and so the peak in X-ray emission was not so bright, this could explain the same substructure being detected in the $1\sigma$ but not the $3\sigma$ catalogues (remember the $\sigma$ values refer to cuts in the residual X-ray surface brightness).
\\

\subsection{Summary}
We have attempted to match every 2D mass map substructure to a 2D substructure in the corresponding X-ray map and have shown that there are numerous occasions when this is not possible, highlighting differences between substructure in the hot gas and DM components. The frequency of matching failures clearly increases with decreasing subhalo mass: a few per cent of cluster cores ($M_{\rm sub}\gtrsim10^{14} h^{-1} {\rm M}_{\odot}$), $\simeq 5$ per cent of groups ($M_{\rm sub} \sim 10^{13} h^{-1} {\rm M}_{\odot}$) and $\simeq 35$ per cent of galaxies ($M_{\rm sub} \sim 10^{12} h^{-1} {\rm M}_{\odot}$) do not have X-ray counterparts. Interestingly, we also find that around a half of the X-ray substructures detected don't have counterparts in the mass maps. As more joint weak lensing and X-ray studies are undertaken, we predict more `dark haloes' will be found, with these discoveries not restricted to rare, merger events involving high mass subhaloes but will occur frequently on the galaxy-mass scale.

\section{Discussion}\label{paramsec}

The benefit of performing a cosmological simulation is that it will best mimic the complicated processes taking place during the formation  of real galaxy clusters. However, by the same token, it can then be difficult to untangle the influence of one parameter or physical process on the conclusions, from that of another. In this section, we investigate the effects of the main selection parameters (Section \ref{selecsec}) and the main model parameters (Section \ref{coolsec}) on Fig. \ref{successvmass_gas_3d}. In addition, we define several broad categories for the fate of a mass substructure's hot gas component, as viewed in 2D. By exploring a Case Study from each category (Section \ref{casestudies}), we attempt to illustrate how the overall picture of the correspondence between the total mass and X-ray maps, shown in Fig. \ref{successvmass_gas_3d}, is built up. In Section \ref{real}, we make a preliminary assessment of the potential impact of analysing maps with noise and observationally achievable resolution on our results.

\subsection[]{Selection parameters}\label{selecsec}

Section \ref{massgassec} dealt with all the cluster maps as one data set, however, in reality there are selection parameters which come in to play when observing clusters. The two most significant are redshift and dynamical state, the effects of which we investigate below. 

\subsubsection{Variations with redshift}\label{zsec}

The role of redshift is generally important when examining any class of astrophysical object, as it cannot be assumed that a population will not evolve significantly. In the case of galaxy clusters, our understanding of this factor is particularly significant if they are to become robust probes of the cosmological parameters. In this section we divide our maps into two samples, $0 \le z \leq 0.2$ and $ 0.5 \le z < 1.0$ (chosen to contain an equal number of snapshots; 11), and test if there is any difference in the results. \\

\begin{figure}
\centering
  \includegraphics[width=0.5\textwidth]{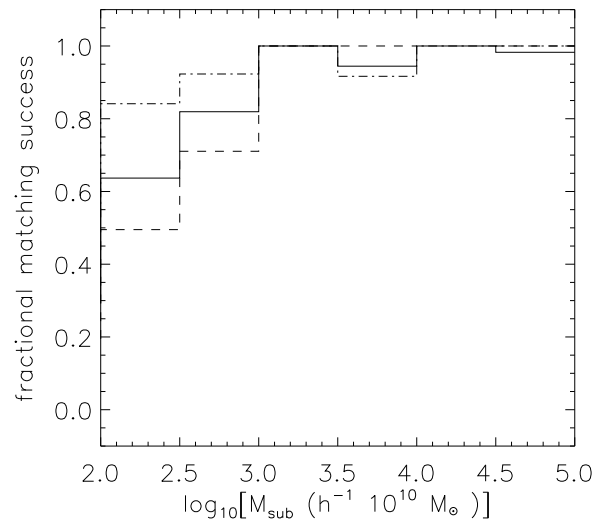}
\caption{Matching success {\it per mass bin} of $3\sigma$ 2D mass substructure catalogue to $1\sigma$  2D X-ray substructure catalogues versus subhalo DM mass. Data divided by redshift:  $0 \le z \le 0.2$ (dashed) and $ 0.5 \le z < 1.0$ (dot-dashed) Solid line shows combined data  from Fig. \ref{successvmass_gas_3d}.}
\label{successvmass_gas_z}
\end{figure}

Fig. \ref{successvmass_gas_z} shows the fractional success of matching 2D mass substructures, that have been associated with a 3D subhalo, to 2D X-ray substructures versus subhalo mass, split into low and high redshift bins.  There is a trend for higher matching success between substructures in the total mass and X-ray surface brightness maps at higher redshifts for $M_{\rm sub}<10^{13} h^{-1} {\rm M}_{\odot}$, most simply explained by the argument that the subhaloes have had less time to suffer the effects of ram pressure stripping.

\subsubsection{Effects of dynamical state}\label{dynamsec}

Since major mergers on the cluster mass scale are such energetic events, it would be unsurprising if disturbed clusters exhibited more discrepancies in their hot gas and DM substructure. Indeed, some of the most extreme observational examples are found in highly disturbed clusters, for example the bullet cluster \citep{bulletclowe} and the `cosmic train-wreck' in Abell 520 \citep{2007ApJ...668..806M}. There is also much debate about the significance of the effect that merger activity has on bulk properties of galaxy clusters potentially making it an important selection effect. Simulations of isolated cluster mergers have suggested that massive, correlated luminosity and temperature boosts are associated with major mergers and that these could cause the masses of high-redshift clusters to be overestimated from  both the $M-T_{\rm X}$ and the $M-L_{\rm X}$ relation \citep{ricksar:apj,sarrick:apj}. It has also been suggested that, if this effect is real, such systems would stand out from scaling relations \citep{ohara:apj}, but in recent high resolution studies \citep{clusmergers_poole06,clusmergers_poole07,clusmergers_poole08} and cosmological simulations where multiple mergers occur, no such simple correlation between scatter in the $ L_{\rm X}-T_{\rm X}$ relation and visible evidence of ongoing major merger activity has been found \citep[e.g.][]{row:mnras,kayetal07}. 

In order to divide our images into just two subsets (major merger or not), it is necessary to have an additional technique to calibrate the centroid shift variance (described in Section \ref{calcdynam}) and determine which value of this statistic marks the threshold between these two states. Since we are interested in separating out the most extreme merging events, as this should make any trend stand out, we choose a value of the centroid shift variance which singles out the highest peaks in this quantity, which we determine to be 0.1. The sample is then split into two - those snapshots with values above the threshold and those with values below and this division is confirmed by examination of X-ray surface brightness contour maps (see Fig. \ref{mh:fig}).

\begin{figure}
\centering
  \includegraphics[width=0.5\textwidth]{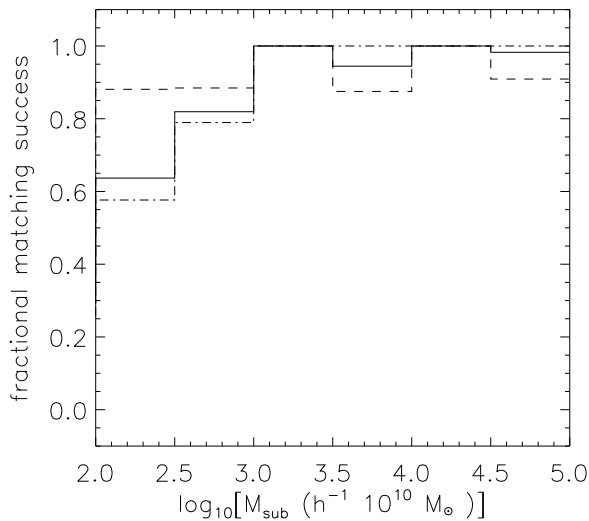}
 
\caption{Matching success {\it per mass bin} of $3\sigma$ 2D mass substructure catalogue to $1\sigma$  2D X-ray substructure catalogues versus subhalo DM mass. Data split into two subsets according to dynamical state (see text for details): relaxed (dot-dashed) and disturbed (dashed). Solid line shows combined data from Fig. \ref{successvmass_gas_3d}.}
\label{successvmass_gas_ds_3d}
\end{figure}

Fig. \ref{successvmass_gas_ds_3d} shows the matching success for the disturbed (dashed) and relaxed (dot-dash) samples, versus subhalo mass. It is clearly evident that all the matching failures above $M_{\rm sub}\simeq3\times10^{13}$ lie in the disturbed sample, and visual inspection of the maps confirms they are all undergoing significant mergers or collisions. The trend is actually reversed for low masses; substructures here have a lower probability of having an X-ray counterpart in relaxed clusters. This can be explained in the same way as the dependency on redshift in this mass range. Substructures in relaxed clusters have been there longer since, by definition, the last merger event was some time ago and therefore have been subject to stripping processes for longer. For comparison with our two redshift samples, we compute the mean redshift of the snapshots in our disturbed and relaxed samples, which we find to be 0.55 and 0.37 respectively. This highlights that there is some degeneracy between the effects of redshift and dynamical state, although these values are much closer than the mean redshifts of our redshift samples (0.72, 0.10; snapshots are equally spaced in time not redshift) and so the fact we get such a dramatic difference suggests looking at redshift alone is not sufficient.

\subsection{Detachment of hot gas from dark matter subhaloes}\label{casestudies}

In the previous section we have investigated the overall probability of finding a 2D X-ray counterpart for the 2D total mass substructures we detected. In order to fully probe all the factors which result in a matching failure in our analysis, a more detailed treatment (e.g. careful tracking of subhaloes between snapshots with higher time resolution or idealised simulations of individual mergers) would be required. Nevertheless, it is very informative to examine some of these matching failures in more detail to gain insight into the variety of scenarios that occur. It is reasonable to assume that ram pressure stripping is the main culprit in the removal of hot gas, however it is interesting to note that there are several distinct realisations of the outcome of this process in the maps. To illustrate these, we choose a small subset of substructures by visual inspection and follow these up in 2D and 3D. We present these {\it case studies} below.  \\

\subsubsection{Evidence of partial ram pressure stripping.}

Although we are primarily concerned  with cases in which matching between the 2D total mass and 2D X-ray surface brightness substructures fails, it is interesting to note that we observe the signatures of ram pressure stripping in objects for which the match is still achieved. \\ 

\begin{figure*}
\centering

  \includegraphics[width=0.3\textwidth]{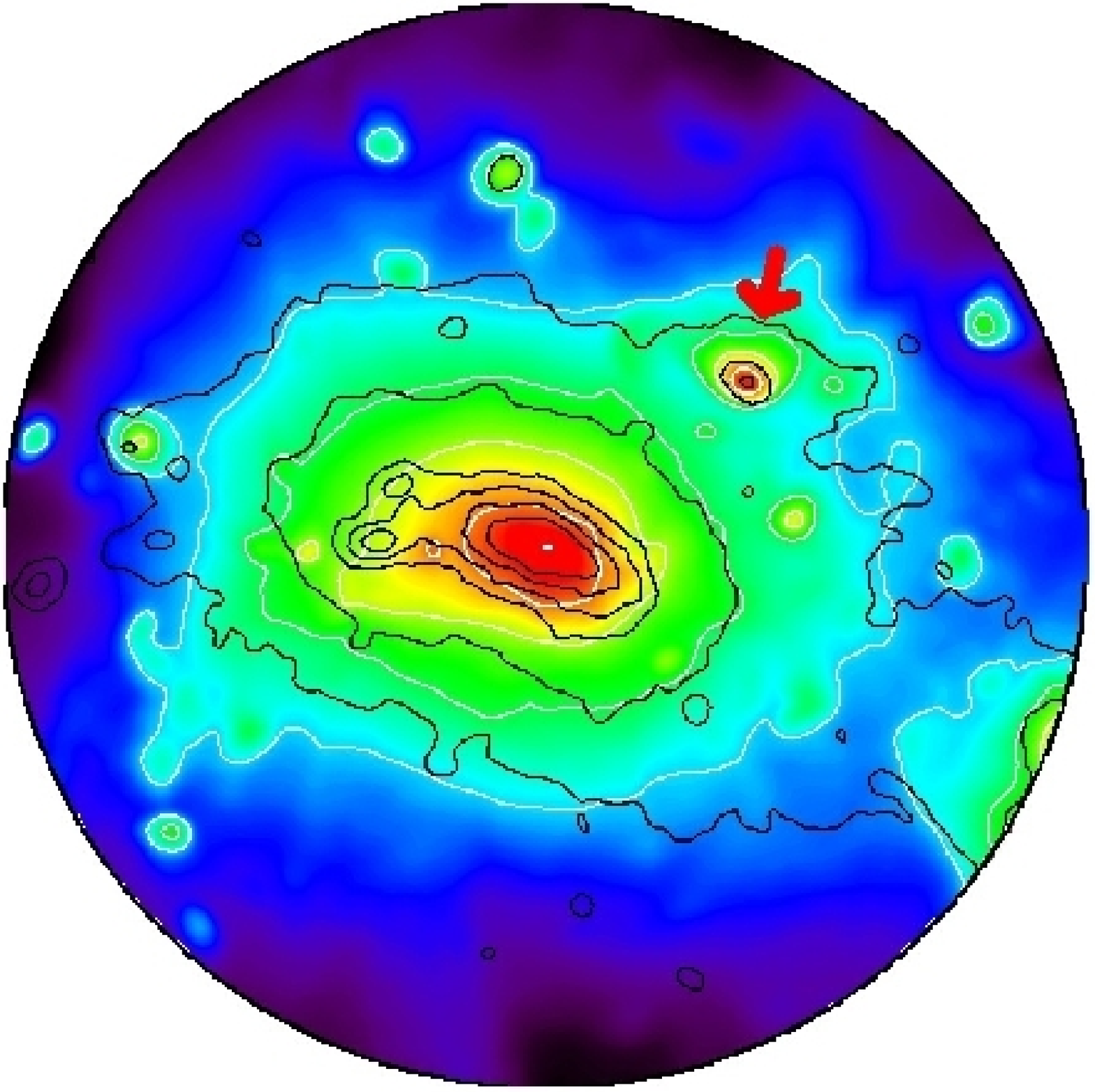}
  \hspace{0.005\textwidth}
  \includegraphics[width=0.3\textwidth]{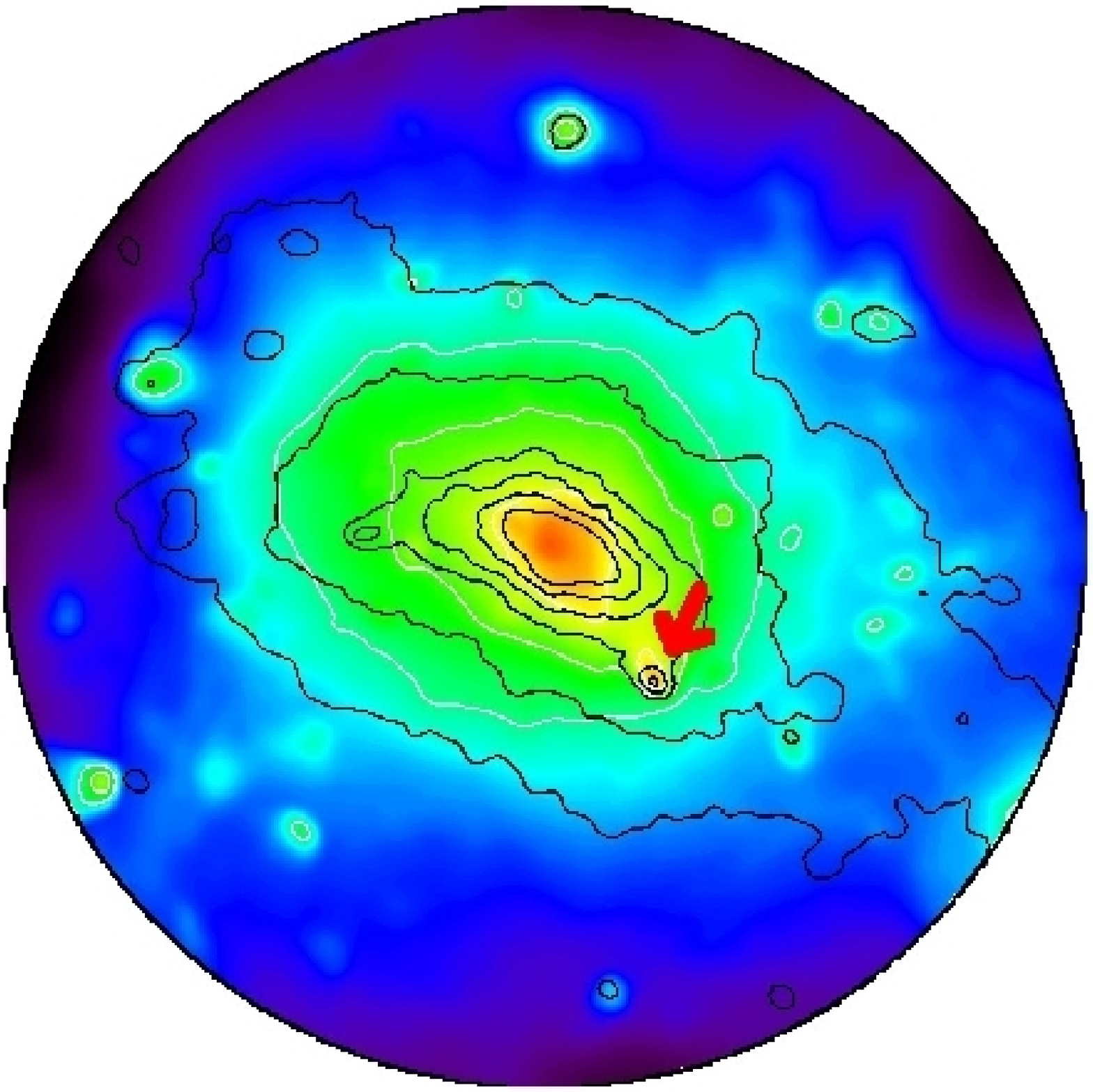}
   \hspace{0.005\textwidth}
 \includegraphics[width=0.3\textwidth]{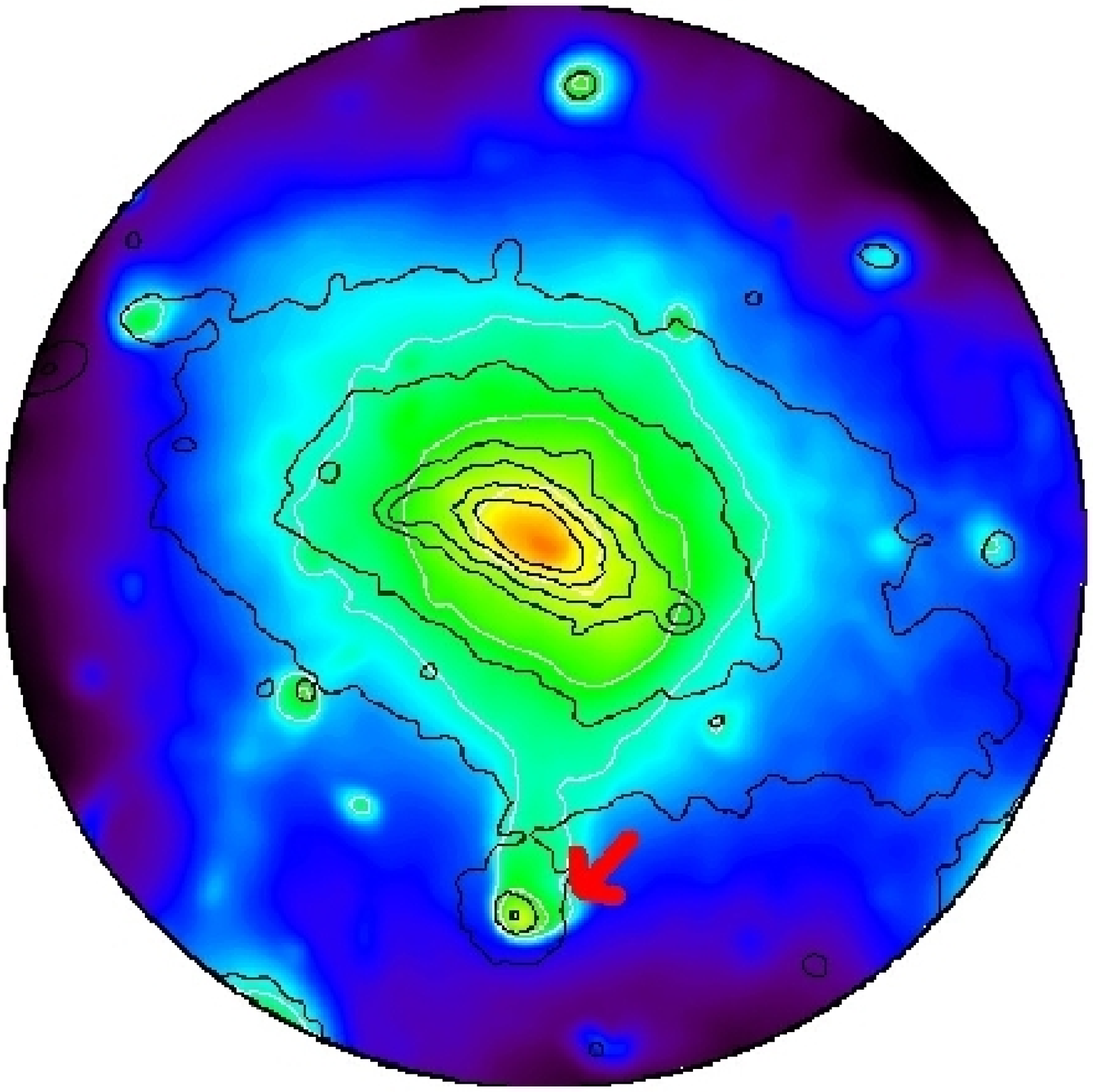}

\caption{Sequence showing strong evidence of stripping in cluster B for $0.16 \leq z \leq 0.21$, but neither detection nor matching procedures fail. Image is composed of the logarithmic X-ray surface brightness map with X-ray surface brightness (white) and surface mass density (black) contours, equally spaced in log, overlaid. Inner contours have half the spacing of outer ones in order to highlight structure in the core region. $M_{\rm sub}\simeq2.5\times10^{12} h^{-1} {\rm M}_{\odot}, 1.3\times10^{12} h^{-1} {\rm M}_{\odot}$ and $2.5\times10^{12} h^{-1} {\rm M}_{\odot}$, respectively.}
\label{gastail}
\end{figure*}

{\bf Case Study 1.} Fig. \ref{gastail} shows the development of a large tail of hot gas which streams behind the main substructure as it passes near to the cluster centre (its 3D physical displacement in the middle panel is approximately $0.5r_{500}$). The substructure is indicated with an arrow and has $M_{\rm sub}\simeq 1-3 \times 10^{12} h^{-1} {\rm M}_{\odot}$. Note the temporary decrease in mass (about a factor of 2) 
in the middle snapshot. This is a well-known issue with {\sc subfind}, where 
subhaloes become more difficult to distinguish when in close proximity
to the cluster's centre, due to their low density contrast \citep{subfindfault}.
In this case, around half of the subhalo's DM particles in the left-hand
image are deemed to belong to the main cluster in the middle image, decreasing
to around 20 per cent in the right-hand image, as the subhalo moves away from
the core region.

The substructure is clearly visible in both the X-ray image and the mass contours and is detected and matched in 2D  at each time shown, regardless of choice of the $1\sigma$ or $3\sigma$ X-ray catalogue. The {\sc subfind} data reveals that the mass of gas associated with the corresponding subhalo is reduced to 60 per cent of its initial value over the course of this time sequence, however. The subhalo may well be on a highly elongated orbit and may eventually be depleted of enough of its hot gas such that it is no longer detected in the X-ray surface brightness map. As such, it is likely that this represents what is, for many substructures, the first stage in a time sequence that eventually leads to matching failure. However, this stage may not be visible for many substructures, depending on the inclination of the gas tail with respect to the line of sight and the amount of X-ray emission from the stripped gas.

This case study also demonstrates that our detection technique is not too sensitive to the stripping of the outer regions of a substructure's hot gas and that matching failures therefore represent an extreme depletion or complete removal (or displacement) of the X-ray emitting component.

\subsubsection{Matching failure due to complete ram pressure stripping.}

\begin{figure*}
\centering

  \includegraphics[width=0.3\textwidth]{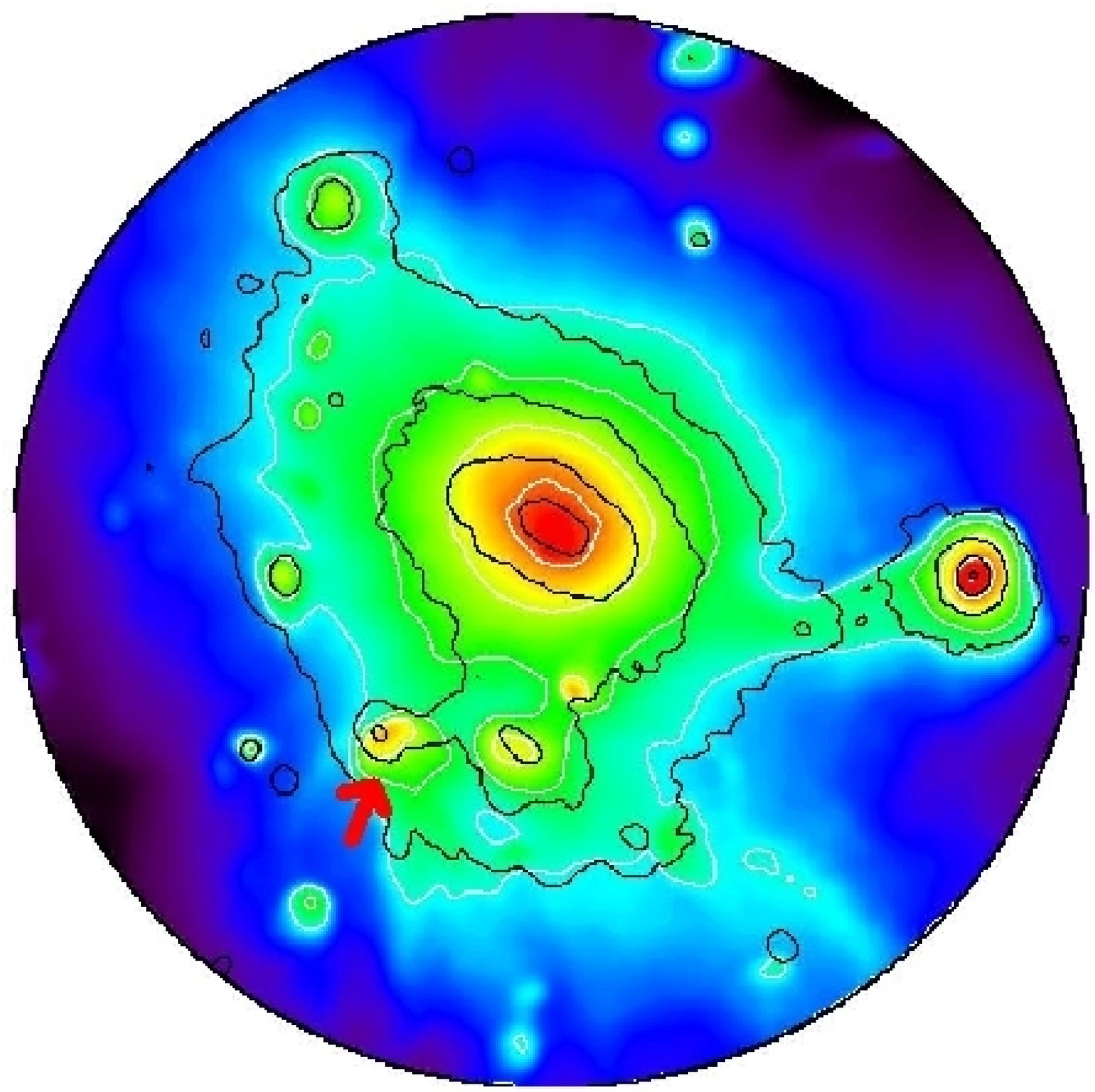}
  \hspace{0.005\textwidth}
  \includegraphics[width=0.3\textwidth]{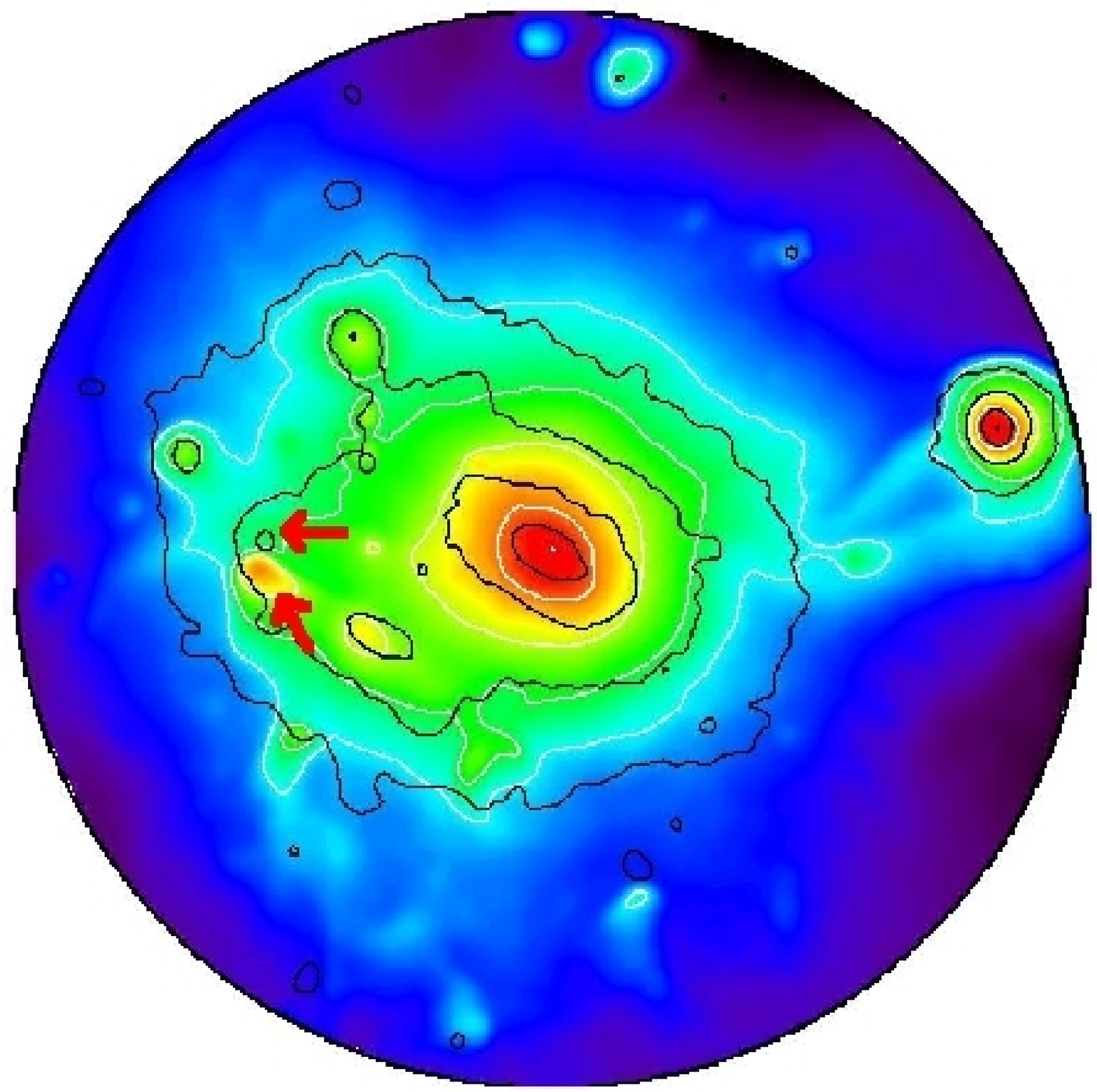}
   \hspace{0.005\textwidth}
  \includegraphics[width=0.3\textwidth]{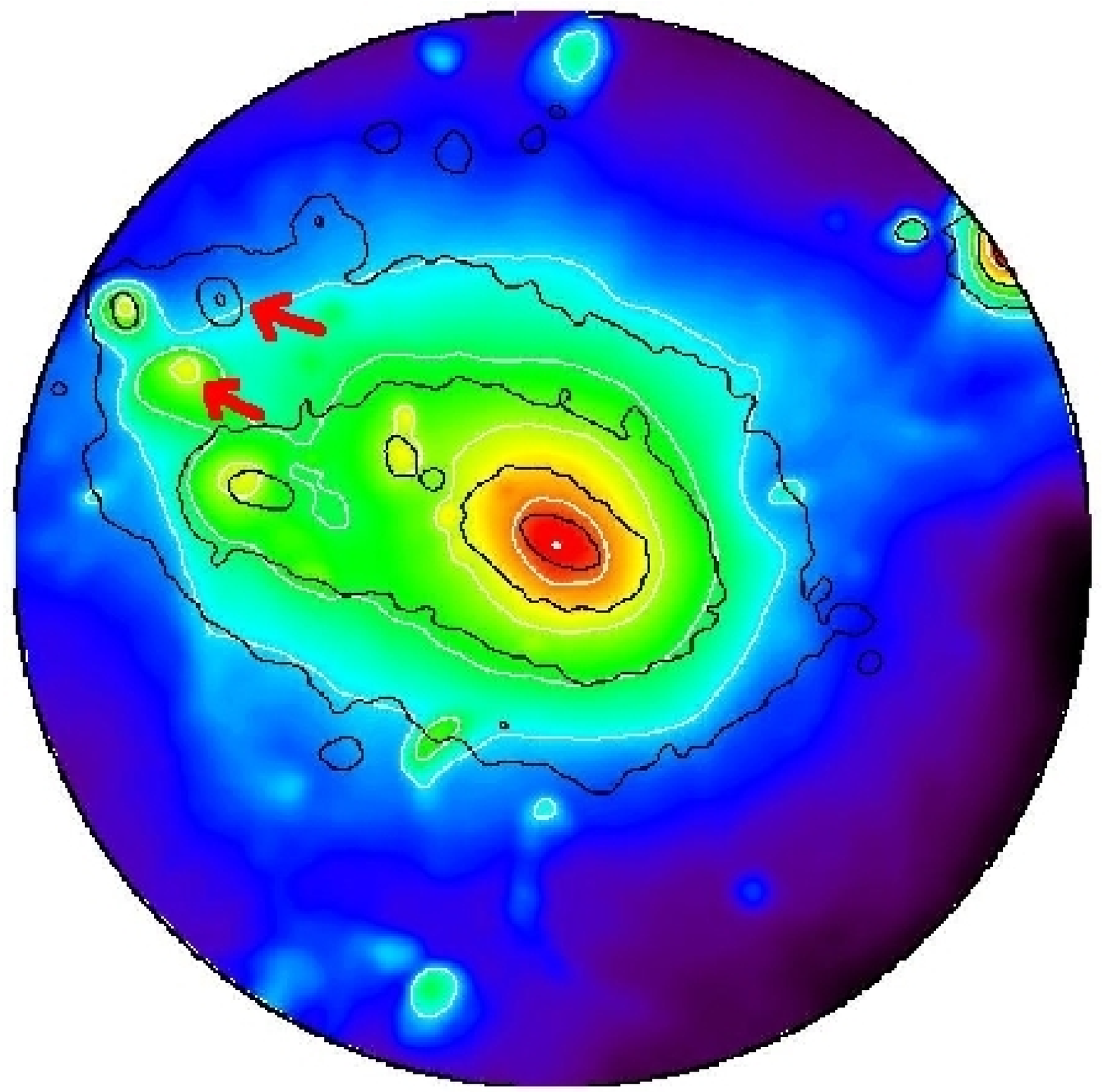}
  \hspace{0.005\textwidth}

\caption{Sequence showing progressive separation of 2D total mass substructure and its 2D X-ray counterpart for cluster B at $0.05 < z < 0.01$, which results in a matching failure in the final map of the sequence. Image is composed of the logarithmic X-ray surface brightness map with X-ray surface brightness (white) and surface mass density (black) contours, equally spaced in log, overlaid. $M_{\rm sub}\simeq2.6\times10^{12} h^{-1} {\rm M}_{\odot}, 2.4\times10^{12} h^{-1} {\rm M}_{\odot}$ and $3.0\times10^{12} h^{-1} {\rm M}_{\odot}$, respectively.}
\label{gasdisplaced}

\end{figure*}

We now examine the scenario whereby stripping of a substructure's hot gas results in the DM component turning up as an unmatched total mass substructure. There are two distinct categories here: matching failure due to the spatial displacement of the hot gas component and matching failure due to the hot gas component being erased completely from the X-ray surface brightness map. \\

{\bf Case study 2}. \citet{2007A&A...474..375P} produce an analytical model describing the increasing separation between the DM and hot gas components of a substructure moving through the ICM. Indeed this type of displacement, where the hot gas substructure remains intact yet is clearly displaced from the DM, is another reason for matching failures in our analysis. Fig. \ref{gasdisplaced} shows a sequence of maps in which an X-ray counterpart is initially found for a total mass substructure, but then clearly becomes spatially separated to the point where matching fails. It is clear that should this type of displacement occur along the line of sight, we would not be aware of it, however this isn't a shortfall of our method; we want to be subject to the same restrictions as observers. In the left panel the substructure is well matched to its X-ray counterpart (indicated by one arrow only as the contours and surface brightness peak are coincident). In the middle panel, due to the effects of ram pressure, the X-ray component is slightly offset, yet matching is still successful for both the $3\sigma$ and $1\sigma$ X-ray catalogues. Finally, in the right panel,  the progression of the X-ray component has been slowed so much that it is significantly displaced from the total mass substructure and is, therefore, not matched regardless of the choice of X-ray catalogue. 

A 3D analysis of this time sequence confirms this picture. The bound gas mass of the corresponding subhalo decreases to almost zero between the left-hand and middle panels of Fig. \ref{gasdisplaced} and the subhalo is completely gas-free by the right-hand panel. We identify, in the right-hand panel, the location of the gas particles which originally belonged to the DM subhalo and confirm that these exist as a clump which is coincident with the substructure in the X-ray surface brightness image. 

This separate hot gas component will eventually be completely disrupted, leaving the mass substructure (which will remain intact longer) with no trace of an X-ray counterpart. This situation is seen in the maps fairly frequently and presumably the separation procedure described above is the precursor to this. However, we also encounter an example of more immediate erasing of substructure from the X-ray surface brightness map, which we describe in the next case study.

It should be noted that the survival time of the gaseous component of subhaloes has been shown to be dependent on the numerical techniques employed and the resulting success with which hydrodynamical instabilities that expedite gas stripping are captured. \citet{amrvsph} show that SPH (with standard artificial viscosity) cannot capture Kelvin-Helmholtz Instabilities (KHI) as well as Adaptive Mesh Refinement codes. Indeed, \citet{newsubfind} demonstrate that using a low-viscosity scheme (less damping of the KHI) in {\sc gadget-2} results in smaller gas fractions for subhaloes inside a cluster's virial radius, suggesting such issues will impact on studies such as this one. More recently, improvements to the SPH methodology have been suggested, which increase its ability to capture KHI \citep{price_newsph, kawata_newsph, read_newsph}. It is therefore a matter for further investigation, whether the concentrations of stripped subhalo gas which give rise to X-ray surface brightness substructures, as illustrated in this case study, would still occur when the growth of KHI are simulated reliably. \\

\begin{figure*}
\centering

  \includegraphics[width=0.3\textwidth]{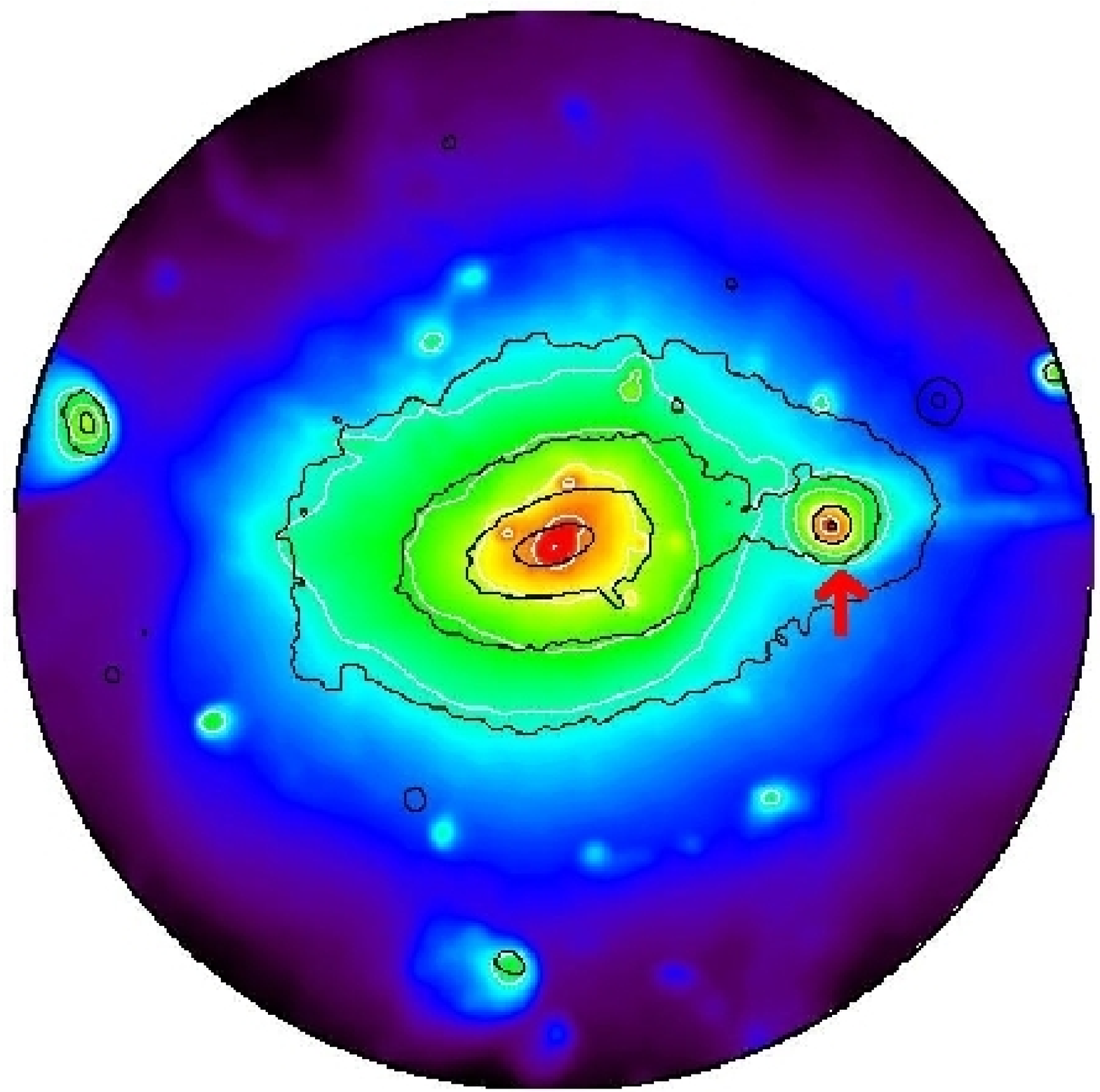}
   \hspace{0.005\textwidth}
  \includegraphics[width=0.3\textwidth]{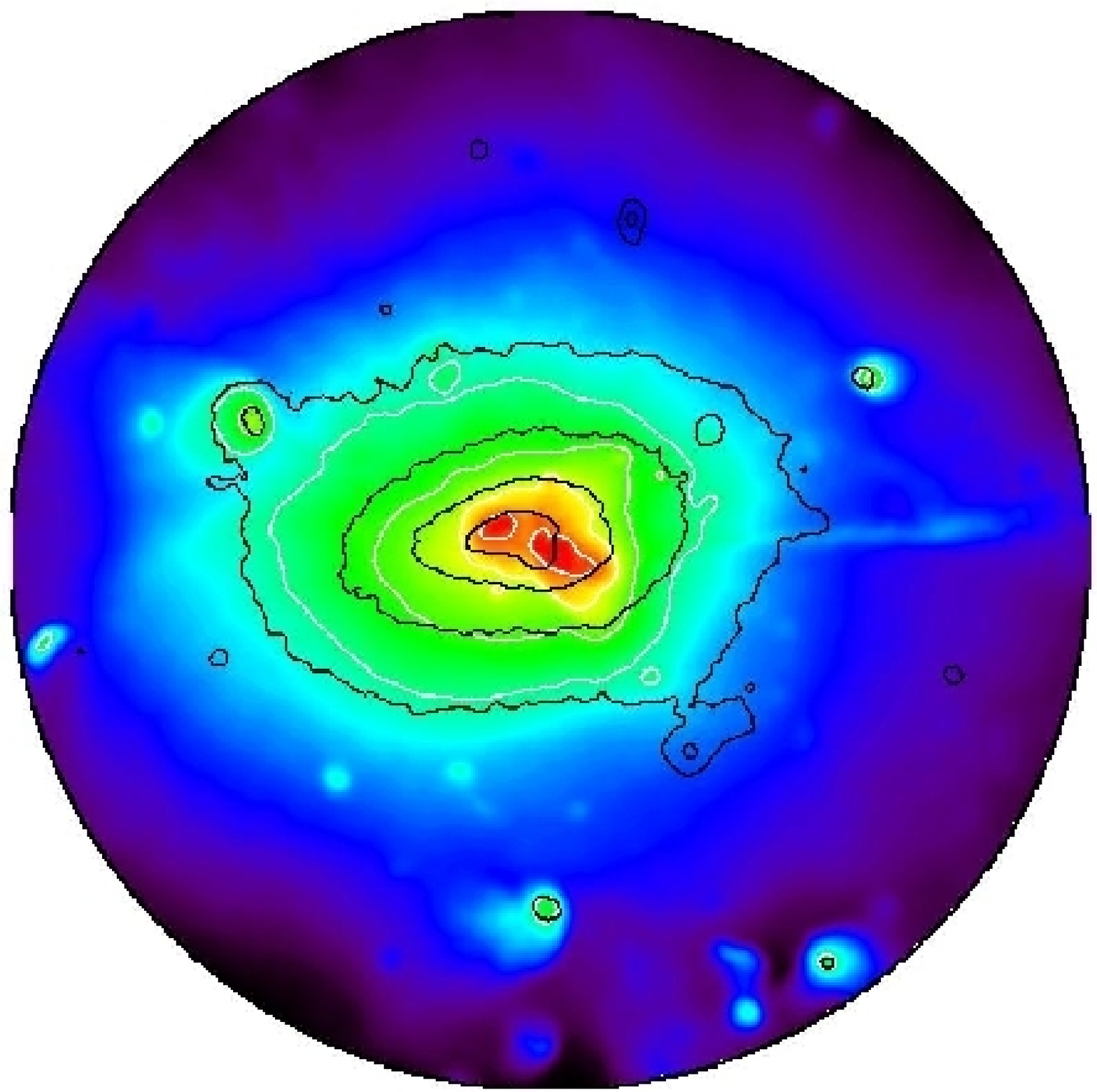}
  \hspace{0.005\textwidth}
  \includegraphics[width=0.3\textwidth]{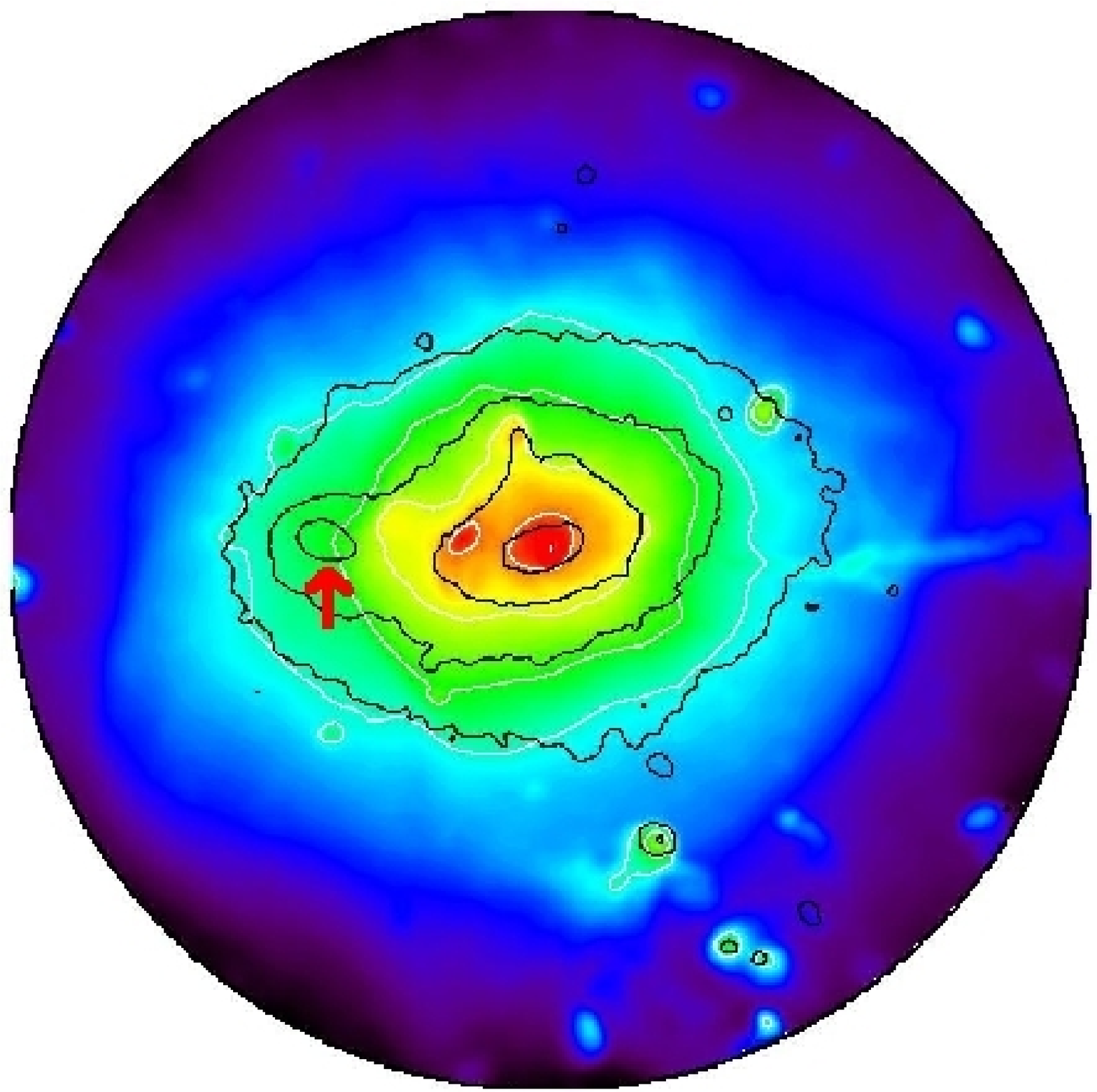}

\caption{Sequence stripping of X-ray gas from a substructure in for cluster A for $0.07\leq z \leq 0$., matching fails in the last image of the sequence. Image and contours as in Fig. \ref{gasdisplaced}. $M_{\rm sub}\simeq1.0\times10^{13} h^{-1} {\rm M}_{\odot}$, undefined (see text for details) and $4.4\times10^{12} h^{-1} {\rm M}_{\odot}$, respectively.}
\label{gaserased}
\end{figure*}

{\bf Case study 3}. Fig. \ref{gaserased} shows the encounter of a substructure ($M_{\rm sub}\simeq 10^{13} h^{-1} \rm{M}_{\odot}$) with the inner regions of the main cluster at $z\approx0$. In the left panel the substructure is detected (and clearly visible by eye, indicated with an arrow) in both the total mass and X-ray surface brightness maps and the matching procedure is successful. In the middle panel there is a clear double-peaked structure in the centre of the X-ray map, yet this is absent in the total mass contours and the substructure is not detected as a separate object from the cluster core in 2D. {\sc subfind} individually resolves the subhalo in 3D, albeit with a significantly reduced mass due to the effect discussed in Case Study 1. Therefore, although we have two 3D haloes (the main cluster and the subhalo), their proximity means both are attributed to the same 2D mass substructure; essentially the resolution limit of our detection procedure has been exceeded here. 

In the right-hand panel the substructure is again detected in the total mass map (its position in the contours is indicated with an arrow) but there is no corresponding detection in this region of the X-ray map. The DM substructure has been completely stripped of its hot gas and is also stripped (tidally) itself; according to {\sc subfind} the corresponding subhalo has no bound gas particles in the final panel and the DM mass has been reduced to roughly 40 per cent of its value in the first panel. Indeed the rest of this subhalo's DM and gas particles are found to belong to the main cluster at the end of this time sequence.

Note, however, the edge-like feature present instead which appears to lag behind the 2D mass substructure (detections of this are made in both X-ray catalogues but are not matched to the mass substructure as the positions are not coincident). A tail of stripped hot gas is also visible on the right-hand side of this map which resulted from the substructure's approach and is also apparent in the left and middle panels. A detailed discussion of the gaseous features that can arise during such interactions appears in \citet{clusmergers_poole06}. 

We also note that there are similar scenarios whereby a collision results in the disruption of the X-ray emitting gas, rather than its complete removal as seen here. In these cases, while there is still evidence of gas in the vicinity of the mass map substructure, a defined peak is no longer visible and the detection of an X-ray substructure can then be catalogue dependent (different catalogues impose different cuts on the residual surface brightness). The issue of catalogue dependence is discussed in the following case study.

If we examine the maps immediately preceding this sequence we observe the same substructure undergoing stripping of its hot gas on an earlier passage through the cluster's inner regions, adding weight to our supposition that Case Study 1 may be the precursor to the hot gas being removed completely. In fact, it is plausible that all the case studies may simply be different moments in a sequence which substructures that continue to orbit within the cluster long enough are subject to: stripping of outer regions of hot gas, displacement of remaining hot gas then complete disruption of the hot gas substructure.\\

\begin{figure*}
\centering
  \includegraphics[width=0.3\textwidth]{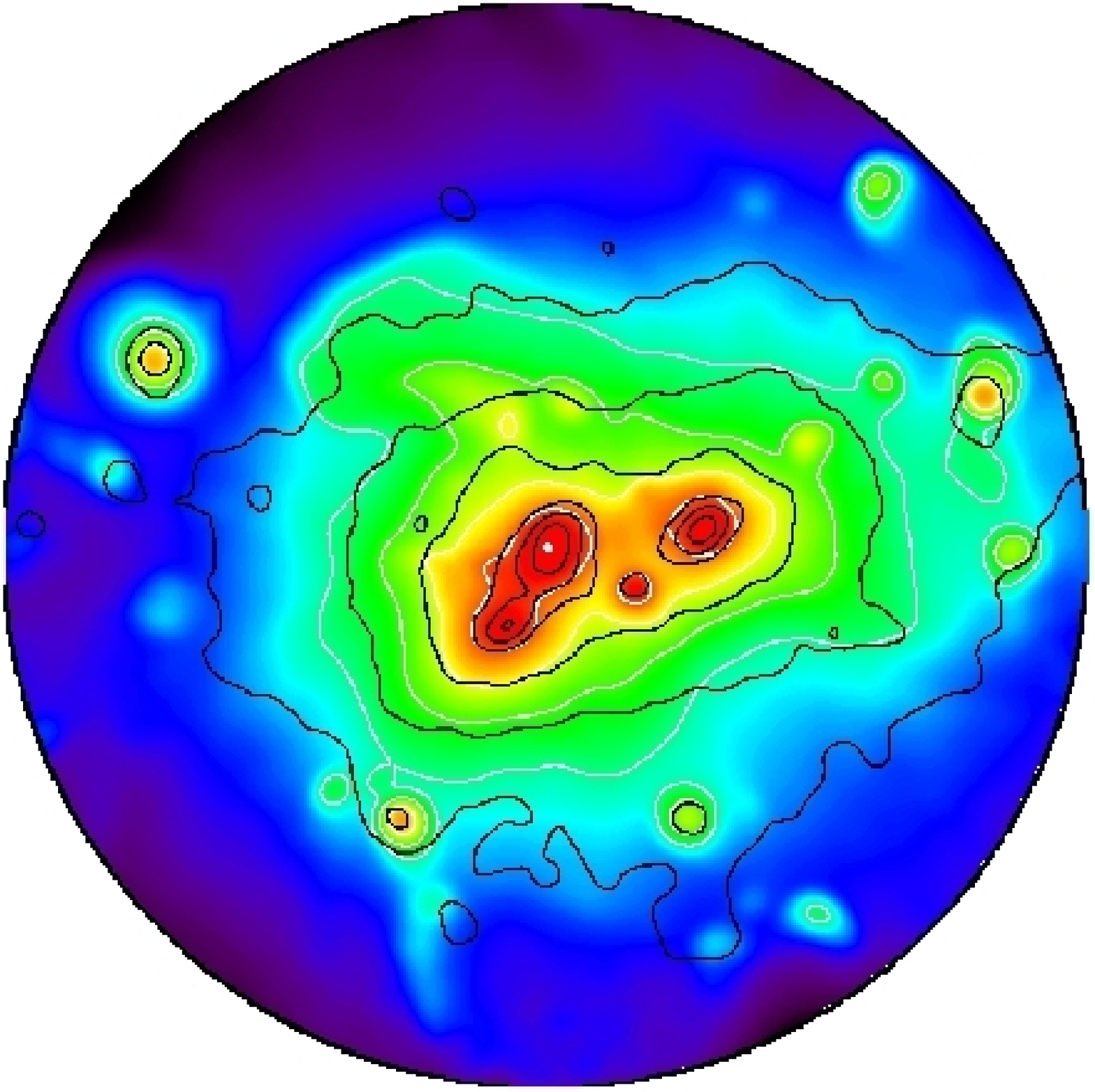}
  \hspace{0.005\textwidth}
 \includegraphics[width=0.3\textwidth]{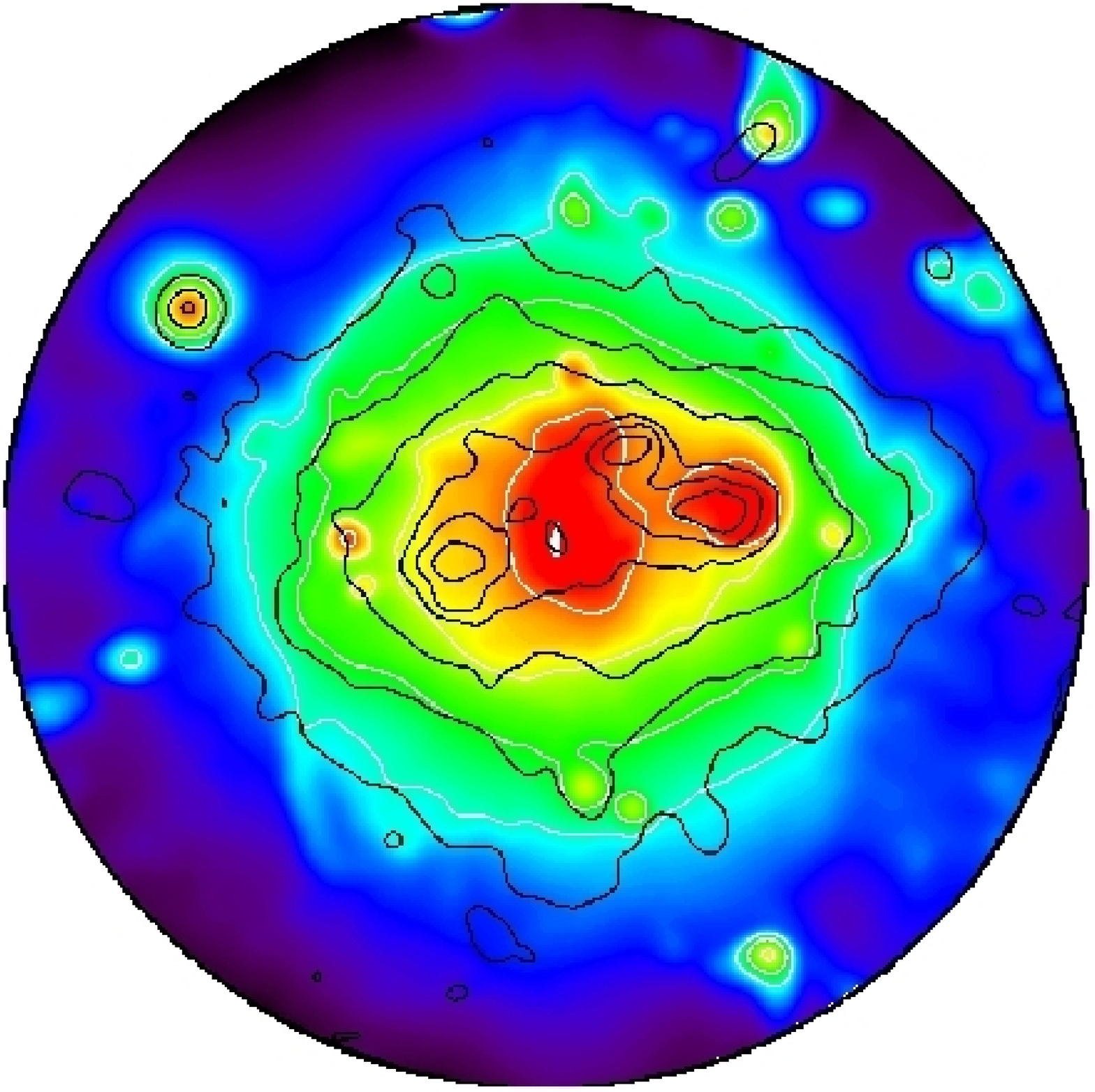}
 \hspace{0.005\textwidth}
  \includegraphics[width=0.3\textwidth]{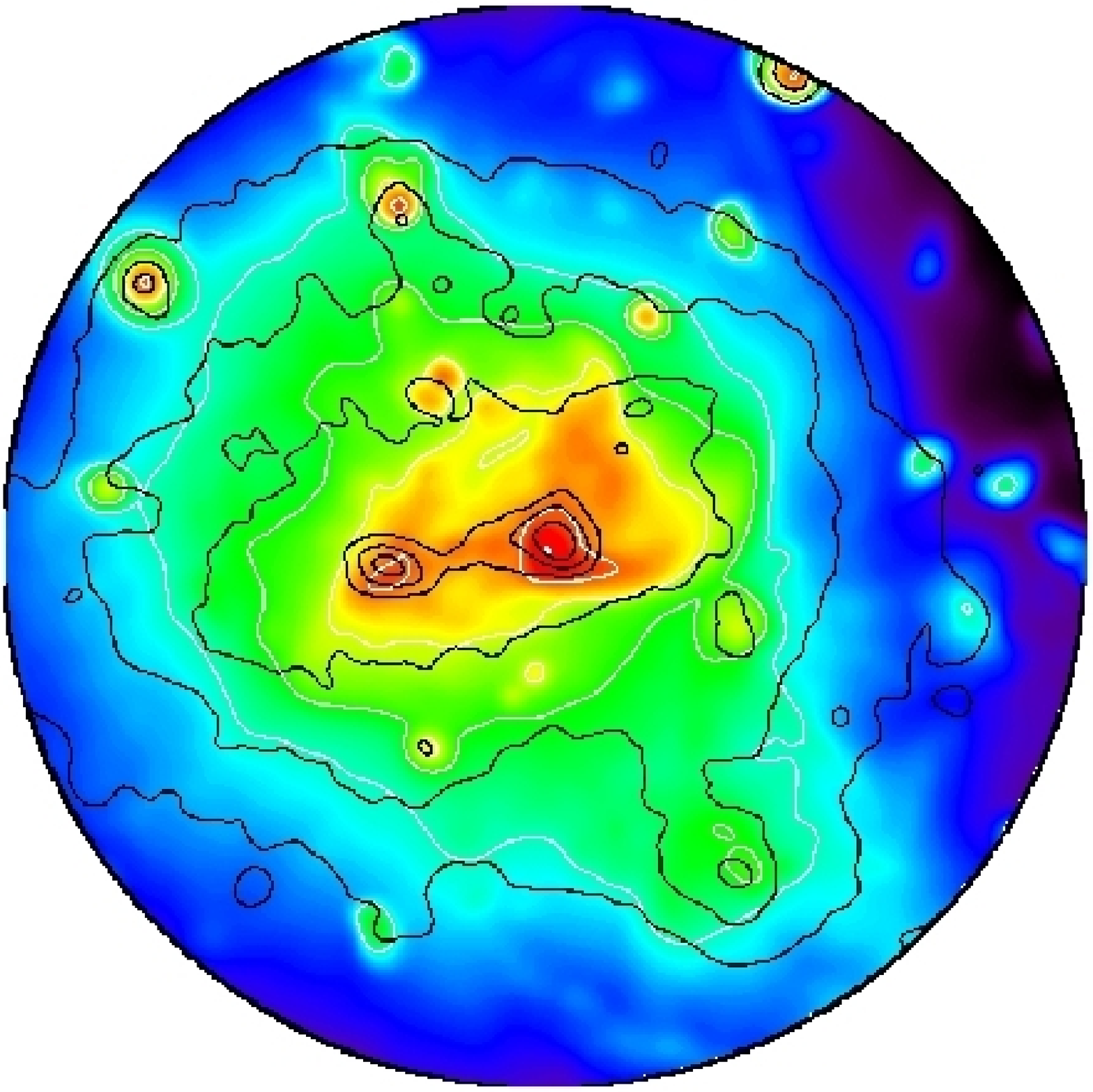}
\caption{Image of cluster A undergoing a merger at  $z \sim 1$, with the middle panel bearing remarkable similarity to that in \citet{bullet2}. Left-hand panel: Masses of three main inner substructures are $2.8\times10^{12}$$h^{-1} \rm{M}_{\odot}$ (smaller peak), $1.8\times10^{14}$$h^{-1} \rm{M}_{\odot}$ (large eastern peak) and $3.7\times10^{12}$$h^{-1} \rm{M}_{\odot}$ (large western peak). Middle panel: The western X-ray peak is coincident with one of the DM substructures yet the eastern X-ray peak lies in-between the other two substructures. The larger of these appears completely devoid of hot gas and is the main cluster as defined by {\sc subfind} ($M_{\rm sub}\simeq 1.9\times10^{14} h^{-1} \rm{M}_{\odot}$). The displacement of this surface mass density peak from the eastern X-ray surface brightness peak is $90 h^{-1}$ kpc.  The smaller DM substructure has $M_{\rm sub}\simeq 1.6 \times 10^{12} h^{-1} \rm{M}_{\odot}$. Right-hand panel: Masses of two main inner substructures are $2.0\times10^{12}$$h^{-1} \rm{M}_{\odot}$ (large eastern peak) and $2.1\times10^{14}$$h^{-1} \rm{M}_{\odot}$ (large western peak). Image and contours as in Fig. \ref{gastail}.  Note that north is up and east is left. See text for further details.}
\label{mybullet}

\end{figure*}

{\bf Case study 4.} We now examine a scenario in which a close encounter between the main cluster and a subhalo has resulted in the hot gas from one object being removed and assimilated into the main cluster's ICM. However, in this case the  conglomerate hot gas does not correspond spatially with a mass map substructure. This case study highlights how challenging it can be to correctly determine which X-ray substructures are associated with which mass substructures when working in 2D and the limitations of our current detection and matching schemes.

In Fig. \ref{mybullet} we present an image of cluster A at $z = 1$ (middle panel) which bears similarity, in terms of its configuration, to the recent observation in \citet{bullet2}. This observation is on a larger scale, however, depicting the merger of two clusters, with nearly equal masses. The left and right-hand panels show the maps for the snapshot directly before and after, respectively, to provide some insight however, due to the complexity of the merger, a detailed subhalo merger tree would be required to unravel the full series of events. Here we focus on the middle panel.

It is clear that the inner region consists of three DM haloes (all detected in 2D by our algorithm) and only two X-ray peaks (also detected). The western X-ray peak is coincident with (and successfully matched to) one of the DM substructures, however, the eastern X-ray peak lies in-between the other two substructures. The smaller of these DM substructures ($M_{\rm sub}\simeq1.6\times 10^{12} h^{-1} \rm{M}_{\odot}$) is matched to the eastern X-ray substructure when using either the $1\sigma$ or $3\sigma$ X-ray catalogue. We can see, however, that this substructure is still offset from the bulk of the hot gas in this region and instead just overlaps with a tail of material extending outwards towards it. Indeed, in 3D this subhalo has a very low gas fraction, suggesting its match with the X-ray substructure is primarily a projection effect. The larger of these DM substructures appears completely devoid of hot gas and is the main cluster as defined by {\sc subfind} ($M_{\rm sub}\simeq 1.9 \times10^{14} h^{-1} \rm{M}_{\odot}$). The displacement of this surface mass density peak from the eastern X-ray surface brightness peak is $90 h^{-1}$ kpc. It is interesting that while the projected mass peak of the main cluster is offset from the bulk of the X-ray emission, its most bound particle actually coincides with the maximum X-ray surface brightness (the white spot in the image). When we use the $1\sigma$ X-ray catalogue, the main cluster core is actually matched to a very small, separate X-ray substructure and so does not show up as an unmatched object in this data set. This pair would add to the scatter in the area-area correlation plot (Fig. \ref{aa_ee}) suggesting that, in future work, poor agreement in area of matched pairs could be used to remove dubious matches. 

This case study illustrates that the use of the $3\sigma$ X-ray catalogue is the most likely to allow retrieval of all high mass substructures with no {\it significant} hot gas component as these will fail to be matched. The $1\sigma$ X-ray catalogue is less stringent and will detect smaller amounts of gas and will also result in the matching of substructures that are slightly offset, since the X-ray substructures in this catalogue are more extended. However, by the same token, it is also more prone to false matches than the $3\sigma$ catalogue.
 
\subsection[]{Gas physics model parameters}\label{coolsec}

Our main results focussed on a set of {\it non-radiative} clusters, the
simplest model for the ICM within a cosmological context.
It is well known, however, that additional physical processes must operate
in clusters; scaling relations such as the X-ray luminosity-temperature
relation are different to what is expected from the so-called 
{\it self-similar} model \citep[e.g.][]{kaiser91}. The most favoured 
explanation for the altered similarity of clusters is that the ICM has undergone an intense, and perhaps extended, period of heating
due to galactic outflows (from stars and active galactic nuclei). Radiative
cooling also plays a role, selectively removing the low entropy gas, although
is completely reliant on subsequent heating to avoid a cooling 
{\it catastrophe} \citep[e.g.][]{babul02,mccarthy04,mccarthy08}.

Investigating the full effects of cooling and heating is beyond the scope
of this paper, but we have performed a preliminary investigation on the
effects of cooling on our results. To avoid over-cooling the gas and motivated
by the observations that stellar populations in clusters are old \citep[e.g.][]{thomasetal_05ApJ}, we only allow the gas to cool radiatively at early
times, until a reasonable fraction of gas has cooled and formed stars. 
We adopted the same procedure as outlined in \citet{kayetal04}, assuming
a metal-free gas ($Z=0$). The simulation of cluster A was repeated and 
cooling was switched on until $z\sim 5$, when around 10 per cent of the gas in the high-resolution region had
formed stars, in agreement with near-infrared observations, which suggest that the stellar mass, as a percentage of the ICM gas mass, is around $10$ per cent on average \citep[e.g.][]{linmohrstanford_03,baloghetal01,2df_coleetal_01}. The cooled fraction in the cluster, within $r_{500}$, at $z=0$ is about 20 per cent, slightly higher than observed by \citet{linmohrstanford_03}. 

\begin{center}
\begin{table}
\caption{Properties of clusters A (NR and CL denote the non-radiative and cooling runs respectively), B and C at $z=0$. $M_{500}$, $T_{\rm sl}$ and $L_{\rm X, sim}$ are calculated from the simulation data directly. $L_{\rm X, obs}$ are the observed X-ray luminosities for clusters with the same value of $M_{500}$ or $T_{\rm sl}$, calculated using the $L_{\rm X}-M_{500}$ or the $L_{\rm X}-T_{\rm X}$ scaling relation from \citet{lmrelation}, denoted by L-M and L-T, respectively.}
\label{clusprops}
\begin{tabular}{p{0.06\textwidth} p{0.09\textwidth} p{0.03\textwidth} p{0.06\textwidth} p{0.05\textwidth} p{0.05\textwidth}}
\hline
\textbf{Cluster} &\textbf{$M_{500}$ } & \textbf{$T_{\rm sl}$} & \textbf{$L_{\rm X, sim}$} &  \multicolumn{2}{l}{\textbf{$L_{\rm X, obs}$}}  \\
& ($10^{14}h^{-1} {\rm M}_{\odot}$) & (keV) & ($10^{44}$ergs) &  ($10^{44}$ergs) & \\
 & & & & \textbf{L-M} & \textbf{L-T} \\
 \hline
A (CL) & 6.0 & 5.7 & 18.1 & 30.3 & 11.5\\
A (NR) & 6.0 & 4.7 & 29.2 & 30.1 & 6.0\\
B & 2.7 & 3.3 & 8.5 & 6.4 & 1.8\\
C & 5.9 & 3.1 & 22.3 & 29.4 & 1.5\\
\hline
\end{tabular}
\end{table}
\end{center}

We use observed $L_{\rm X}-M_{500}$ and $L_{\rm X}-T_{\rm X}$ relations \citep{lmrelation} in order to compare the simulated X-ray luminosity ($L_{\rm X, sim}$) of cluster A at $z=0$ with that expected from observations ($L_{\rm X, obs}$) based on both its $M_{500}$ value and its spectroscopic-like temperature, $T_{\rm sl}$ \citep{slt}. Table \ref{clusprops} summarises these properties, including the values for the other clusters as a point of comparison. The non-radiative version of cluster A has a similar X-ray luminosity to that observed based on its mass, but a much higher luminosity than observed ($L_{\rm X, sim}/L_{\rm X, obs}\approx 5$) based on $T_{\rm sl}$. Similarly, clusters B and C also exhibit X-ray luminosities close to those observed for their masses at $z=0$, yet are very over-luminous for their temperatures. Assuming the observed mass determinations are accurate, then the main difference is that the simulated $T_{\rm sl}$ is too low in the non-radiative model due to the presence of too much cold gas \citep[see][]{sztemp}. Turning on high-redshift cooling in cluster A preferentially removes this cool gas, bringing down the luminosity ($L_{\rm X, sim}/L_{\rm X, obs}\approx 0.6$, based on mass) but increasing $T_{\rm sl}$ such that $L_{\rm X, sim}/L_{\rm X, obs}\approx 1.6$, based on temperature. So, overall, the cooling model is closer to the observed (best-fit) $L_{\rm X}-T_{\rm X}-M_{500}$ plane. In future, we will also consider the additional effects of heating from supernovae and active galactic nuclei. It will be interesting to see how the competing effects of cooling and heating affect the structure of the subhaloes.

The cluster identification procedure is the same as described in Section \ref{simsec}, except that in order to ensure we follow the same object in all resimulations of the same cluster, the list of cluster candidates in the cooling run is searched for the best match to the selected object in the non-radiative run. In Fig. \ref{zeq0mapfig} the surface mass density (left) and X-ray surface brightness (right) maps for the cooling run (first row) and non-radiative run (second row) of cluster A at $z=0$ can be compared. Qualitatively, the two sets of maps appear similar, suggesting that cooling at high redshift does not strongly affect our main results.

\begin{figure}
\centering
\includegraphics[width=0.5\textwidth]{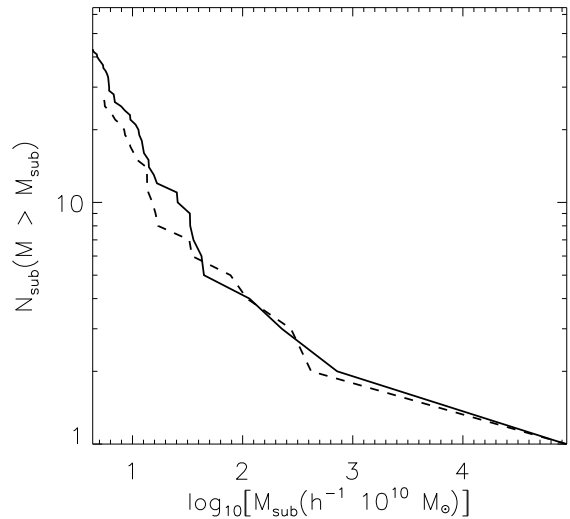}
\caption{Cumulative subhalo DM mass functions for cluster A cooling run (solid line), 
and cluster A non-radiative run (dashed line), at $z=0$. Data are for subhaloes with their most bound particle within $r_{500}$.}
\label{mfcoolfig}
\end{figure}

We can once again examine the properties of our 3D subhalo sample, to see  the role cooling has played. Fig. \ref{mfcoolfig} shows the $z=0$ subhalo DM mass function for the cooling (solid line) and non-radiative (dashed line) runs of cluster A. The DM mass functions agree well, although it is apparent there are more low DM mass subhaloes (a similar effect is also seen at $z=0.5$ and $1$), suggesting that the central condensation of the baryons deepen the potential wells and reduce the amount of disruption of the DM.

\begin{figure}
\centering
\includegraphics[width=0.5\textwidth]{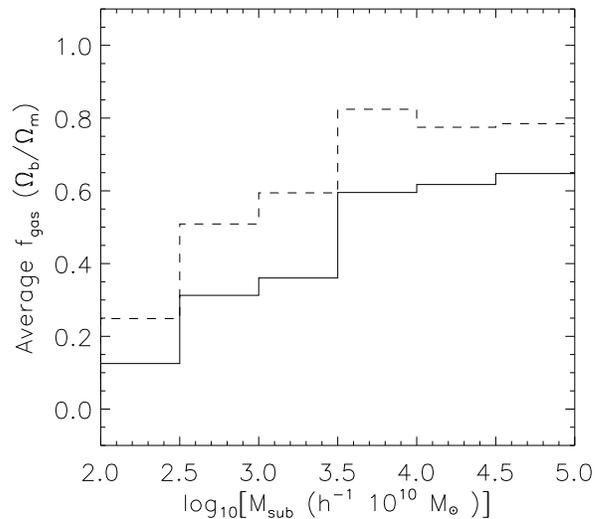}
\caption{Average hot gas fraction per mass bin (in units of $\Omega_{\rm b}/\Omega_{\rm m}=0.15$) for cooling run (solid) and non-radiative run (dashed) of cluster A.}
\label{gasfraccool}
\end{figure}

We examine the effect of cooling on the hot gas within the 3D subhaloes, by computing the average hot (T$>10^{6}$K) gas fractions within each mass bin. The results are shown in Fig~\ref{gasfraccool}, for the non-radiative (dashed) and cooling (solid) versions of 
cluster A (expressed in units of the global value, $\Omega_{\rm b}/\Omega_{\rm m}=0.15$). In the non-radiative cluster, the hot gas fraction increases with subhalo mass, reflecting the increasing ability of the subhaloes to retain their hot gas as their potential wells deepen (note the main haloes are also shown). In the cooling run, the same trend is seen, but the gas fraction is lower at all masses (compared to the non-radiative run). This is due to cooling causing the additional depletion of the hot gas reservoir by transforming it into cold gas (and eventually, stars).  

As a result of their shallower potential, ram pressure stripping of hot gas is most effective in the low mass ($\sim 10^{12} \, h^{-1} {\rm M}_{\odot}$) subhaloes, as indicated by their very low average gas fractions ($\approx 0.2$) in both runs. It is at this mass range, however, that cooling is also most effective because these objects form at high redshift and have short cooling times. We see the result of this effect when comparing the average subhalo total baryon fractions in the two runs; the cooling cluster has around 40 per cent more baryons at $M_{\rm sub}\sim 10^{12} \, h^{-1} {\rm M}_{\odot}$, yet the total baryon fractions agree well between the runs at higher masses.

The procedure to detect 2D substructures in maps of X-ray surface brightness and total mass density (described in Section \ref{2dsec}) is applied, with identical parameters, to the cooling run of cluster A.
As could be expected from the discussion of the 3D subhalo data for this run, more 2D total mass substructures are found; the $3\sigma$ 2D mass catalogue contains 611 substructures compared with 473 in the non-radiative run of the same cluster. The number of substructures in the $1\sigma$ X-ray catalogue is very similar, with 395 compared to 363 in the non-radiative cluster. 

The matching of 2D total mass substructures to the 3D subhalo data was undertaken for the cooling run as described in Section \ref{2d3dsec}, and an assessment of completeness and purity of the 2D catalogue was repeated. The choice of the $3\sigma$ 2D total mass catalogue was again deemed the most appropriate and the purity and completeness limits already established were found to be valid for application to the cooling run data.

\begin{figure}
\centering
\includegraphics[width=0.5\textwidth]{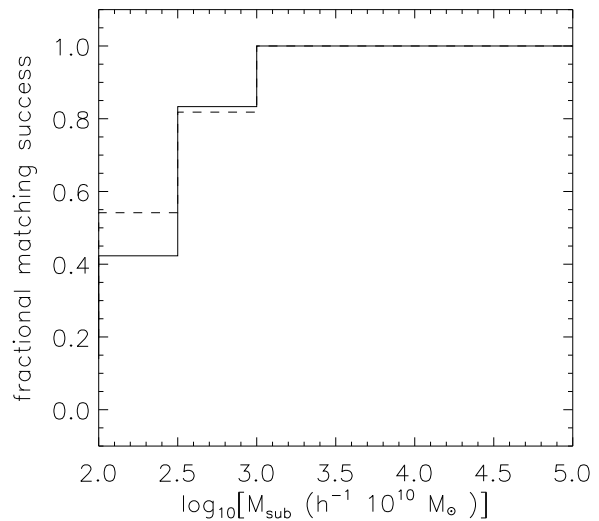} 
\caption{Matching success {\it per mass bin} of $3\sigma$ 2D mass substructure catalogue to $1\sigma$ 2D X-ray substructure catalogues for cooling run (solid) and non-radiative run (dashed) versus subhalo mass. }
\label{successvmass_gas_cool}
\end{figure}

We now consider the likelihood of finding an X-ray counterpart for the substructures in the calibrated 2D total mass catalogue (i.e. the catalogue containing only those substructures that were successfully identified  with a 3D subhalo).
Fig. \ref{successvmass_gas_cool} shows the matching success to X-ray for total mass substructures. Surprisingly, the results for the cooling run (solid line) match those for the non-radiative run (dashed line) very closely. In particular, although Fig.~\ref{gasfraccool} shows subhaloes of all masses are more depleted of hot gas in the cooling run, this doesn't seem to translate into a significant decrease in the likelihood of finding a 2D X-ray counterpart for 2D mass substructures, except for $M_{\rm sub}\sim 10^{12} h^{-1} {\rm M}_{\odot}$.
Overall, it seems that the introduction of high-redshift cooling does not affect the main results and therefore that our non-radiative results are not too sensitive to the gas physics model employed (although further investigation into the effects of cooling plus feedback would be desirable).

\subsection{Towards realistic observations}
\label{real}

The main results of this paper have focussed on `perfect' observations. Although a detailed analysis of all the potential observational and instrumental effects is beyond the scope of this paper, we now consider the impact on our results of using an observationally achievable map resolution and including basic noise in the maps. We defer a more detailed treatment of noise and instrumental effects to future work. This analysis is undertaken on the non-radiative simulations as this provides us with a larger sample of clusters and we have shown that the impact of high-redshift cooling is minimal. 

\begin{figure}
\centering

  \includegraphics[width=0.23\textwidth]{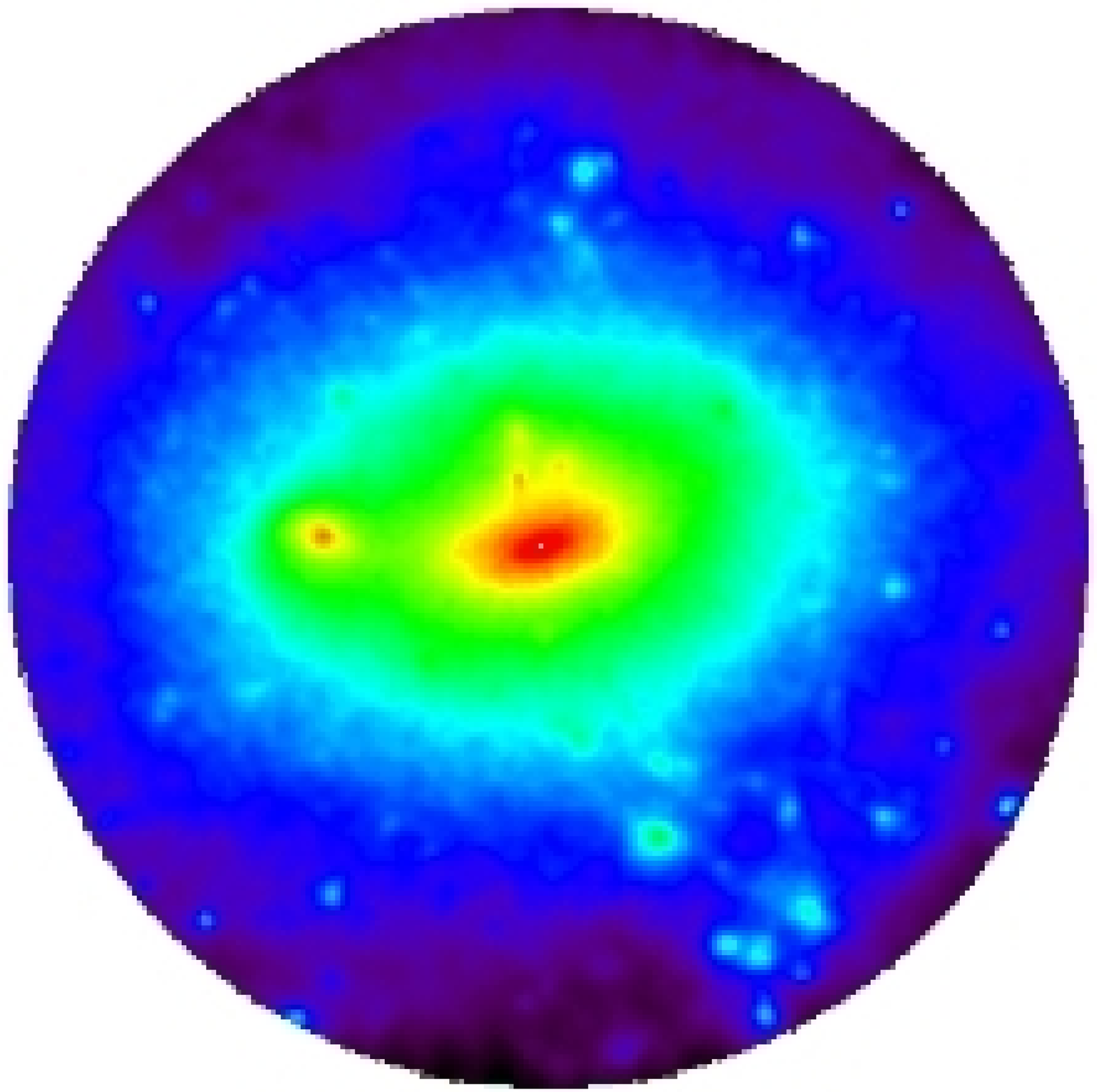}
  \hspace{0.005\textwidth}
  \includegraphics[width=0.23\textwidth]{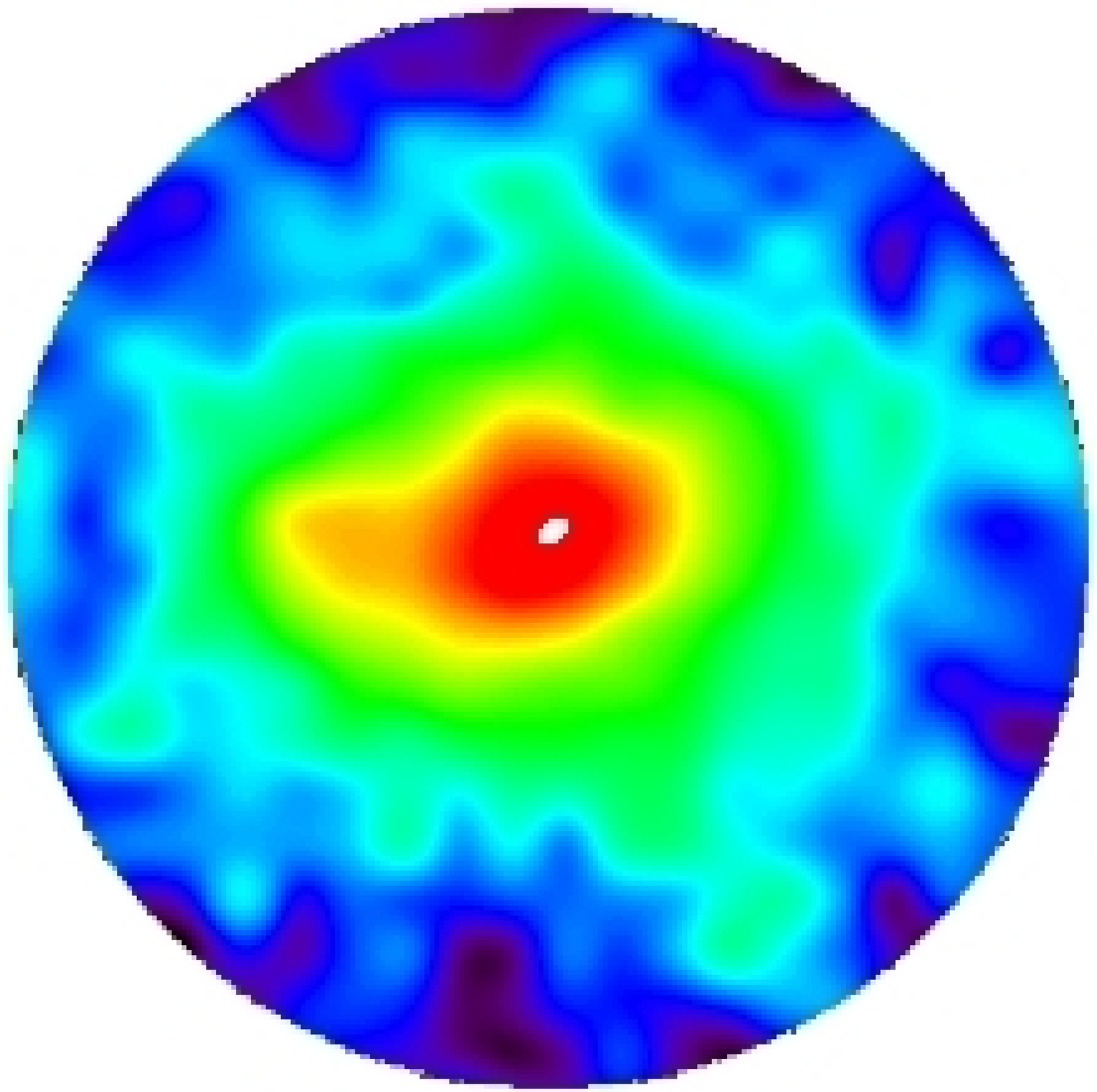}\\

    \includegraphics[width=0.23\textwidth]{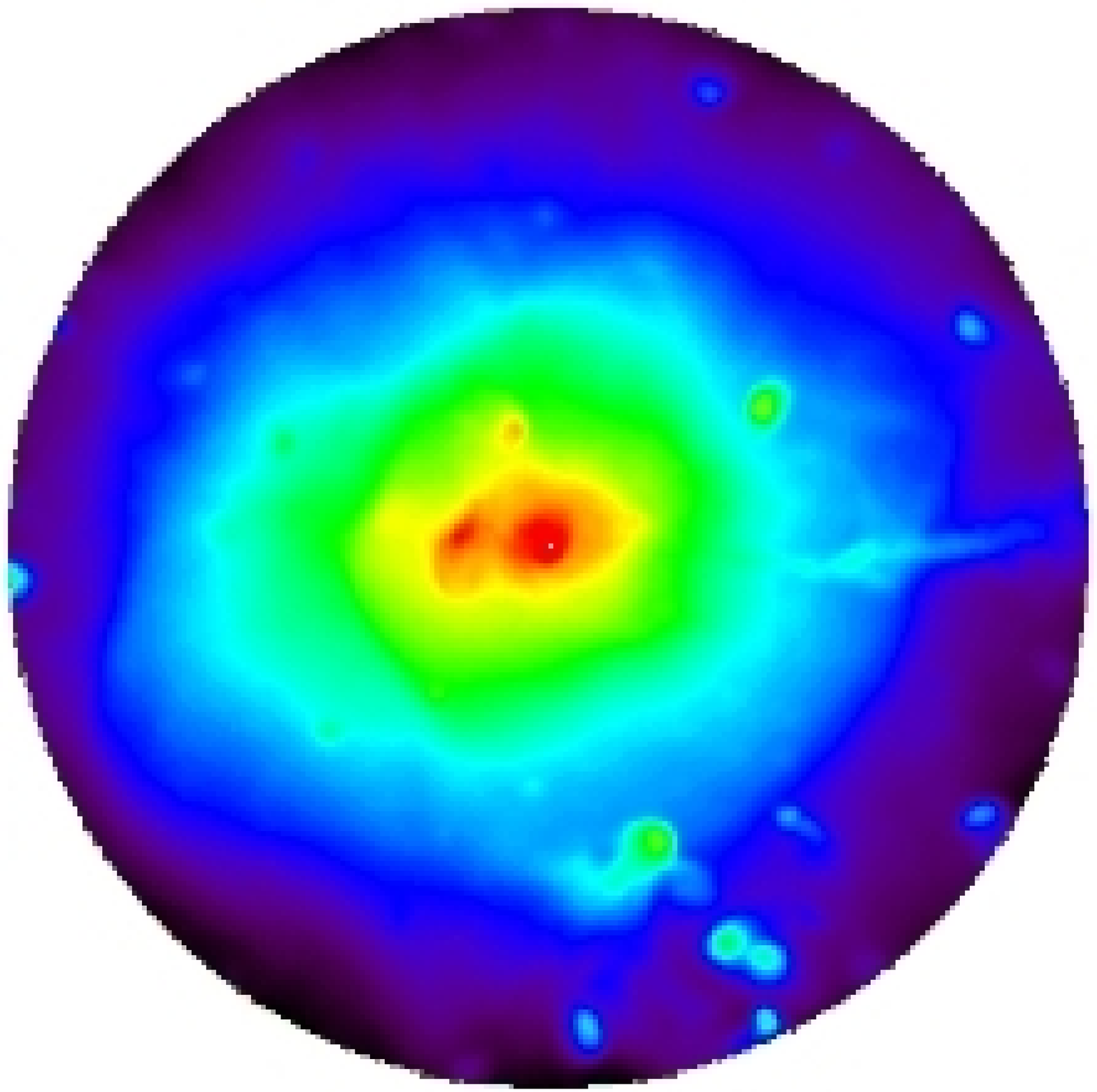}
  \hspace{0.005\textwidth}
  \includegraphics[width=0.23\textwidth]{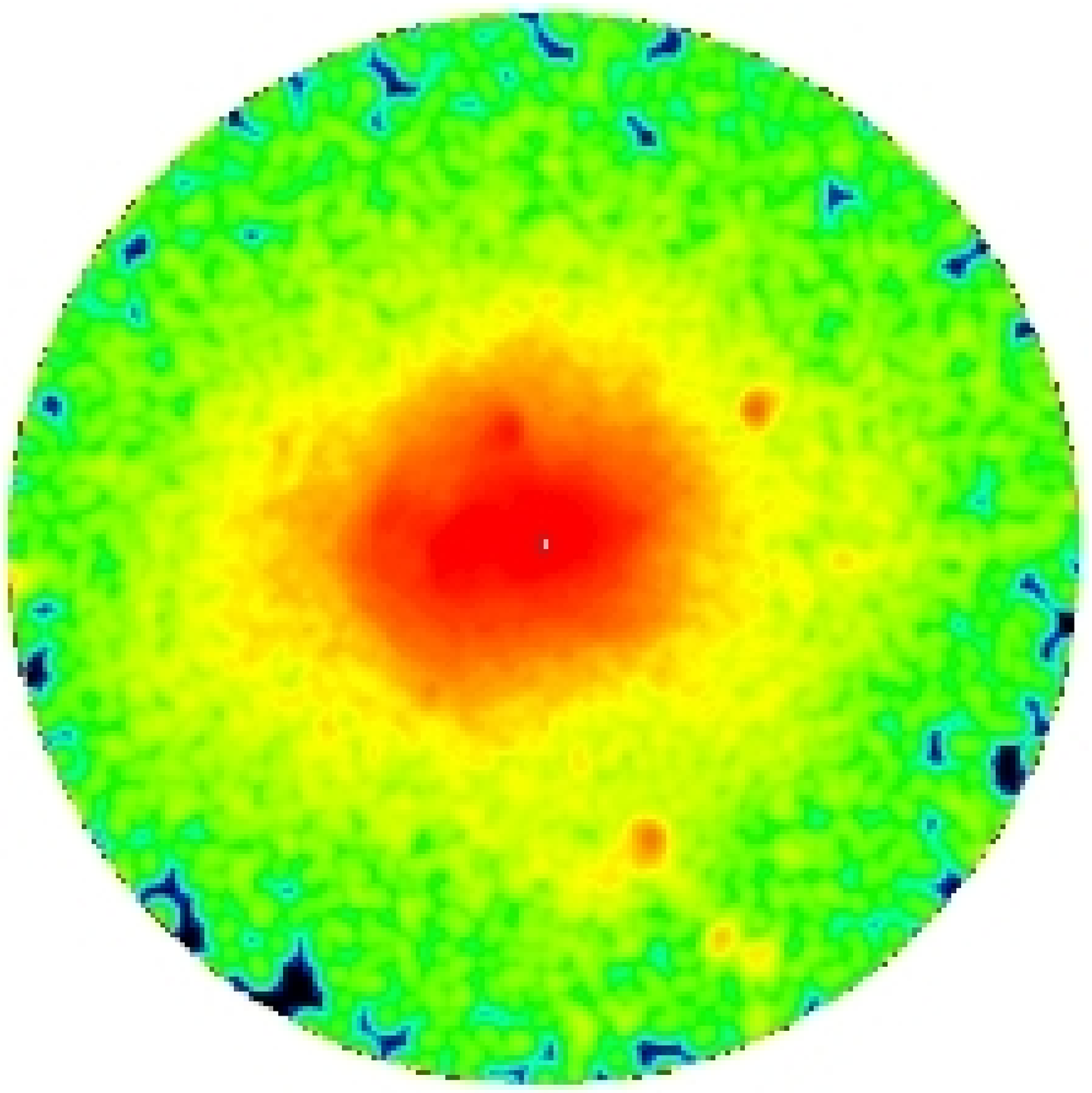}

\caption{Images of cluster A at $z\approx0$. First row: Projected mass map with resolution of $15 h^{-1}$kpc (left) and mass map with Gaussian noise and resolution degraded $100 h^{-1}$Kpc (right). Second row: X-ray surface brightness image with resolution of $15 h^{-1}$kpc (left) and X-ray surface brightness image with $25 h^{-1}$kpc resolution and Poisson noise (right). Note that the noisy X-ray image would typically be more heavily smoothed for presentation purposes. See text for details of the noise models. This image is featured in Case Study 3, Fig.\ref{gaserased}.}

\label{realistic}

\end{figure}

\subsubsection{Introducing noise and degrading the resolution.}
For the purposes of adding noise and adopting a realistic resolution, we opt to place the clusters at $z=0.2$ as this is both the redshift at which our X-ray map resolution can be achieved by {\it XMM} and is also a redshift representative of recent observations (the bullet cluster  is at $z\approx0.3$  \citep{bulletclowe} and Abell 520 is at $z\approx0.2$  \citep{2007ApJ...668..806M}). 

As described in Section \ref{ussec} maps of both types were first smoothed with a Gaussian kernel with FWHM $=15 h^{-1}$kpc, equal to the spatial resolution of the simulations. While this resolution is potentially achievable in X-ray observations, it is necessary for us to increase the smoothing slightly (to FWHM $=25 h^{-1}$kpc) in the presence of noise, but note that this resolution is still much higher than currently achievable with weak lensing analyses. In this case, the resolution that can be obtained is dependent on the number density of background galaxies: for ground-based data, this angular resolution is typically $1$ arcminute, yet for space-based data this can be improved to around $45$ arcseconds \citep[see][for example]{heymans_superclus}. To investigate the impact of this decreased resolution, we now adopt a preliminary Gaussian kernel with FWHM$=100 h^{-1}$kpc  when analysing the projected mass maps (corresponding to approximately $45\arcsec$ angular resolution at $z=0.2$) and increase $\sigma_{2}$ accordingly. 

We add Poisson noise to the X-ray maps by making the crude approximation that the photon number is proportional to the X-ray surface brightness. We find that $10^{5}$ photons (corresponding roughly to an exposure of $20$ ks) allows us to recover the majority of substructures that were detected in the absence of noise. We add Gaussian noise to the mass maps with zero mean and with a variance determined by \citet{vw2000}. The latter is given by,

\begin{eqnarray}
\nonumber \Delta M^{2} & = & 4\sigma^{4}(\frac{\sigma_{\epsilon}^{2}}{4\pi\sigma^{2}n_{\rm g}}) \Sigma_{\rm crit}^{2} \\
\nonumber & & [1-\exp{(-\frac{a^{2}}{2\sigma^{2}})} - \sqrt{\frac{\pi}{2}}\frac{a}{\sigma} \erf{(\frac{a}{\sqrt{2}\sigma})}]^{2} \\
\label{wlvar}
\end{eqnarray}

 for a pixel of size $a$ in a weak lensing mass reconstruction and is due to the intrinsic ellipticities (with rms, $\sigma_{\epsilon}$) of background galaxies with an average density of $n_{\rm g}$. $\Sigma_{\rm crit}$, the critical surface density for lensing to occur, is given by,

\begin{equation}
\Sigma_{\rm crit}= \frac{c^{2}}{4\pi G} \frac{D_{\rm os}}{D_{\rm ol},D_{\rm ls}}
\end{equation}

where $D_{\rm ol}$ is the angular diameter distance between the observer and the lens, $D_{\rm os}$ is that between the observer and the galaxies and $D_{\rm ls}$ is that between the lens and the galaxies. Note that we fix the redshift of the lens to be $z=0.2$, as outlined above. We use typical values of $n_{\rm g}= 100$ galaxies ${\rm arcmin}^{-2}$ (for space-based data) and $\sigma_{\epsilon} = 0.3$ \citep[e.g.][]{starcketal06} and,  we assume that the galaxy ellipticities were smoothed with a Gaussian of standard deviation $\sigma=a$ prior to reconstruction, as in \citet{P&B2007}. We note that the pre-smooth damps the noise a little, but our aim in this preliminary investigation into the impact of noise is simply to make an estimate of the noise level.

Fig.~\ref{realistic} illustrates how the original maps (left-hand column) are affected by the smoothing and addition of noise (right-hand column). While some small substructures are still visible in the X-ray map (albeit made less distinct by the noise), all but the largest substructure has been erased from the mass map. 

\subsubsection{The impact of noise and resolution.}

We now review our main findings in order to make a preliminary assessment of how they are affected by noise and degraded map resolution. 

First, we re-evaluate the relationship between the 2D mass map substructures and the underlying distribution of 3D subhaloes. The impact of just degrading the map resolution is significant and reduces the number of subhaloes detected above $10^{12} h^{-1} \rm{M}_{\odot}$ to around 60 per cent of its value in the original maps. When noise is also included, there are further detection failures, most frequently below $5\times10^{12} h^{-1} \rm{M}_{\odot}$, which reduce the total number of detections above $10^{12} h^{-1} \rm{M}_{\odot}$ by an additional 5 per cent.

\begin{figure}
\centering
\includegraphics[width=0.5\textwidth]{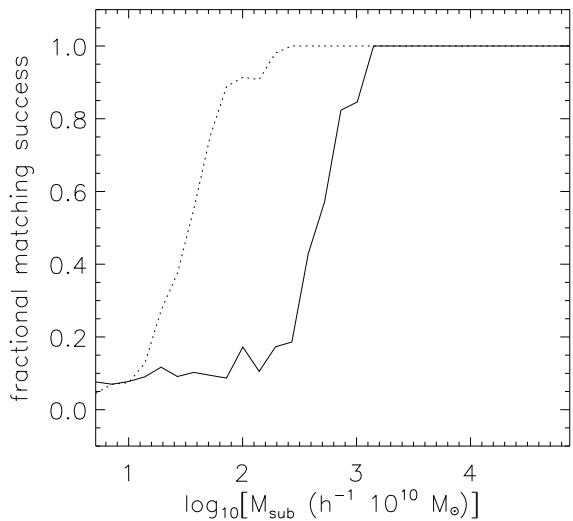}
\caption{Fractional matching success {\it per mass bin} of 3D subhaloes to 2D substructures in the noisy, degraded resolution mass maps (solid) and the original mass maps (dotted) as a function of 
subhalo DM mass for $3\sigma$ 2D catalogue. Bins are equally spaced in $\log(M_{\rm sub})$.}
\label{successmass_obs}
\end{figure}

Fig.~\ref{successmass_obs} compares the completeness of the $3\sigma$ 2D mass substructure catalogue obtained from the degraded resolution, noisy maps (solid line) with that obtained from our high resolution, noise-free maps (dotted line, taken from Fig.~\ref{successmass}). We can see that the mass threshold for $90$ per cent completeness (per mass bin) is now around an order of magnitude higher. For the $3\sigma$ catalogue, $90$ per cent completeness is now achieved only above $10^{13} h^{-1} \rm{M}_{\odot}$ and a factor $\approx 8$ fewer substructures are detected in total (159 {\it cf.} 1233). 

\begin{figure}
\centering
\includegraphics[width=0.5\textwidth]{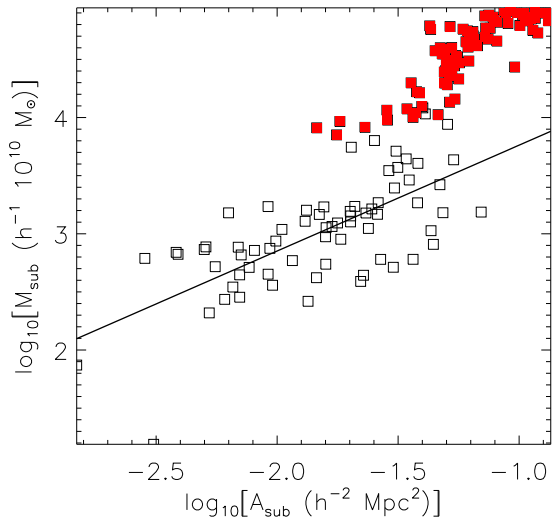}
\caption{Correlation between physical area, $A_{\rm sub}$, of 2D substructure in noisy, degraded resolution mass map ($3\sigma$ catalogue) and DM mass of 3D subhalo, $M_{\rm sub}$, to which it is matched. Filled squares show {\sc subfind} background 
haloes (see text for details). See text for slope and normalisation of best-fitting line.}
\label{mafig_obs}
\end{figure}

Despite the impact of noise, a correlation between the area of the 2D substructure, $A_{\rm sub}$, and the DM mass of its 3D subhalo counterpart, $M_{\rm sub}$, is still evident in Fig.~\ref{mafig_obs}. The reduction in the number of detections (immediately apparent in Fig.~\ref{mafig_obs}) translates into higher $1\sigma$ errors on the best-fitting line. The purity of the $3\sigma$ substructure catalogue is very high (only $1$ per cent are false detections), so we do not define a purity threshold here, but simply include all substructures in the fit. 

The normalisation and slope are now $4.7\pm0.2$ and $0.9\pm0.1$, respectively. The change in the former is most significant and can be attributed to the use of a larger kernel, resulting in a given 2D substructure having a larger area than before.

We also re-examine the likelihood of finding an X-ray counterpart for all 2D mass map substructures in the $3\sigma$ catalogue (the results for our fiducial data set are presented in Fig.~\ref{successvmass_gas_3d}). Above $10^{13} h^{-1} \rm{M}_{\odot}$ (our completeness limit) we now find X-ray counterparts in the $1\sigma$ catalogue for all of the 2D mass map substructures. Using the $3\sigma$ X-ray catalogue, however, we fail to find matches for a few high mass mass map substructures. The difference between the two X-ray catalogues here arises in situations where there is hot gas in the vicinity of the mass substructure, but it has been disrupted and so does not have a defined peak; the $1\sigma$ catalogue detects this whereas the $3\sigma$ does not. 

The other scenarios in which we no longer find matching failures (but did previously) is an effect of the reduced mass map resolution. At low subhalo masses we are now unable to resolve the 2D mass map substructure. At high masses, matching failures typically occurred in complex merging cores which now, in some cases, cannot be individually resolved. This can result in two merging cores being detected as one extended mass substructure, facilitating matching with an X-ray substructure. This is an issue that requires further investigation. The current criterion for a match is any degree of overlap between the mass and X-ray substructure. Since substructures in the realistic mass map substructures have a much greater spatial extent due to the lower resolution, they could be associated with an X-ray substructure which is significantly offset from the mass peak. A more detailed follow-up of X-ray-mass matches in the context of more realistic maps would be an interesting extension to current work.

While we have shown that only a few discrepancies between substructure in the X-ray and the mass maps could be observed currently, our fiducial results show there is an abundance of these to be uncovered. A detailed substructure comparison, such as the one undertaken here, will yield a wealth of interesting results when predicted improvements in lensing mass map resolution are achieved. For example, a resolution of $\sim 10 h^{-1}$kpc is forecast by \citet{whitepaper_coe} based on a novel strong lensing analysis technique.

\subsection{Summary of Discussion}

In this Section, we have discussed in detail the reasons for failing to find an X-ray counterpart for all of our 2D mass substructures. We have demonstrated two distinct scenarios that give rise to a 2D total mass substructure not being matched to a 2D X-ray surface brightness substructure: spatial separation of the X-ray component and destruction (or disruption) of the X-ray component. We have also highlighted the dependence of the matching procedure on choice of X ray catalogue.

We have examined how several factors affect the likelihood of finding X-ray counterparts for substructures in the total mass maps.  The inclusion of high-redshift cooling in the simulations does not have a dramatic effect on the correspondence between X-ray and total mass. We show there is a higher probability of finding an X-ray counterpart at high redshift, which can be attributed to a shorter time-scale on which ram pressure stripping could occur. By dividing our sample based on dynamical state we find subhaloes with $M_{\rm sub}>3\times10^{13} h^{-1} {\rm M}_{\odot}$ only lack an X-ray counterpart when the cluster is highly disturbed (in agreement with recent observations), however, relaxed clusters exhibit many deviations from the basic picture that light traces mass at lower subhalo masses. Joint weak lensing and X-ray analyses of relaxed systems are therefore also valuable and will yield much information about the physics of the ICM. 

A review of our main results in the presence of observational noise and degraded resolution reveals many of these interesting mismatch scenarios are not currently observable, yet predicted improvements in lensing mass map resolution suggest these will be revealed in the coming decade, unveiling frequent deviations from the simple assumption that light traces mass.

\section[]{Conclusions}\label{conclusionsec}

In this paper, we have used resimulations of three non-radiative galaxy clusters in order to investigate the discrepancies between substructure in the hot gas and DM components, evident from recent comparisons of X-ray and weak lensing observations. We developed a simple technique to detect 2D substructures in simulated surface mass density and X-ray surface brightness maps of the clusters, without any reliance on circular symmetry or dynamical state. The resulting catalogues of 2D mass and 2D X-ray substructures were matched and we investigated how the success of this matching procedure varied with redshift, dynamical state and choice of gas physics employed. By utilising information about the underlying 3D subhalo distribution (obtained with {\sc subfind}) we have assigned subhaloes to the 2D mass substructures, allowing us to characterise the efficiency of our 2D substructure detection technique and reveal the effect of subhalo mass on the 2D mass to 2D X-ray substructure matching success. 

Our main results can be summarised as follows:

\begin{itemize}
\item Having undertaken a thorough assessment of the properties of the 2D substructure catalogues resulting from our novel detection procedure, we have ensured that any selection effects or biases the technique may have introduced are understood. By attempting to match all 2D substructures detected in the surface mass density map with the 3D subhaloes (identified with {\sc subfind}), we have concluded that our 2D substructure catalogue is $90$ per cent complete per mass bin ($98$ per cent overall) down to a 3D subhalo DM mass of $\sim10^{12} h^{-1} {\rm M}_{\odot}$ and $100$ per cent complete down to a DM mass of $10^{13} h^{-1} {\rm M}_{\odot}$. We are confident therefore that, in the 3D subhalo mass range currently probed by weak lensing, the 2D substructure catalogues provide an accurate representation of the true 3D picture. We also establish that the 2D mass substructure catalogue is pure and complete for $A_{\rm sub,TM}>10^{-3}$ $h^{-1}$ Mpc, i.e. all 2D mass substructures with areas above this limit are successfully matched to a 3D subhalo and are, therefore, genuine. This purity threshold should allow the same detection procedure to be reliably applied to other simulated surface mass density maps in future, without the need for 3D subhalo data with which to compare.

\item We present a correlation between $A_{\rm sub,TM}$, the area of a 2D mass substructure, and $M_{\rm sub}$, the DM mass of the 3D subhalo to which it is matched. The correlation is still apparent upon the introduction of basic observational noise, suggesting it could provide a quick estimate of the mass of a subhalo responsible for a peak in a weak lensing mass reconstruction, after accurate calibration. A measurement of the intrinsic scatter suggests such an estimate would be out by a factor of $\sim2$. 

\item The results of the matching between 2D mass substructures and 2D X-ray substructures are surprising. We do not find X-ray counterparts for $23-33$ per cent (depending on choice of X-ray catalogue) of all 2D mass substructures in the pure catalogue. Below $M_{\rm sub}\sim 10^{13}h^{-1} {\rm M}_{\odot}$ the matching success per mass bin begins to decrease significantly with decreasing subhalo mass. For the $1\sigma$ X-ray catalogue, a few per cent of cluster cores, $5$ per cent of group-size 2D mass substructures and $35$ per cent of galaxy-size 2D mass substructures are not associated with a 2D X-ray substructure. The reasons for a matching failure are: 1) displacement of hot gas, where the X-ray substructure is intact yet spatially distinct from the DM, 2) depletion of hot gas, where so much gas has been stripped that detection of the 2D X-ray substructure fails or 3) complete disruption of the hot gas, where all hot gas appears to have been removed such that no 2D X-ray substructure is evident, even on visual inspection. We have conducted a detailed follow-up of examples of these scenarios with a set of case studies.

\item The dynamical state of the clusters (characterised by measuring the centroid shift variance in the X-ray surface brightness maps), is found to play a role in determining the fraction of 2D mass substructures without X-ray counterparts. Substructures with $M_{\rm sub}> 3\times10^{13} h^{-1} {\rm M}_{\odot}$ without an X-ray counterpart are restricted to the disturbed sample, suggesting major merger events are the cause. Substructures below this mass are less likely to have X-ray counterparts in relaxed systems, suggesting ram pressure stripping plays an important role on this scale; by definition, a long time has elapsed since the last merger, so these substructures have had most time to experience its effects. Similarly, the low redshift sample ($0 \leq z \leq 0.2$) contains more 2D mass substructures, in this mass range, that are unmatched to X-ray substructures than the high redshift sample ($0.5 \leq z <1$).

\item The inclusion of high-redshift (until $z\simeq5$) cooling has only a mild impact on our results. It has little effect on the matching success between 2D mass substructures and 2D X-ray substructures for $M_{\rm sub}> 3\times10^{12} h^{-1} {\rm M}_{\odot}$, but reduces it, compared to the non-radiative case, below this due to the reduction of hot gas in these objects.

\item We have demonstrated that our simple 2D detection technique is still successful when noise which approximates that in real observations is added to the maps and the map resolution is degraded. As could be expected, the subhalo mass at which high completeness is achieved for the mass substructure catalogues is around an order of magnitude higher than in the fiducial set of maps. We have shown that this increase means many of the interesting mismatches which occur at lower mass scales cannot currently be observed. If the resolution of lensing mass maps can be improved by a factor of 10, to $\sim 10 h^{-1}$kpc, we predict that many more discrepancies between the hot gas and dark matter components of clusters will be observed and that these will not be restricted to rare, extreme merger events such as the Bullet cluster. Such an improvement, while dramatic, has been predicted for the coming decade and authors are already developing new observational analysis techniques to allow this, such that comparisons in the spirit of the present work can be undertaken \citep{whitepaper_coe}. These future observations will provide a wealth of information about the physics of the ICM, the dynamical state of galaxy clusters and will allow us to probe the properties of the DM substructure directly.

\end{itemize}

In future work, we will assess the impact of introducing heating processes, e.g. from galactic winds and the effects of active galactic nuclei into the resimulations, as well as the effect of including more realistic noise in the maps.

\section[]{Acknowledgments}

We are grateful to the referee for helpful comments that improved this paper. We thank Volker Springel for providing the new version of {\sc subfind} we use here. LCP was supported by an STFC studentship. AB would like to acknowledge the Levehulme Trust for awarding him a Leverhulme Visiting Professorship during the tenure of which this study was initiated.   He is also grateful to Astrophysics at the University of Oxford for hospitality during this period.   This work has been partially funded a NSERC (Canada) Discovery Grant to AB. 
\bibliographystyle{mn2e}

\label{lastpage}

\end{document}